\documentclass[prx,aps, twocolumn, amsmath,amssymb,superscriptaddress,longbibliography]{revtex4-2}

\usepackage{graphicx}
\usepackage{dcolumn}
\usepackage{bm}
\usepackage[sort&compress]{natbib}
\usepackage{xspace}
\usepackage{amsmath}
\usepackage{epstopdf}
\usepackage{amssymb}
\usepackage{epsfig}
\usepackage{color} 
\usepackage[dvipsnames]{xcolor}
\usepackage[english]{babel} 
\usepackage{siunitx} 
\usepackage{upgreek} 
\usepackage{soul} 

 \newcommand{\ez}[1]{\textcolor{OliveGreen}{#1}}

\begin{document}

\title{Microwave-to-optical transduction using a mechanical supermode for coupling piezoelectric and optomechanical resonators}
\author{Marcelo Wu}\email{marcelo.wu@nist.gov}
\affiliation{Physical Measurement Laboratory, National
Institute of Standards and Technology, Gaithersburg, MD 20899,
USA}\affiliation{Department of Chemistry and Biochemistry, University of Maryland,
College Park, MD 20742, USA}

\author{Emil Zeuthen}\email{zeuthen@nbi.ku.dk}
\affiliation{Niels Bohr Institute, University of Copenhagen, 2100 Copenhagen, Denmark}

\author{Krishna Coimbatore Balram}\email{krishna.coimbatorebalram@bristol.ac.uk}
\affiliation{Department of Electrical and Electronic Engineering, University of Bristol, Bristol BS8 1UB, United Kingdom}

\author{Kartik Srinivasan} \email{kartik.srinivasan@nist.gov}
\affiliation{Physical Measurement Laboratory, National
Institute of Standards and Technology, Gaithersburg, MD 20899, USA}
\affiliation{Joint Quantum Institute, NIST/University of Maryland, College Park, MD 20742, USA}

\date \today

\begin{abstract}
\noindent The successes of superconducting quantum circuits at local manipulation of quantum information and photonics technology at long-distance transmission of the same have spurred interest in the development of quantum transducers for efficient, low-noise, and bidirectional frequency conversion of photons between the microwave and optical domains. We propose to realize such functionality through the coupling of electrical, piezoelectric, and optomechanical resonators. The coupling of the mechanical subsystems enables formation of a resonant mechanical supermode that provides a mechanically-mediated, efficient single interface to both the microwave and optical domains. The conversion process is analyzed by applying an equivalent circuit model that relates device-level parameters to overall figures of merit for conversion efficiency $\eta$ and added noise $N$. These can be further enhanced by proper impedance matching of the transducer to an input microwave transmission line. The performance of potential transducers is assessed through finite-element simulations, with a focus on geometries in GaAs, followed by considerations of the AlN, LiNbO$_3$, and AlN-on-Si platforms. We present strategies for maximizing $\eta$ and minimizing $N$, and find that simultaneously achieving $\eta>50~\%$ and $N < 0.5$ should be possible with current technology. We find that the use of a mechanical supermode for mediating transduction is a key enabler for high-efficiency operation, particularly when paired with an appropriate microwave impedance matching network. Our comprehensive analysis of the full transduction chain enables us to outline a development path for the realization of high-performance quantum transducers that will constitute a valuable resource for quantum information science.
\end{abstract}

\maketitle

\setcounter{figure}{0}
\makeatletter
\renewcommand{\thefigure}{\@arabic\c@figure}

\setcounter{equation}{0}
\makeatletter
\renewcommand{\theequation}{\@arabic\c@equation}

\section{Introduction}
\label{sec:Intro}

\begin{figure*}
\includegraphics[width=0.7\linewidth]{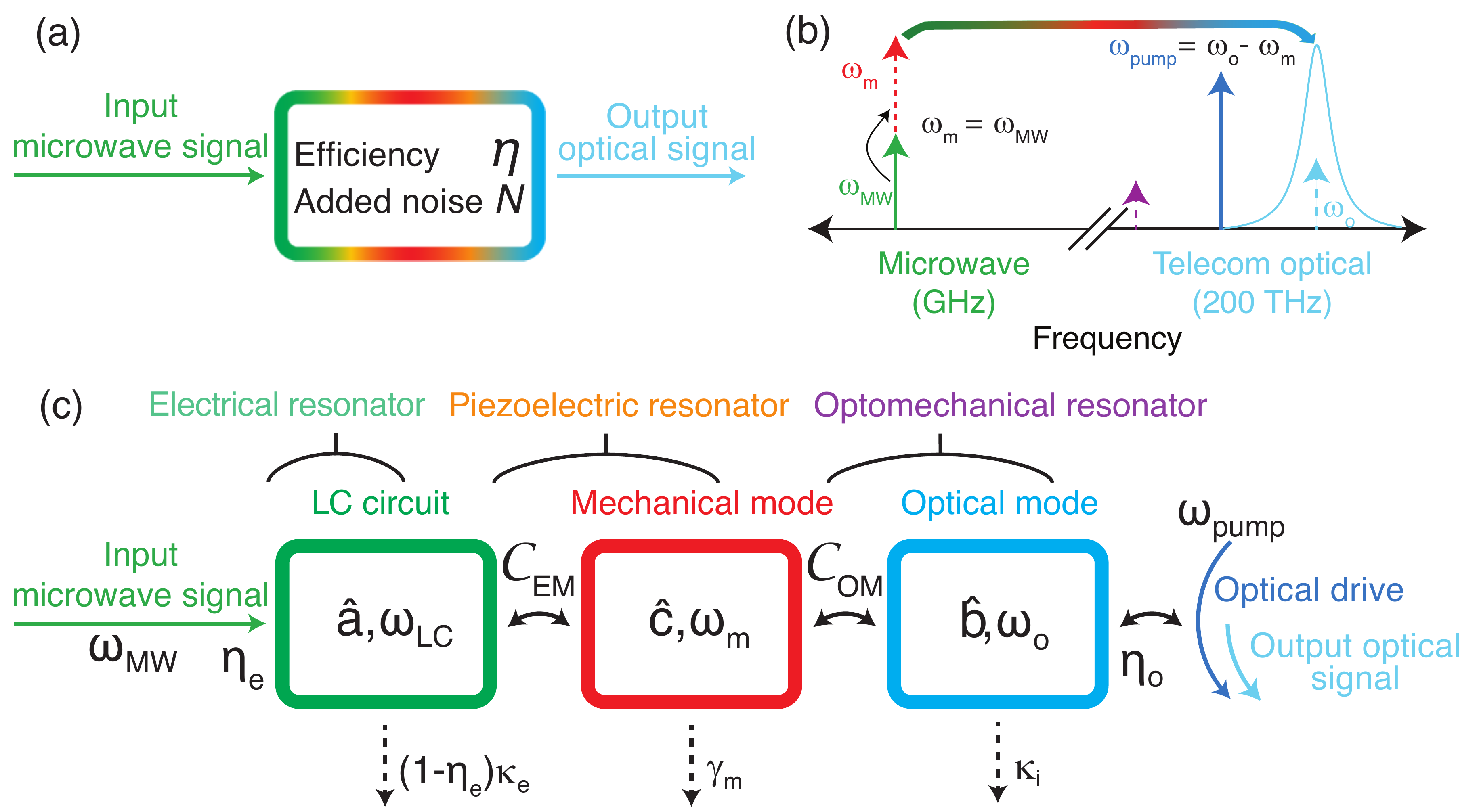}
\caption{(a) General schematic for microwave-to-optical quantum transduction with efficiency ($\eta$) and added noise ($N$) as figures of merit. (b) Frequency-domain depiction of the transduction process, where an input microwave signal on resonance with the mechanical mode of a piezo-optomechanical resonator (so that $\omega_\text{MW}=\omega_\text{m}$) is up-converted to the optical domain through an optical pump at frequency $\omega_\text{pump}$.  The pump is red detuned with respect to the optical mode (frequency $\omega_\text{o}$) of the piezo-optomechanical resonator so that $\omega_\text{pump}=\omega_\text{o}-\omega_\text{m}$. The optomechanical interaction creates upper (dashed blue arrow) and lower (dashed purple arrow) sidebands with respect to the pump; the target output signal for quantum transduction is the upper sideband, which is filtered by the optical mode .  (c) Block diagram of the piezo-optomechanical transduction process, indicating the electrical, mechanical, and optical modes, relevant frequencies ($\omega_\text{MW}$, $\omega_\text{LC}$, $\omega_\text{m}$, $\omega_\text{o}$, $\omega_\text{pump}$), decay channels ($(1-\eta_e)\kappa_\text{e}$, $\gamma_\text{m}$, $\kappa_{i}$), cooperativities ($\mathcal{C}_\text{EM}$, $\mathcal{C}_\text{OM}$), and electrical and optical coupling efficiencies ($\eta_\text{e}$ and $\eta_\text{o}$). While the schematics in (a)-(c) indicate conversion from a microwave input to an optical output, the system is bidirectional, and frequency down-conversion is also possible. }
\label{fig:schematic}
\end{figure*}

\noindent Quantum information science requires a wide range of physical resources to store, manipulate, process, and transmit quantum states. Superconducting quantum circuits operating at microwave (MW) and radio frequencies (rf) have made great strides in quantum computation~\cite{clarke_superconducting_2008,devoret_superconducting_2013}, while systems based on optical-wavelength photons are the dominant approach for quantum communication~\cite{nicolas_gisin_quantum_2007,yin_satellite-based_2017}. As a result, there has been significant interest in connecting microwave (or RF, alternatively) and optical domains with high efficiency $\eta$ and low added noise $N$ (Fig.~\ref{fig:schematic}(a)) to enable, for example, distributed quantum computing and quantum networks based on superconducting quantum nodes~\cite{regal_cavity_2011,safavi-naeini_proposal_2011}. This approach is also a key enabling method for low-noise optical detection of weak microwave signals~\cite{bagci_optical_2014}, e.g., in the context of nuclear magnetic resonance~\cite{Takeda_Electro_2018,Simonsen2019a,Simonsen2019b}.

While coherent interfaces between the microwave and optical domains already exist, for example, using telecommunication electro-optic modulators~\cite{yariv_photonics_2006, zhang_broadband_2019}, their transduction efficiency is likely too small to be of practical benefit for quantum applications. To address this challenge, many approaches have been explored~\cite{lambert_coherent_2019}, based on doubly enhanced electro-optics~\cite{Javerzac-Galy_chip_2016,rueda_efficient_2016,Soltani_Efficient_2017, fan_superconducting_2018}, magneto-optic effects in doped crystals~\cite{williamson_magneto-optic_2014}, ferromagnetic magnons~\cite{hisatomi_bidirectional_2016}, and mechanically mediated processes~\cite{regal_cavity_2011,safavi-naeini_proposal_2011,Wang_quantum-state-transfer_2012,wang_using_2012_njp,Tian_Optoelectromechanical_2014}.  The latter approach utilizes phonons as an intermediary that can couple to both microwave and optical photons. One implementation of this mechanics-based approach is a thin membrane that capacitively couples an electromechanical (EM) circuit to the optical field in a Fabry-Perot cavity. This has proven to be quite effective, with up to $47\,\%$ conversion efficiency and as few as 38 added noise photons demonstrated~\cite{Higginbotham_Harnessing_2018}. So far, this type of approach has only been demonstrated using planar MHz-frequency electromechanics coupled to hand-assembled free-space optical cavities. In parallel, fully chip-integrated versions with mechanical frequencies in the 100 MHz to 10 GHz range are being developed~\cite{Midolo_Nano_2018,davanco_slot-mode-coupled_2012, grutter_slot-mode_2015, fink_quantum_2016}.

Piezoelectric platforms provide another approach for mechanically mediated microwave-to-optical conversion~\cite{zou_cavity_2016}. Piezoelectric devices such as filters based on interdigitated transducers~\cite{campbell_surface_1998} and film bulk acoustic resonators~\cite{ruby_review_2007, Han_Multimode_2016, valle_high-frequency_2019} directly couple GHz-frequency electromagnetic and acoustic waves. These GHz-frequency acoustic modes have a micrometer-scale wavelength consistent with the localization scale of the optical mode in highly confined nanophotonic resonators~\cite{bochmann_nanomechanical_2013, fong_microwave_2014, vainsencher_bi_2016, balram_coherent_2016, forsch_microwave_2018}. The mechanical vibration can then modulate the optical signal via the optomechanical (OM) interaction (Fig.~\ref{fig:schematic}(b)).  This has led to the realization of several integrated platforms combining piezoelectricity and optomechanics~\cite{bochmann_nanomechanical_2013, fong_microwave_2014, vainsencher_bi_2016, balram_coherent_2016, forsch_microwave_2018}. Building an efficient piezo-optomechanical transducer requires optimization of each step of the conversion process from electrical to mechanical to optical, with emphasis on effective interactions between each element and low losses. However, conversion efficiencies in recent demonstrations have been low, due to factors such as weak piezoelectric coupling, geometric size and impedance mismatch between acoustic elements, inefficient single-pass electroacoustic transfer, and a host of other design, technical, and material difficulties.

Here, we propose a piezo-optomechanical approach that overcomes many of the aforementioned challenges by mediating transduction through a mechanical supermode that results from coupling piezoelectric and optomechanical resonators. A supermode is formed by the hybridization of two modes of the system when their resonant frequencies are near to each other (see Ref.~\cite{Soltani_Efficient_2017} for an example of an optical supermode). This piezo-optomechanical transducer combines the low insertion loss of piezoelectric resonators~\cite{piazza_piezoelectric_2006,gong_design_2013,olsson_high_2014,Wang_Design_2015} with the large optomechanical coupling exhibited by nanoscale cavity-optomechanical resonators~\cite{chan_optimized_2012}. The combined enhancement of both resonators alongside strong mechanical interaction between the two integrated subsystems opens the door toward efficient and reversible coherent transfer of quantum states. The addition of an electrical impedance-matching network further enhances the efficiency by tuning the electromechanical interaction to match that of the optomechanical system. Moreover, the resonant signal enhancement provided by the matching circuit serves to diminish the relative size of the mechanical thermal noise. In practice, however, the coordination, performance, and matching of all the elements into an efficient and low-noise transducer is difficult, and the ensuing design trade-offs are a central topic of this paper.

Our proposal and supporting theory are discussed in the sections below. In Section~\ref{sec:Coupled_resonator_approach}, we outline the basic coupled resonator concept. In Section~\ref{sec:Piezo_optomechanical_transduction}, we apply the equivalent circuit analysis of optoelectromechanical systems proposed in Ref.~\cite{Zeuthen_Electrooptomechanical_2018} to establish formulas for key metrics based on physical parameters that characterize the component elements. After laying down the theoretical groundwork for transduction efficiency $\eta$, added noise $N$, and conversion bandwidth $\Delta \omega$, two optimization scenarios are addressed: maximizing $\eta$ (Section~\ref{sec:maxEta}) and minimizing $N$ (Section~\ref{sec:minN}). We then present in Section~\ref{sec:ApplicationMaterials} finite-element simulations of coupled piezoelectric and optomechanical resonator geometries in GaAs, from which we extract estimates for device-level physical parameters such as the piezoelectric and optomechanical couplings. This information is combined with recent data from experiments on GaAs optomechanical crystals operating at $T<100$~mK~\cite{forsch_microwave_2018,ramp_elimination_2018} to yield estimates of $\eta$ and $N$. We discuss these metrics in terms of what is currently achievable in practice and what advances need to be realized to improve performance.  Within this context, we also consider the potential of stronger piezoelectric material systems such as AlN, LiNbO$_3$, and AlN on Si.

\section{Coupled resonator approach}
\label{sec:Coupled_resonator_approach}

\begin{figure*}
\includegraphics[width=0.8\linewidth]{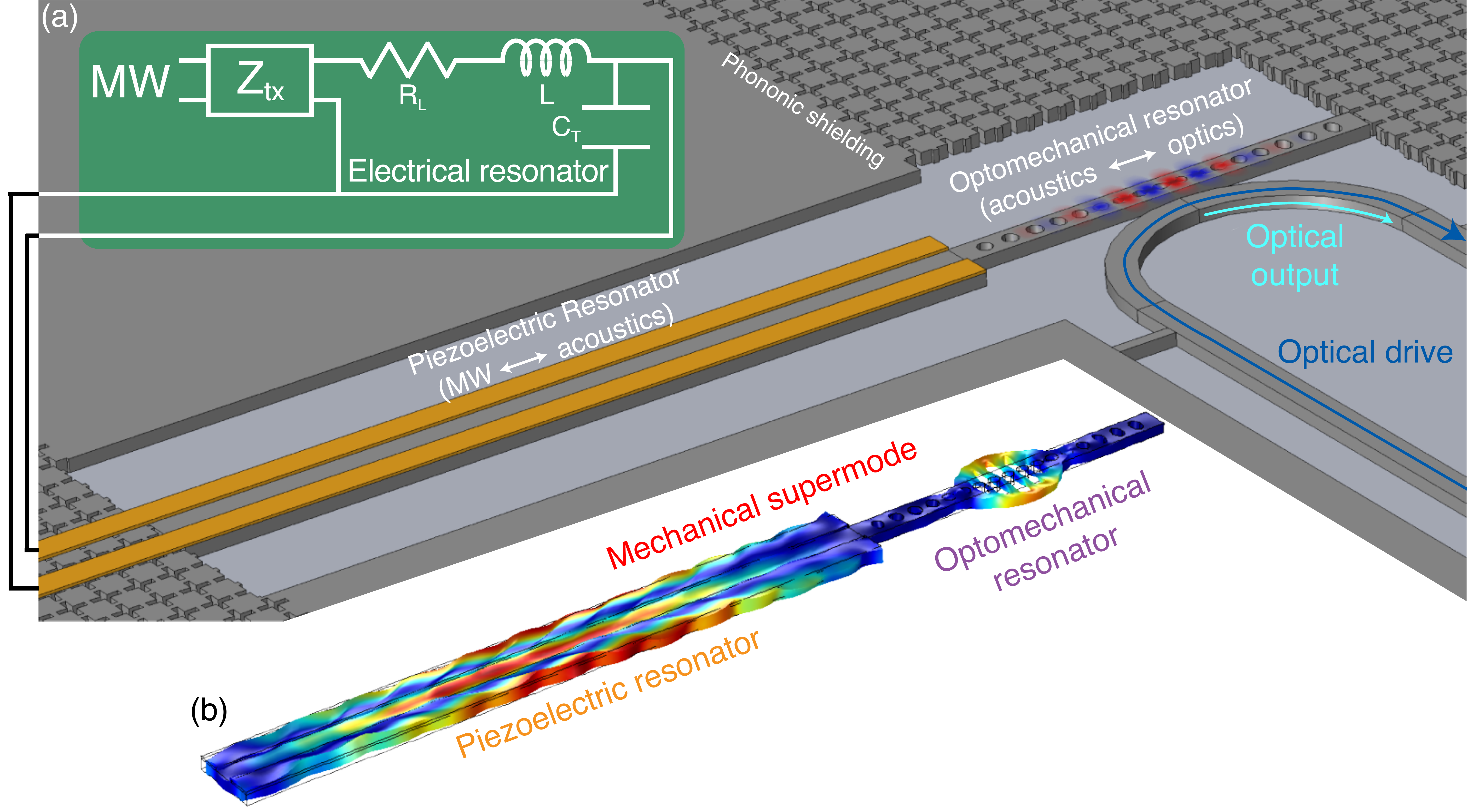}
\caption{Representative coupled piezoelectric and optomechanical resonator system. (a) Illustration of a physical implementation of the piezo-optomechanical transducer showing the direct coupling of the piezoelectric resonator (bottom left) to the optomechanical resonator (top right).  The gray region corresponds to a suspended GaAs layer while the gold traces depict the metal-electrode configuration, which is connected to an $LC$ circuit shown in the green box as a schematic. Finite-element-method simulations show the optical field (overlaid on top of the optomechanical resonator) confined in the photonic crystal nanobeam-cavity portion of the transducer, as well as (b) the mechanical displacement profile for a hybrid supermode of the coupled resonator device. }
\label{fig:coupled_res_example}
\end{figure*}

\noindent Figure~\ref{fig:schematic}(c) presents an overview of the microwave-to-optical transduction scheme. Briefly, an input microwave signal at frequency $\omega_{\text{MW}}$ is coupled into an $LC$ circuit with resonant frequency $\omega_{\text{LC}} = \omega_{\text{MW}}$ and coupling efficiency $\eta_{\text{e}}$. Embedded in the $LC$ circuit is the piezo-optomechanical transducer, which has a mechanical frequency $\omega_{\text{m}}$ that is equal to the input microwave field, $\omega_{\text{m}}=\omega_{\text{MW}}$. The mechanical excitation driven by the input microwave field is up-converted to the optical domain ($\omega_{\text{pump}}+\omega_{\text{MW}}$) using an optical drive at frequency $\omega_{\text{pump}}$. For low-noise quantum transduction applications, $\omega_{\text{pump}}$ is typically red detuned from the resonant frequency $\omega_{\text{o}}$ of the optical cavity by $\omega_{\text{o}}-\omega_{\text{pump}}=\omega_{\text{m}}$ in order to enhance the optomechanical beam-splitter interaction associated with the upper sideband (light blue arrow in Fig.~\ref{fig:schematic}(b)) while suppressing the unwanted amplification effects from the two-mode-squeezing interaction of the lower sideband. Sideband-scattered intracavity photons at $\omega_{\text{o}}$ are finally outcoupled into an output optical waveguide with efficiency $\eta_{\text{o}}$.

Figure~\ref{fig:coupled_res_example} shows an example of the piezo-optomechanical transducer geometry we propose in this paper. The transducer (Fig.~\ref{fig:coupled_res_example}(a)) consists of a piezoelectric resonator that is directly coupled to a nanobeam optomechanical crystal resonator. The mechanical coupling between these two resonators can be made sufficiently large so as to hybridize their mechanical modes, resulting in an effective mechanical supermode that can be coupled to, piezoelectrically and optomechanically (see Appendix~\ref{sec:piezo_modelling}).  Figure~\ref{fig:coupled_res_example}(b) provides an example finite-element method (FEM) simulation of the coupled resonator geometry, indicating such a mechanical supermode that is a hybrid of the modes of the individual piezoelectric and optomechanical resonators. The optical field profile is confirmed to be confined within the nanobeam portion of the transducer (top right of Fig.~\ref{fig:coupled_res_example}(a)), with its optical quality factor $Q_\text{o}$ depending on the specifics of the connection to the piezoelectric resonator.  For the optical output, a waveguide, either built in (Fig.~\ref{fig:coupled_res_example}(a)) or an optical fiber taper, couples to and from the optical cavity created by the photonic crystal in the nanobeam.

There are several benefits of this coupled resonator approach.  First, it separates the metallic electrodes from the optical field, important for maintaining high $Q_\text{o}$ and avoiding optical absorption-induced heating of the electrical circuit (which may be superconducting). Second, it supports the GHz mechanical mode frequencies associated with nanobeam optomechanical crystals that have been implemented in piezoelectric platforms such as GaAs~\cite{balram_moving_2014}, AlN~\cite{fan_AlN_OMC,bochmann_nanomechanical_2013}, and LiNbO$_3$~\cite{liang_LN_OMC, jiang_lithium_2019}. High mechanical mode frequencies enable lower thermal phonon numbers at a given temperature and thus allow lower added noise. On the cavity-optomechanical side, good sideband resolution ($(4 \omega_\text{m}/\kappa_\text{o})^2 > 1$, where $\kappa_\text{o}$ is the total optical cavity-mode decay rate, typically hundreds of MHz to GHz for most integrated cavity optomechanical systems) is desirable to suppress scattering into the lower frequency sideband induced on the optical drive by the mechanics, which acts as a source of parametric amplification noise. In addition, our approach takes advantage of the large optomechanical coupling rate [$g_0/(2\pi) > 1$\,MHz] that has been demonstrated in piezoelectric nanobeam optomechanical crystals, especially in GaAs, due to its high refractive index and large photoelastic coefficients~\cite{balram_moving_2014}. Since the optomechanical interaction scales with intracavity pump photon number and the square of its single-photon coupling rate, the latter is extremely important when the former is limited to prevent excess heating in cryogenic experiments~\cite{meenehan_silicon_2014,meenehan_pulsed_2015,forsch_microwave_2018,ramp_elimination_2018}. 

Finally, the coupled resonator approach successfully addresses acoustic wave impedance-matching challenges.  Such challenges arise in developing a platform that can simultaneously couple traveling acoustic waves to both microwaves and optical waves, while maintaining spatial separation of optical fields and metal electrodes.  For example, interdigitated transducers (IDTs) used for generating surface acoustic waves are straightforward to design and fabricate~\cite{balram_coherent_2016,vainsencher_bi_2016}, and can easily be spatially separated from the optomechanical resonator.  However, the geometry in which they are incorporated introduces two problems. First, their efficiency in converting a microwave input signal to an acoustic wave can be limited, particularly in materials such as GaAs with a relatively weak piezoelectric effect. Second, the traveling surface acoustic wave that is typically generated by an IDT suffers from acoustic impedance mismatch. The main challenge is satisfying simultaneously a microwave transmission line impedance of $50\,\Omega$, which requires IDTs tens of micrometers in width, and coupling the laterally wide acoustic waves into a thin 500-nm-wide nanobeam cavity where the localized mechanical mode resides. Our approach addresses both of these challenges.

\section{Piezo-optomechanical transduction theory}
\label{sec:Piezo_optomechanical_transduction}

\begin{figure}
\begin{center}
\includegraphics[width=\linewidth]{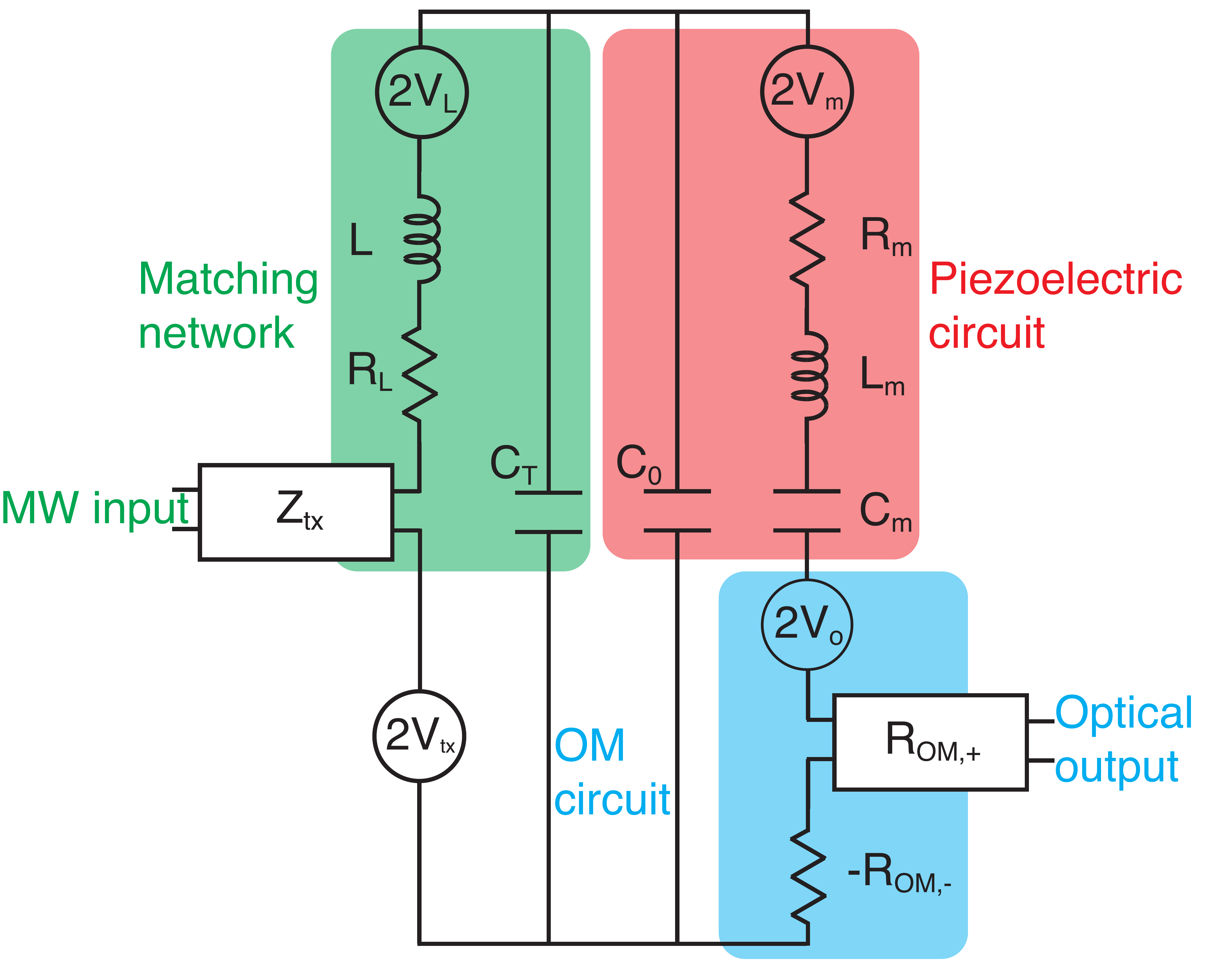}
\end{center}
\caption{Piezo-optomechanical circuit: a transmission line with characteristic impedance $Z_\text{tx}$ is piezoelectrically coupled, through an $RLC$ matching network (green region), to a mechanical mode represented by a Butterworth-van Dyke (BVD) circuit (red region). The BVD circuit model is composed of a motional arm including a motional resistance $R_\text{m}$, capacitance $C_\text{m}$, and inductance $L_\text{m}$ in parallel with the static arm that includes a static capacitance $C_0$. The optomechanical cavity (blue region) is coupled in series to the motional arm of the piezoelectric circuit via equivalent impedances in which the upper sideband, whose coupling is represented by the transmission line with (positive) characteristic impedance $R_{\text{OM,}+}$, is the target optical output for low-noise transduction, whereas the negative resistance $-R_{\text{OM,}-}<0$ accounts for the amplification effect of the lower sideband. This simple equivalent circuit is a valid description of the optomechanical coupling in the adiabatic regime where the sidebands are much narrower than the optical cavity linewidth $\kappa_\text{o}$. All resistive elements, including the transmission line impedance, are accompanied by a voltage source $2V_i$ accounting for their associated signal or noise inputs. In particular, the incoming transmission line signal is $V_\text{tx}$, and the fraction of its power dissipated in $R_\text{OM,+}$ is the signal transfer efficiency $\eta$.}
\label{fig:BVD-LCcircuit-Tx-Opto}
\end{figure}

\noindent Several schemes for quantum transduction have been put forward in the literature (which we do not attempt to exhaustively review here). In this work, we focus on linear phase-insensitive transducers. This is a meaningful approach when transducing signals for which the arrival time is unknown. In contrast, if the arrival time (and temporal mode) is known, various schemes in which, e.g., coupling rates or detunings are varied in time, may be advantageous~\cite{Tian_sequential-swap-transduction_2010,Wang_quantum-state-transfer_2012,wang_using_2012_njp,Tian_adiabatic-state-conversion_2012,McGee_EOM-storage-transfer_2013,Zhang_OM-MW-sensor_2015}. 
A central performance metric for a transducer is its ability to convert an input signal into the desired output channel, represented by the transfer efficiency $\eta$. Another essential figure of merit concerns the unwanted incoherent noise quanta injected by the transducer into the output channel. We quantify this contribution by the number of noise photons per unit time per unit bandwidth, $N$; we reference this number to the input of the transducer, rendering it the inverse signal-to-noise ratio. Lastly, the spectral profile for the efficiency has bandwidth $\Delta \omega$.

In this section, the full transduction scheme from input to output is laid out. To start, the input and output ports are defined and some assumptions about the detection method are made. Then, all essential elements of the piezo-optomechanical transducer are modeled through an equivalent circuit approach using methodology from Ref.~\cite{Zeuthen_Electrooptomechanical_2018} (summarized for the present purposes in Appendix~\ref{sec:equiv-circ}), resulting in the joint equivalent circuit shown in Fig.~\ref{fig:BVD-LCcircuit-Tx-Opto}. Based on this circuit, we finally calculate the transducer figures of merit $\eta$, $N$, and $\Delta \omega$.

\subsection{Input-output theory and detection scheme}
\label{sec:input-output}

A familiar concept from circuit analysis is the scattering matrix $\mathbf{S}$ that links the incoming and outgoing fields $\vec{v}_{\text{in(out)}}$ of the various signal and noise ports of a circuit in the frequency domain,
\begin{equation}
    \vec{v}_{\text{out}}(\omega)= \mathbf{S}(\omega)\vec{v}_{\text{in}}(\omega),
    \label{eq:scattering-matrix}
\end{equation}
where each port is represented by an entry in the vectors $\vec{v}_{\text{in(out)}}$. The action of a linear piezo-optomechanical quantum transducer can be described by such a formalism provided that the following generalizations are made: (1) different ports can have different carrier frequencies, in order to account for the up-conversion brought about by the optomechanical pump field, and (2) the itinerant fields $\vec{v}_{\text{in(out)}}$ are quantized. Note that the scattering matrix $\mathbf{S}$ of a linear system is the same in the classical and quantum cases, hence explicit quantization of the internal transducer modes is not required.

When characterizing the performance of a transducer, not all elements of $\mathbf{S}$ are of equal importance, and this permits a more economical description. Two interconnected types of simplification are applied. First, if a particular output port is of interest (i.e., a particular element of $\vec{v}_{\text{out}}$), we may choose to focus on the corresponding row in $\mathbf{S}$. Second, concerning the noise inputs, assumed to be uncorrelated with the input signal, we are interested only in their net noise statistics in the output. These two aspects are connected because the noise statistics depends on the type of measurement performed on the output field, e.g., photon counting or homodyne detection, which, in turn, is reflected in the basis choice for $\vec{v}_{\text{out}}$.

We now apply these considerations in the present context of piezo-optomechanical transducers introduced in Section~\ref{sec:Coupled_resonator_approach}. In this work we assume for definiteness and simplicity that the upper sideband of the outgoing optical field is measured by photon counting while the residual outgoing lower sideband is discarded (as can be achieved using a sufficiently narrow filter cavity). While this detection strategy is clearly suboptimal, since the information in the lower sideband is lost, it is reasonable in the resolved-sideband regime and makes for a straightforward practical interpretation of our results. Based on these considerations we can write an effective scattering relation for electrical-to-optical conversion ($\omega>0$),
\begin{equation}
\hat{b}_\text{out}(\omega_{\text{pump}}+\omega) = \sqrt{\eta(\omega)} [\hat{a}_\text{in}(\omega) + \sqrt{N(\omega)}],\label{eq:scattering_relation}
\end{equation}
in terms of the incoming field $\hat{a}_\text{in}$ of the transmission line and the outgoing field $\hat{b}_\text{out}$ of the upper-sideband optical field obeying commutation relations $[\hat{a}_\text{in}(\omega),\hat{a}^\dagger_\text{in}(\omega')]=[\hat{b}_\text{out}(\omega),\hat{b}^\dagger_\text{out}(\omega')]=\delta(\omega-\omega')$ (see Appendix~\ref{sec:deriveEtaN} for a derivation of Eq.~\eqref{eq:scattering_relation} for our piezo-optomechanical system).
Equation~\eqref{eq:scattering_relation} gives definite meaning to $\eta$ as the signal (power) transfer efficiency and $N$ as the transducer dark-count noise flux per unit bandwidth referenced to the input, thus completing the black-box transducer description illustrated in Fig.~\ref{fig:schematic}(a). 

Even if solely interested in electrical-to-optical conversion, the outgoing field of the electrical (input) port contains information of interest due to signal reflection and noise cross-correlations, as can be exploited in feed-forward~\cite{Higginbotham_Harnessing_2018} and adaptive~\cite{Zhang_adaptive-transduction_2018,Lau_imperfect-transducers_2019} transduction schemes. However, we choose to discard the outgoing field of the input port in order to characterize the performance of our transducer within the simplest possible scheme. In this sense, Eq.~\eqref{eq:scattering_relation} suffices to describe the transducer, but it is straightforward to extend our analysis to evaluate signal reflection and noise cross-correlations if desired.

For quantum transduction, high transfer efficiency ($\eta \rightarrow 100\,\%$) and low added noise ($N \rightarrow 0$) are desired. In practice, this ideal limit cannot be attained and a trade-off between large $\eta$ and small $N$ must be made. Their relative importance, and hence the optimal trade-off, depends on the application and the method of detection involved~\cite{Zeuthen_Figures_2016}. Our analysis focuses, to some extent, on transduction of microwave quantum signals to the optical domain as captured by Eq.~\eqref{eq:scattering_relation}. Even though $\eta$ is the same in both directions due to symmetry considerations, the same does not hold for $N$ due to the nonequilibrium nature of the system (see Appendix~\ref{sec:optical2electrical} for details). So while our optimization of $\eta$ in a subsequent section automatically applies to both directions, in the main text our noise analysis focuses on electrical-to-optical conversion since only for this direction do all of the transducer implementations considered here perform well in terms of $N$ (although some exhibit good bidirectional performance).

\subsection{Piezoelectric circuit}
\label{sec:Piezoelectric_circuit}

\noindent The electromechanical behavior of a piezoelectric resonator can be modeled effectively using an equivalent electrical circuit. One conventional lumped-element circuit model is the Butterworth-van Dyke (BVD) model as shown in the red box of Fig.~\ref{fig:BVD-LCcircuit-Tx-Opto}~\cite{Larson_Modified_2000, Wang_Design_2015}. The motional arm of the circuit (resistance $R_\text{m}$, inductance $L_\text{m}$, capacitance $C_\text{m}$) is the equivalent-circuit representation of the mechanical susceptibility (defined in terms of, e.g., mass, spring constant, and relaxation rate). The proportionality factors between mechanical and equivalent electrical parameters encode the conversion strength of electrical energy to mechanical energy. The equivalent charge on the mechanical capacitance $C_\text{m}$ is proportional to the excursions in position of the mechanical mode relative to its equilibrium configuration. The static arm (capacitance $C_0$) forms the electrical capacitance of the physical device.

The strength of the piezoelectric interaction in a particular device is commonly quantified by the effective electromechanical coupling coefficient $k^2$, which relates the energy conversion between electrical and mechanical subsystems. This coefficient is purely a material and geometric parameter~\footnote{$k^2$ is often denoted $k_\text{eff}^2$ in literature but we adopt the simple form $k^2$ in this work.}. Although there are many ways to define $k^2$~\cite{chang_analysis_1995, ikeda_fundamentals_1990}, we choose a definition that relates back to the BVD circuit parameters,
\begin{equation}
k^2 = \frac{C_\text{m}}{C_\text{m}+C_0}, 
\label{eq:k-square-def}
\end{equation}
illustrating the fact that the magnitude of the equivalent mechanical capacitance $C_\text{m}$ encodes the electromechanical interaction strength.
Having defined the coupling, the motional resistance can then be related as
\begin{equation}
R_\text{m} = \gamma_\text{m} L_\text{m} = \frac{\gamma_\text{m}}{\omega_\text{s}^2}\frac{1/k^2-1}{C_0}, 
\label{eq:R-m}
\end{equation}
where $\gamma_\text{m}$ is the mechanical energy loss rate and $\omega_\text{s} = 1/\sqrt{L_\text{m} C_\text{m}}$ is the mechanical series resonance frequency (the latter expression fixes $L_\text{m}$ for given $\omega_\text{s}$ and $C_\text{m}$). The final expression in Eq.~\eqref{eq:R-m} shows that the equivalent resistance $R_\text{m}$ for a given $\gamma_\text{m}$ decreases with increasing piezoelectric coupling strength $k^2$.

\subsection{Matching network}
\label{sec:matchingNetwork}

\noindent The impedance $Z$ of a bare nanoscale piezoelectric device can be difficult to impedance match to a $Z_\text{tx} = 50\,\Omega$ transmission line, as $|Z|$ can vary between a few ohms to thousands of ohms when taking into account the range of possible parameters that enter into Eqs.~\eqref{eq:k-square-def} and~\eqref{eq:R-m}. A suitably designed electrical network appropriately transforms $\text{Re}[Z]$ and $\text{Im}[Z]$ to form a natural bridge between the piezoelectric device and the input. There are several options for such a matching network~\cite{pozar_microwave_1990, OConnell_piezo-quantum-limit_2014, harabula_measuring_2017, woolley_quartz-superconductor_2016, Santos_Optomechanical_2017}, including our own suggested design in Appendix~\ref{sec:network}. For simplicity in our current analysis, here we consider a simple $RLC$ network (green box in Figs.~\ref{fig:coupled_res_example} and~\ref{fig:BVD-LCcircuit-Tx-Opto}), which consists of a tuning capacitor with capacitance $C_\text{T}$ in parallel with $C_0$ and a tuning inductor $L$ in series. A resistor $R_\text{L}$ is also added to account for inductor resistive loss as well as any additional Ohmic loss at the transmission line input. The impedance transformation provided by the matching network can be viewed as being due to the resonant signal enhancement according to its (loaded) quality factor $Q_\text{LC}=\sqrt{L/(C_\text{T} + C_0)}/(Z_\text{tx}+R_\text{L})$. The desired transformation depends on the optical loading and is discussed in a subsequent section.

The above points to the fact that the electrical resonance $\omega_\text{LC} = 1/\sqrt{L(C_\text{T} + C_0)}$ must be aligned with a suitable mechanical resonance frequency. The mechanical resonance of the piezo-optomechanical resonator shifts to a new effective frequency, due its coupling to the electrical network,
\begin{equation}
\omega_\text{m} = \sqrt{ \frac{1}{L_\text{m}} \left ( \frac{1}{C_\text{m}} + \frac{1}{C_\text{T} + C_0} \right ) },
\label{eqn:effectiveResonance}
\end{equation}
which can be interpreted as the resonance obtained by lumping the tuning capacitance $C_\text{T}$ together with $C_0$ and forming a loop current with the mechanical arm (see Appendix~\ref{sec:deriveEtaN} for details). At $C_\text{T} \rightarrow \infty$, this resonance approaches the series resonance $\omega_\text{s}$, while for $C_\text{T} \rightarrow 0$, the resonance shifts to the parallel resonance $\omega_\text{p} = \omega_\text{s}/\sqrt{1-k^2}$. We assume a negligible frequency shift due to optical forces, as is typically the case for the high-frequency, large-stiffness resonators we consider. 

With a suitable choice of $L$ and $C_\text{T}$ (see Appendix~\ref{sec:network}), the electrical resonance can be matched to the mechanical resonance frequency $\omega_\text{LC}=\omega_\text{m}$ while simultaneously achieving the desired enhancement $Q_\text{LC}$ (provided that it does not exceed the maximal value $1/[(Z_\text{tx}+R_\text{L})C_0 \omega_\text{p}]$). When these frequencies match, the imaginary part of the impedance of the piezo-optomechanical transducer is zero at $\omega=\omega_\text{m}$ as seen from the transmission line (provided that the corresponding optomechanical resonance matching $\omega_\text{o} = \omega_\text{m}+\omega_\text{pump}$ is ensured); this is a necessary requirement for impedance matching.

With the electrical and piezoelectric circuit elements defined, electrical input parameters can now be calculated. Since the resistance $R_\text{L}$ is in series with the transmission line, it simply results in a finite electrical coupling efficiency:
\begin{eqnarray}
\eta_\text{e} = \frac{Z_\text{tx}}{Z_\text{tx} + R_\text{L}}.
\label{eq:eta_electrical}
\end{eqnarray}
The resonantly enhanced electrical loading of the mechanical mode can be expressed as 
\begin{equation}
R_\text{EM} =Q_{\text{LC}}^2(Z_\text{tx} + R_\text{L})= \frac{Z_\text{LC}^2}{Z_\text{tx} + R_\text{L}},\label{eq:R-EM_LC}
\end{equation}
where $Z_\text{LC} = \sqrt{L/(C_0 + C_\text{T})}$, or alternatively, in terms of the electromechanical cooperativity
\begin{equation}
\mathcal{C}_\text{EM} \equiv \frac{R_\text{EM}}{R_\text{m}} = \frac{Z_\text{LC}^2}{R_\text{m} (Z_\text{tx} + R_\text{L})} = \frac{4g^2_\text{EM}}{\gamma_\text{m} \kappa_\text{e}}, 
\label{eq:C-EM-def}
\end{equation}
where $\kappa_\text{e} = (Z_\text{tx}+R_\text{L})/L$ is the electrical decay rate (FWHM) and $g_\text{EM} = \sqrt{k_\text{T}^{2}} \omega_\text{m}/2$ is the electromechanical coupling rate in terms of the reduced piezoelectric coupling strength $k_\text{T}^{2}=C_{\text m}/(C_{\text m}+C_0+C_{\text T})$ [cf.~Eq.~\eqref{eq:k-square-def}] assuming matching frequencies $\omega_\text{m} = \omega_\text{LC}$ (see Appendix~\ref{sec:gEM} for derivation).

\subsection{Optomechanical equivalent circuit}

\noindent The last element of the equivalent circuit concerns the optomechanical coupling (blue box in Fig.~\ref{fig:BVD-LCcircuit-Tx-Opto}), represented by the frequency-independent effective resistances $R_{\text{OM},\pm}$. This simple description of the optomechanical coupling is valid for signals that are narrowband compared to the optical cavity linewidth $\kappa_\text{o}$ (FWHM). It results as a limiting case of a more general equivalent circuit derived in Ref.~\cite{Zeuthen_Electrooptomechanical_2018} and summarized in Appendix~\ref{sec:equiv-circ}. 

For the simple quantum transduction scheme specified in Section~\ref{sec:input-output}, the desired optical output port is the upper sideband (see Fig.~\ref{fig:schematic}(b)); this is represented by the positive transmission line characteristic impedance $R_{\text{OM,}+}>0$, which plays a role analogous to that of the electrical transmission line impedance $Z_\text{tx}$. The value of the optomechanical impedance $R_{\text{OM,}+}$ encodes the optomechanical coupling strength and the optical resonant enhancement, in analogy to what is discussed for the electromechanical coupling above, and hence these are knobs for engineering the transducer circuit.
The residual coupling to the lower sideband, owing to finite sideband resolution, is represented by the negative resistance $-R_{\text{OM,}-}<0$, indicative of the ability to amplify motion through the optical drive. In the present context of quantum transduction, it is typically desirable to suppress this amplification effect, as can be achieved by operating in the resolved-sideband regime $(4 \omega_\text{m}/\kappa_\text{o})^2 \gtrsim 1$ with a red-detuned pump $\omega_\text{pump}=\omega_\text{o}-\omega_\text{m}$. However, we retain the residual amplification (and associated noise) due to nonzero $R_{\text{OM,}-}$ in our description to account for its impact on our transducer figures of merit, $\eta$ and $N$. Note that our depiction in Fig.~\ref{fig:BVD-LCcircuit-Tx-Opto} of $R_{\text{OM,}+}$ as being associated with a transmission line but $-R_{\text{OM,}-}$ with a resistor is consistent with the simple transduction scheme analyzed here; it hinges on the equivalence between a resistor and an unmonitored semi-infinite transmission line with a suitable thermal input field~\cite{nyquist_thermal_1928}.

The definition for the optomechanical impedances for the upper ($+$) and lower ($-$) sidebands are
\begin{equation}
R_{\text{OM},\pm} = R_\text{m} \mathcal{C}_\text{OM} \mathcal{L}_\pm^2,
\label{eq:R-OM-def}
\end{equation}
where the well-known optomechanical cooperativity is defined as
\begin{eqnarray}
\mathcal{C}_\text{OM} = \frac{4g^2_\text{OM}}{\gamma_\text{m} \kappa_\text{o}},
\label{eq:Com}
\end{eqnarray}
with the pump-enhanced optomechanical coupling rate $g_\text{OM} = g_0 \sqrt{n_\text{phot}} $ proportional to the square root of the number of intracavity drive photons $n_\text{phot}$ and the single-photon optomechanical coupling rate $g_0$, and $\kappa_\text{o}$ is the energy decay rate of the optical mode. The optical-cavity Lorentzian sideband amplitudes are expressed as
\begin{equation}
\mathcal{L}_\pm^2 = \frac{(\kappa_\text{o}/2)^2}{(\kappa_\text{o}/2)^2 + (\Delta \pm \omega_\text{m})^2}
\label{eq:Lorentzian}
\end{equation}
in terms of the laser detuning from cavity resonance $\Delta = \omega_\text{pump} - \omega_\text{o}$. Finally, to complete the picture at the output, the optical cavity is coupled to an external channel, for example a waveguide, with efficiency
\begin{eqnarray}
\eta_\text{o} = \frac{\kappa_\text{ext}}{\kappa_\text{ext} + \kappa_\text{i}},
\label{eq:eta_optical}
\end{eqnarray}
where $\kappa_\text{o} = \kappa_\text{ext} + \kappa_\text{i}$ consists of waveguide coupling $\kappa_\text{ext}$ and intrinsic loss $\kappa_\text{i}$ contributions. $Q_\text{o} = \omega_\text{o}/\kappa_\text{o}$ is the loaded optical quality factor of the optical mode.

\subsection{Signal transfer efficiency $\eta$}\label{sec:eta}

\noindent With all the pieces in place, we now turn to the signal transfer efficiency $\eta$ of our piezo-optomechanical transducer, which is the probability that an incoming signal photon is converted to an outgoing photon in the desired output channel. Though the transfer efficiency is the same for the two conversion directions (as shown in Appendix~\ref{sec:optical2electrical}), the flow here is described as going from the microwave regime to the optical regime.

Overall, the peak signal transfer $\eta_\text{peak} \equiv \eta(\omega_\text{m})$ from the microwave transmission line to the upper optical sideband for the piezo-optomechanical transducer in Fig.~\ref{fig:BVD-LCcircuit-Tx-Opto} is (see derivation in Appendix~\ref{sec:deriveEtaN})
\begin{align}
\eta_\text{peak}  & = \eta_\text{e} \eta_\text{o} \frac{4 R_\text{EM} R_{\text{OM,}+}}{(R_\text{m} + R_\text{EM} + R_{\text{OM,}+} - R_{\text{OM,}-})^2}  \nonumber\\
& =  \eta_\text{e} \eta_\text{o} \frac{4 \mathcal{C}_\text{EM} \mathcal{C}_\text{OM} \mathcal{L}^2_+}{(1 + \mathcal{C}_\text{EM} + \mathcal{C}_\text{OM} (\mathcal{L}^2_+ - \mathcal{L}^2_-))^2},
\label{eq:peakEta}
\end{align}
which is the mainstay equation for optoelectromechanical efficiency~\cite{ safavi-naeini_proposal_2011, Tian_adiabatic-state-conversion_2012,Wang_quantum-state-transfer_2012,Andrews2014}. The two prefactors in this expression represent incoupling and outcoupling of the microwave and optical signals, respectively, while the third term is an internal efficiency of conversion, which takes into account losses due to mechanical dissipation and lack of impedance matching (see further below and Appendix~\ref{sec:deriveEtaN}).

\subsection{Added noise $N$}\label{sec:N}

\noindent In this section, we consider the second figure of merit, added noise $N$, as referenced to the signal input in the sense of Eq.~\eqref{eq:scattering_relation}. We focus below on two contributions to the noise arising in our transduction platform: optical noise and thermomechanical noise. We assume our electrical circuit to be in the ground state under thermal conditions, in which case the Ohmic losses $R_{\text{L}}$ of the matching network only lead to the injection of vacuum noise, which will not contribute to $N$ under the chosen detection scheme.

To start, assuming that the optical pump field is in a coherent state such that its fluctuations are of those of vacuum, for finite optomechanical sideband resolution the two-mode-squeezing interaction produces a nonzero outgoing flux of noise quanta in the upper sideband (even in absence of signal input). This noise contribution can be suppressed by appropriately squeezing the incoming pump field so as to counteract (unwanted) squeezing due to finite sideband resolution of the cavity~\cite{Lau_parametric-drive-transduction_2019}. However, in the transducer optimization presented below, we do not explicitly invoke this technique. The two-mode-squeezing interaction also gives rise to a lower (Stokes) sideband in the optical output (relative to the carrier $\omega_\text{pump}$). But, as discussed in Section~\ref{sec:input-output}, we consider the output port to be the upper sideband while the lower sideband is a source of noise.

Moreover, under realistic conditions, the mechanical mode has a finite thermal occupation due to the ambient mechanical bath and also injects noise into the output port. The total added noise flux per unit bandwidth referenced to the input signal is the sum of these two contributions, Raman scattering noise $N_\text{o}$ and mechanical thermal noise $N_\text{m}$, so that
\begin{equation}
N = N_\text{o} + N_\text{m},
\label{eq:Noise}
\end{equation}
where $N$ is defined as ($\omega>0$)
\begin{equation}
    N(\omega) \delta(\omega-\omega')=\frac{1}{\eta(\omega)} \langle \hat{b}^\dagger_\text{out} (\omega_\text{pump}+\omega) \hat{b}_\text{out} (\omega_\text{pump}+\omega') \rangle,
\end{equation}
in accordance with the choice of measurement scheme described in Section~\ref{sec:input-output}, that is photon counting of the upper optical sideband.
In the present section we assume the regime of adiabatic transduction where the signal bandwidth is small compared to the linewidths of both electrical ($\kappa_{\text{e}}$) and optical ($\kappa_{\text{o}}$) resonators, within which $N(\omega)$ is essentially flat (the behavior outside this regime is discussed in the next subsection).

\subsubsection{Optical amplification noise $N_\text{o}$ (Raman noise)}

\noindent The amplification noise due to the Stokes process leads to added noise contribution in the optical output,
\begin{equation}
N_\text{o} = \frac{1}{\eta_\text{e}} \frac{\mathcal{C}_\text{OM}\mathcal{L}^2_-}{\mathcal{C}_\text{EM}},
\label{eq:Nopt1}
\end{equation}
which is independent of the Fourier frequency $\omega$ within the adiabatic regime of narrow signal bandwidths compared to the electrical and optical linewidths. This contribution arises from the fluctuations in the lower sideband which, via two-mode squeezing, create phonons in the optomechanical cavity which, in turn, are transduced into the upper sideband. 

\subsubsection{Mechanical thermal noise $N_\text{m}$}

\noindent The mechanical thermal noise is proportional to the thermal occupancy of its bath, as given by the Bose-Einstein distribution $n_\text{m}(\omega)=(e^{\hbar\omega/(k_{\text{B}}T)}-1)^{-1}$, and inversely proportional to $\mathcal{C}_{\text{EM}}$,
\begin{eqnarray}
N_\text{m} = \frac{1}{\eta_\text{e}} \frac{n_\text{m}}{\mathcal{C}_\text{EM}}, 
\label{eq:Nm}
\end{eqnarray}
capturing the enhancement in the ratio of electrical signal to mechanical noise brought about by the electrical resonator. The dependence of $n_{\text{m}}$ on Fourier frequency is typically negligible over the signal bandwidth and can hence be approximated by setting $\omega\approx\omega_{\text{m}}$. 
The quantity $\mathcal{C}_\text{EM}/n_\text{m}$ appearing in Eq.~\eqref{eq:Nm} is known as the electromechanical quantum cooperativity; it is (approximately) the ratio of coherent electromechanical coupling to the thermal decoherence induced by the mechanical bath. The desired regime for quantum-level transduction $N_\text{m} \ll 1$ thus requires $\mathcal{C}_\text{EM}/n_\text{m} \gg 1$. 

\subsection{Transduction bandwidth $\Delta \omega$}

\noindent In our discussion of the transduction efficiency $\eta$ in Section~\ref{sec:eta} we focus on its peak value, achieved at the transducer resonance $\omega_\text{MW}=\omega_\text{m}$. However the finite bandwidth of any realistic signal requires us, in general, to consider the full frequency profile of the transfer efficiency $\eta(\omega)$ and added noise $N(\omega)$. Nevertheless, we generally focus on the adiabatic regime of signals that are narrowband compared to the electrical and optical resonator linewidths, $\kappa_\text{e}$ and $\kappa_\text{o}$, therefore $N(\omega)$ is approximately constant around the frequency of interest as mentioned previously. Hence, the noise bandwidth (approximately equal to $\kappa_\text{e}$) is effectively infinite. 

In this adiabatic regime, the transducer bandwidth can be meaningfully characterized as that of $\eta(\omega)$ and is simply given by the dynamically broadened mechanical linewidth (FWHM),
\begin{align}
\Delta \omega &= (R_\text{m} + R_\text{EM} + R_{\text{OM,}+} - R_{\text{OM,}-} )/L_\text{m}\nonumber\\
&= \gamma_\text{m}(1 + \mathcal{C}_\text{EM} + \mathcal{C}_\text{OM} (\mathcal{L}^2_+ - \mathcal{L}^2_-)),\label{eq:BW}
\end{align}
which is the quantity appearing in the denominator of Eqs.~\eqref{eq:peakEta}. 
Narrow intrinsic mechanical linewidths $\gamma_{\text{m}}$ are inherent in high-$Q_\text{m}$ resonators, but Eq.~\eqref{eq:BW} shows that transducer bandwidth can be significantly enhanced beyond this value in the regime in which at least one of the cooperativities is large, $\mathcal{C}_\text{EM}\gtrsim 1$ and/or $\mathcal{C}_\text{OM}\gtrsim 1$. Since this regime is compatible with large transduction efficiencies $\eta$, as discussed in the next section, we do not delve into a specific optimization of bandwidth in this work.

Equation~\eqref{eq:BW} for the transduction bandwidth of $\eta(\omega)$ is valid as long as its result is much smaller than the electrical linewidth, $\Delta \omega \ll \kappa_\text{e}$.  Beyond the simple adiabatic regime, the full spectrum of $\eta(\omega)$ and $N(\omega)$ must be considered, each with its associated bandwidth (see Appendix~\ref{sec:deriveEtaN}).

\section{Maximizing efficiency $\eta$}
\label{sec:maxEta}

\noindent In the preceding sections, we introduce the essential transducer metrics, signal transfer efficiency $\eta$, added noise $N$, and bandwidth $\Delta \omega$. As mentioned previously, the relative importance of these depends on the specific transducer application~\cite{Zeuthen_Figures_2016}. To keep our analysis general, we do not delve into optimizing the transducer for specific applications, but instead discuss maximization of $\eta$ and minimization of $N$. 
This serves to identify the performance limits of our platform and provides a good starting point for application-specific optimization.

Our first optimization scheme seeks to maximize conversion efficiency $\eta$. However, we make the implicit assumption that $N$ should be kept reasonably small. In fact, it is possible to reach the regime where $\eta > 1$ due to amplification by decreasing the optomechanical sideband resolution $(4 \omega_\text{m}/\kappa_\text{o})^2 < 1$, but this is accompanied by more added noise (see Eq.~\eqref{eq:Nopt1} and further derivation in Appendix~\ref{sec:amplification}). We therefore refrain from employing this effect to boost $\eta$ in our optimization by assuming a fixed degree of sideband resolution. We provide some heuristic optimization principles after our analysis, taking into account typical experimental limitations.

\subsection{Analysis}

\noindent Assuming the optomechanical and mechanical parameters to be fixed, the peak signal transfer efficiency $\eta_\text{peak}$ [Eq.~\eqref{eq:peakEta}] reaches an optimal point as a function of $\mathcal{C}_\text{EM}$ at
\begin{equation}
\mathcal{C}^\text{opt}_\text{EM} = 1 + \mathcal{C}_\text{OM} (\mathcal{L}_+^2 - \mathcal{L}_-^2),
\label{eq:Copt}
\end{equation}
which amounts to choosing the electromechanical broadening of the mechanical mode to be equal to the intrinsic mechanical linewidth plus the net optomechanical broadening. 
Note that only for $\eta_\text{e}=1$ does this amount to exact impedance matching of the microwave transmission line to the transducer so that no reflection occurs. To reach the cooperativity matching of Eq.~\eqref{eq:Copt}, the elements of the matching network must be correctly chosen (refer to section~\ref{sec:matchingNetwork} and see Appendix~\ref{sec:network} for details). The maximized peak efficiency, achieved at the matching condition $\mathcal{C}_{\text{EM}} = \mathcal{C}^\text{opt}_{\text{EM}}$, is 
\begin{align}
\eta_\text{peak}^\text{opt} & = \eta_\text{e} \eta_\text{o} \frac{ \mathcal{C}_\text{OM} \mathcal{L}^2_+}{1 + \mathcal{C}_\text{OM} (\mathcal{L}^2_+ - \mathcal{L}^2_-) } \nonumber\\
& \xrightarrow{\Delta \rightarrow-\omega_\text{m}} \eta_\text{e} \eta_\text{o}  \frac{\mathcal{C}_\text{OM} 
\left [1+\left(\frac{4\omega_\text{m}}{\kappa_\text{o}}\right)^2 \right ] }{1+\left(\frac{4\omega_\text{m}}{\kappa_\text{o}}\right)^2 (1+\mathcal{C}_\text{OM})}.
\label{eq:etaCom}
\end{align}
The final expression assumes the most common operating point for low-noise quantum transduction, where the laser is red detuned with respect to the optical resonance by $\omega_\text{m}$. Moreover, in the limit of good sideband resolution, the peak efficiency is approximately given by 
\begin{eqnarray}
\eta_\text{peak}^\text{opt} 
& \xrightarrow[(4 \omega_\text{m}/\kappa_\text{o})^2 \gg 1]{\Delta \rightarrow-\omega_\text{m}}
\eta_\text{e} \eta_\text{o}  \frac{\mathcal{C}_\text{OM}}{1+\mathcal{C}_\text{OM}}.
\label{eq:etaCom_resolved_sideband}
\end{eqnarray}
It is easy to see that in this amplification-free limit $\eta_\text{peak}^\text{opt}\leq1$.

For the optical amplification noise $N_\text{o}$, evaluating Eq.~\eqref{eq:Nopt1} under the cooperativity matching condition $\mathcal{C}_\text{EM}=\mathcal{C}^\text{opt}_\text{EM}$ [Eq.~\eqref{eq:Copt}] that maximizes $\eta_\text{peak}$, we find
\begin{align}
N_\text{o} & = \frac{1}{\eta_\text{e}} \frac{\mathcal{C}_\text{OM}\mathcal{L}^2_-}{1+\mathcal{C}_\text{OM}(\mathcal{L}_+^2 - \mathcal{L}_-^2)}.
\end{align}
Assuming that the laser drive is red detuned by the mechanical frequency, $\Delta = -\omega_\text{m}$, this becomes
\begin{align}
N_\text{o} & \xrightarrow{\Delta = -\omega_\text{m}} \frac{1}{\eta_\text{e}}  \frac{\mathcal{C}_\text{OM}}{1+\left(\frac{4\omega_\text{m}}{\kappa_\text{o}}\right)^2(1+\mathcal{C}_\text{OM})}.
\end{align}
For sufficiently good sideband resolution, this is approximately
\begin{align}
N_\text{o} & \xrightarrow[(4 \omega_\text{m}/\kappa_\text{o})^2 \gg 1]{\Delta = -\omega_\text{m}} \frac{1}{\eta_\text{e}}\left(\frac{\kappa_\text{o}}{4\omega_\text{m}} \right)^2  \frac{\mathcal{C}_\text{OM}}{1+\mathcal{C}_\text{OM}}.
\end{align}
This noise contribution increases with $\mathcal{C}_\text{OM}$, although it saturates for $\mathcal{C}_\text{OM}\gg1$.  Moreover, it can be suppressed by the factor $\left (4\omega_\text{m}/\kappa_\text{o} \right )^2$ by increasing the sideband resolution.

Finally, we note that the thermal noise $N_\text{m}$ is suppressed by a factor $1/\mathcal{C}_\text{EM}=1/\mathcal{C}_\text{EM}^\text{opt}$ from Eq.~\eqref{eq:Nm}. Hence suppression of thermal noise is sacrificed by the present choice of $\mathcal{C}_\text{EM}=\mathcal{C}_\text{EM}^\text{opt} < \mathcal{C}_\text{EM}^\text{max}$ below the maximum electromechanical cooperativity, which is discussed in Section~\ref{sec:minN}.

\subsection{Discussion}

\begin{figure}
\includegraphics[width=1\linewidth]{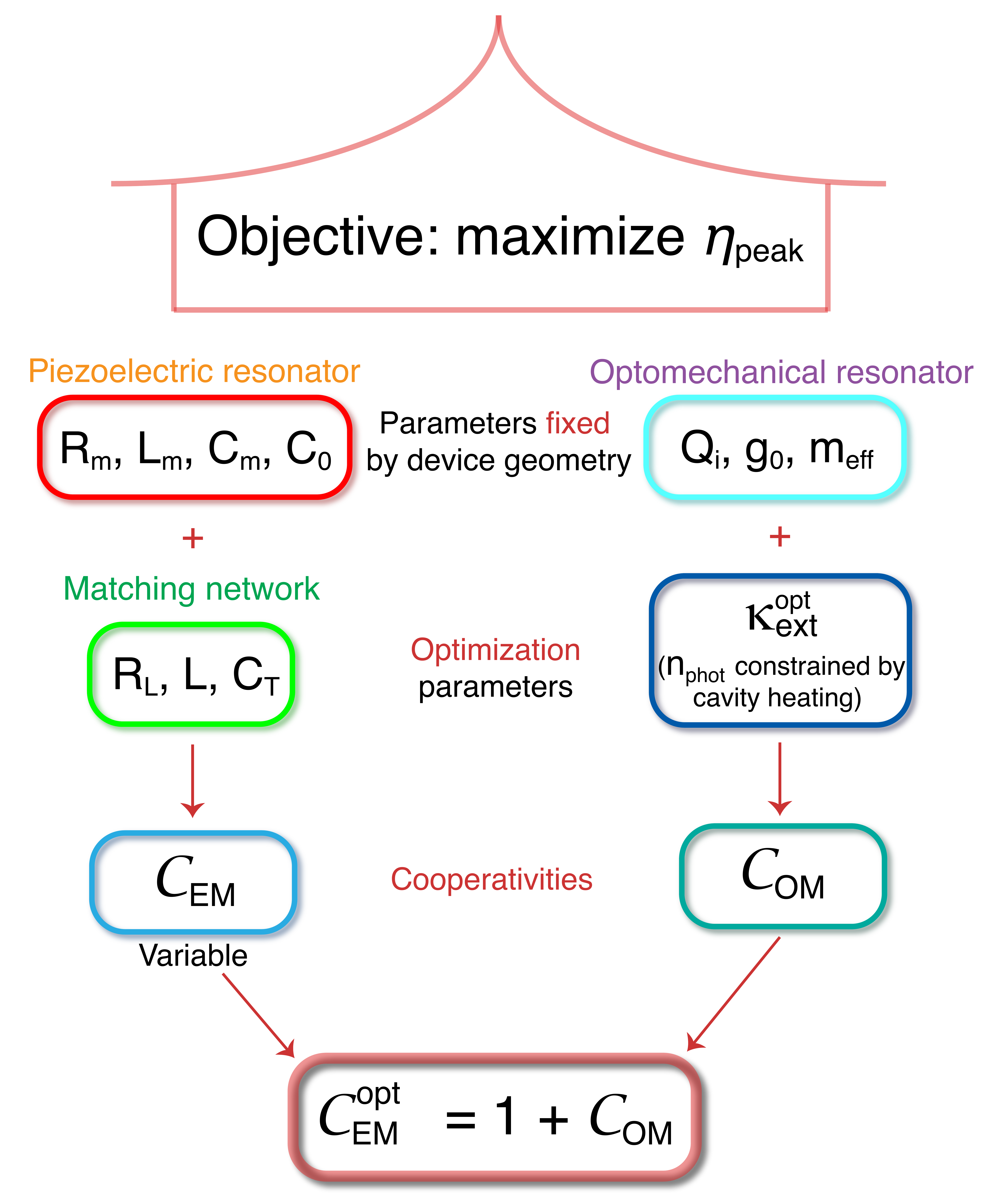}
\caption{Flow chart (from top to bottom) detailing some of the important dependencies for microwave-to-optical transduction in order to maximize efficiency $\eta_\text{peak}$.}
\label{fig:MaximizeEta}
\end{figure}

\noindent The theoretical optimization analysis in the previous subsection is now discussed in view of the experimental constraints of our platform. To this end, we summarize the procedure using the flow chart in Fig.~\ref{fig:MaximizeEta}. We optimize $\eta_\text{peak}$ for a given piezoelectric circuit and given optomechanical system, assuming $\mathcal{C}_\text{EM}$ can be optimized by constructing the right matching circuit so that $\mathcal{C}_\text{EM} = \mathcal{C}_\text{EM}^\text{opt}$ is realized [Eq.~\eqref{eq:Copt}]. We note that $\mathcal{C}_\text{OM}$ can, in principle, be tuned by injecting more photons $n_\text{phot}$ into the optical cavity to increase $g_\text{OM}$. However, to prevent excessive heating or nonlinear losses in some materials, $n_\text{phot}$ should be kept low, especially when attempting to reduce added noise $N$ (mainly thermal phonons) by lowering the effective temperature $T$. Recent experiments working at dilution refrigerator temperatures indeed give us insight that $n_\text{phot}$ should be restricted to around $n_\text{phot} \approx 280$~\cite{forsch_microwave_2018, ramp_elimination_2018}. Moreover, since the external optical coupling $\kappa_\text{ext}$ can tune both $\eta_\text{o}$ and $\mathcal{C}_\text{OM}$, its value can be optimized to obtain a trade-off between them that maximizes $\eta_\text{peak}$ [Eq.~\eqref{eq:etaCom}]. This optimal optical coupling $\kappa_\text{ext}^\text{opt}$ amounts to adjusting the \textit{optical} matching network (further discussion in Appendix~\ref{sec:kappa_ext_opt}). Therefore, in this work, $\mathcal{C}_\text{OM}$ is treated as a quasi-fixed value due to the capped value of $n_\text{phot}$ and the optimization of $\kappa_\text{ext}$, while the electromechanical cooperativity $\mathcal{C}_\text{EM}$ can be more easily adjusted via the electrical matching network.

\begin{figure*}
\includegraphics[width=1\linewidth]{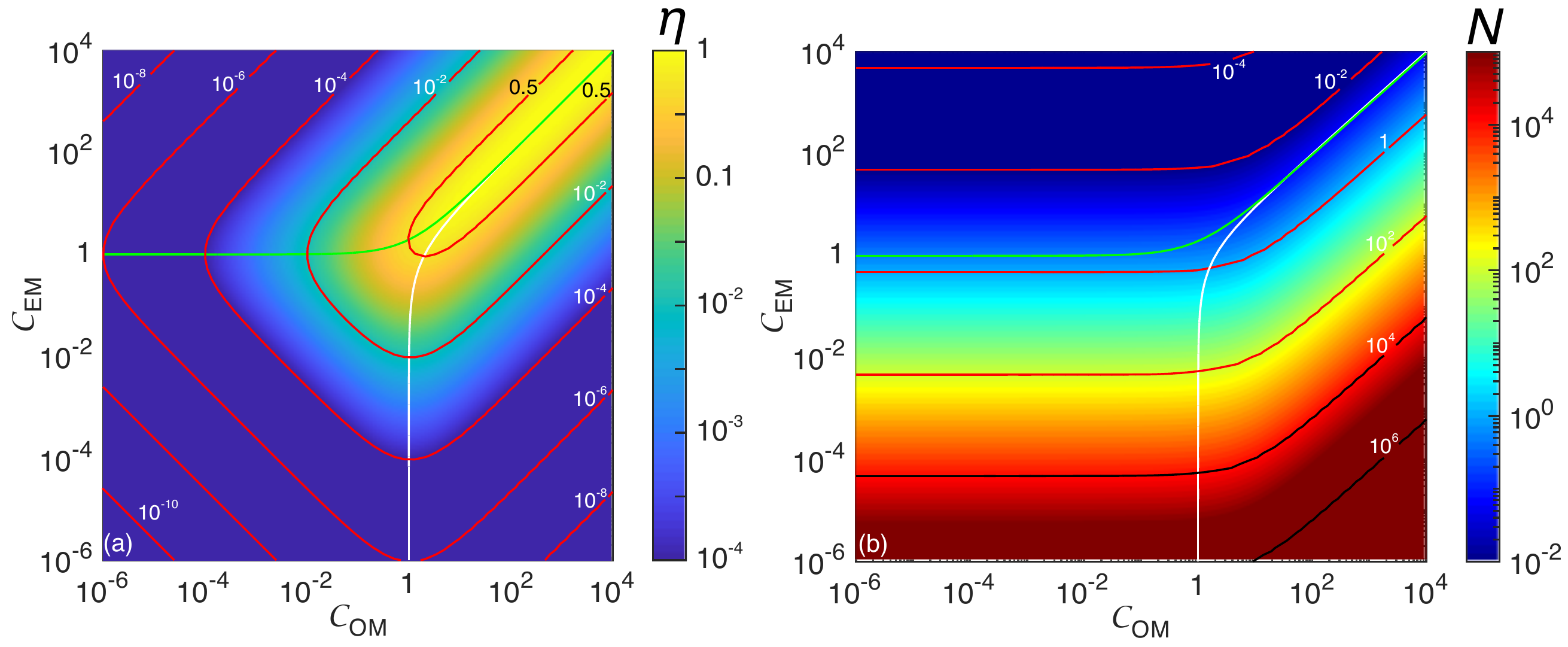}
\caption{(a) Peak efficiency $\eta_\text{peak}$ and (b) added noise $N$ as a function of cooperativities $\mathcal{C}_\text{EM}$ and $\mathcal{C}_\text{OM}$. Common parameters are $\eta_\text{e} = \eta_\text{o} = 1$, $\kappa_\text{o} = \omega_\text{m}$, $\Delta = -\omega_\text{m}$, effective temperature $T$ = 100 mK.  The green line represents $\mathcal{C}^\text{opt}_\text{EM} = 1 + \mathcal{C}_\text{OM} (\mathcal{L}_+^2 - \mathcal{L}_-^2)$ while the white line represents $\mathcal{C}^{(\text{opt,N})}_\text{OM} = (1 + \mathcal{C}_\text{EM})/(\mathcal{L}_+^2 - \mathcal{L}_-^2)$.}
\label{fig:etaNoise}
\end{figure*}

Looking at a higher level, perfect couplings $\eta_\text{e} = \eta_\text{o} = 1$, high matched cooperativities $\mathcal{C}_\text{OM} \approx \mathcal{C}_\text{EM} > 1$, and decent sideband resolution evidently lead to higher efficiencies $\eta_\text{peak} \approx 1$ and low added noise $N < 1$ as long as thermal phonons are suppressed (upper right regions in Fig.~\ref{fig:etaNoise}). In this high cooperativity regime, strong sideband resolution is the main focus for optical noise, while low thermal occupation is the main requirement for low thermal noise.

In the low cooperativity regime ($\mathcal{C}_\text{OM} < 1$ and $\mathcal{C}^\text{opt}_\text{EM} \approx 1$), achieving high $\eta$ involves maximizing $\mathcal{C}_\text{OM}$ (see green line in Fig.~\ref{fig:etaNoise}(a)). Based on typical performance of existing piezo-optomechanical systems, large optomechanical cooperativity $\mathcal{C}_\text{OM} \gg 1$ is generally more difficult to achieve in chip-integrated optomechanics, particularly in cryogenic environments due to constraints on $n_\text{phot}$ to avoid heating. By exploiting the resonant enhancement discussed in Section~\ref{sec:matchingNetwork}, large $\mathcal{C}_\text{EM} \gtrsim 1$ can easily be achieved even for weak piezoelectric coupling $k^2 \ll 1$ for the material platforms we assess.

\section{Minimizing added noise $N$}
\label{sec:minN}

\noindent Having discussed the maximization of the signal transfer efficiency $\eta$ in the previous section (while also evaluating the resulting noise $N$), we now turn to minimizing $N$. Noting that $N$ is essentially the ratio of noise to signal photons, this optimization strategy is particularly relevant to transducer applications that employ postselection (conditioned on the detection of a photon). In such scenarios, it is largely $N$ that determines the protocol fidelity whereas $\eta$ mainly sets the success rate, and hence the number of repetitions of the protocol required to detect a photon in the output. Thus, our primary focus in this section is on minimizing $N$, and subject to this constraint we seek secondarily to make $\eta$ as large as possible.

\subsection{Analysis}

\noindent In the present context of electrical-to-optical conversion, minimization of $N$ is achieved with the matching network that provides maximal resonant signal enhancement and thus the maximal $\mathcal{C}_\text{EM}=\mathcal{C}_\text{EM}^\text{max}$. This is achieved with $C_\text{T}=0$ while choosing $L$ so as to achieve a joint resonance $\omega_\text{LC}=\omega_\text{m}$ as previously. From Eq.~\eqref{eq:C-EM-def} we have,
\begin{equation}
\mathcal{C}_\text{EM}^\text{max} = \left.\frac{ Z_\text{LC}^2}{R_\text{m} (Z_\text{tx} + R_\text{L})}\right|_{C_\text{T}=0} = \frac{k^2}{\gamma_\text{m}C_{0}(Z_\text{tx}+R_\text{L})},
\label{eq:Cem_max}
\end{equation}
having used Eq.~\eqref{eq:k-square-def} to achieve an expression in terms of the native piezoelectric device parameters. This matching network is optimal for noise suppression insofar as the Ohmic resistance $R_\text{L}$ introduced by the inductor does not excessively degrade $\eta_\text{e}$.

That $\mathcal{C}_\text{EM}=\mathcal{C}_\text{EM}^\text{max}$ leads to minimal $N$ follows directly from Eqs.~\eqref{eq:Nopt1} and~\eqref{eq:Nm}. It remains to decide on the optomechanical parameters $\mathcal{C}_\text{OM}$ and $\mathcal{L}_{-}^2$. In the limit $\mathcal{L}_{-}^2\rightarrow 0$ (while maintaining $\mathcal{L}_{+}^2=1$) the optical amplification noise vanishes ($N_\text{o}\rightarrow 0$) and $C_\text{OM}$ enters $\eta_\text{peak}$ [Eq.~\eqref{eq:peakEta}] only, hence uniquely determining its optimal value in this limit,
\begin{equation}
    \mathcal{C}_\text{OM}^{(\text{opt},N)} = \frac{1 + \mathcal{C}_\text{EM}^\text{max}}{\mathcal{L}_+^2 - \mathcal{L}_-^2} \rightarrow 1 + \mathcal{C}_\text{EM}^\text{max},
    \label{eq:C-OM-optN}
\end{equation}
cf.~Eq.~\eqref{eq:Copt}, resulting in the peak signal transfer efficiency
\begin{equation}
    \left.\eta_\text{peak}\right|_{\mathcal{C}_\text{OM}=\mathcal{C}_\text{OM}^{(\text{opt},N)}}= \eta_\text{e} \eta_\text{o} \frac{\mathcal{C}_{\text{EM}}^\text{max} }{1 + \mathcal{C}_{\text{EM}}^\text{max}},
\end{equation}
cf.~Eq.~\eqref{eq:etaCom_resolved_sideband} (see white line in Fig.~\ref{fig:etaNoise}).

However, the required smallness of $\mathcal{L}_{-}^2$ is typically intractable and, as discussed in previous sections, the piezo-optomechanical transducers considered here are typically in a parameter regime where $\mathcal{C}_\text{OM} \ll 1+\mathcal{C}_\text{EM}^\text{max}$ so that $\mathcal{C}_\text{OM}^{(\text{opt,}N)}$ [Eq.~\eqref{eq:C-OM-optN}] cannot be achieved. We observe that within this regime, the signal transfer efficiency $\eta_\text{peak}$ [Eq.~\eqref{eq:peakEta}] is independent of the optical amplification $\mathcal{L}_{-}^2 > 0$ to leading order in $\mathcal{C}_\text{OM}$,
\begin{equation}
     \eta_\text{peak} \sim \eta_\text{e} \eta_\text{o} \frac{4\mathcal{C}_\text{EM}^\text{max} \mathcal{C}_\text{OM} \mathcal{L}_{+}^2}{(1+\mathcal{C}_\text{EM}^\text{max})^2}.
     \label{eq:eta-peak-smallCOM}
\end{equation}
Consequently, in this regime, the optical Stokes process essentially only adds noise while the amplification in $\eta_\text{peak}$ is negligible.

To proceed, we make the heuristic restriction that the optical amplification noise must be kept below the mechanical thermal noise,
\begin{equation}
N_\text{o} \lesssim N_\text{m}\Leftrightarrow \mathcal{C}_\text{OM} \mathcal{L}_{-}^2 \lesssim n_\text{m}.    
\label{eq:N-o_bound}
\end{equation}
Within this constraint, the product $\eta_{\text{o}}\mathcal{C}_\text{OM}\mathcal{L}_{-}^2$ should be made as large as possible in order to make $\eta_\text{peak}$ [Eq.~\eqref{eq:eta-peak-smallCOM}] large. If operating deeply in the regime of Eq.~\eqref{eq:N-o_bound}, this implies increasing $n_{\text{phot}}$ as much as is permissible and optimizing the optical outcoupling rate $\kappa_{\text{ext}}$ (see Appendix~\ref{sec:kappa_ext_opt} for details).

If optical noise is larger than thermal noise, then Eq.~\eqref{eq:N-o_bound} prompts us to ensure $\mathcal{C}_\text{OM}=n_\text{m}/\mathcal{L}_{-}^2 (\ll 1 + \mathcal{C}_\text{EM}^\text{max})$ by either decreasing $n_\text{phot}$ or $\kappa_\text{ext}$ (in order to lower $\mathcal{L}_{-}^2$), whereby $N=2N_\text{m}$ and Eq.~\eqref{eq:eta-peak-smallCOM} reads
\begin{gather}
    \eta_\text{peak} \sim \eta_\text{e} \eta_\text{o} \frac{4\mathcal{C}_\text{EM}^\text{max} n_\text{m} \mathcal{L}_{+}^2/\mathcal{L}_{-}^2}{(1+\mathcal{C}_\text{EM}^\text{max})^2}\label{eq:eta-peak-Nmatch}\\
    \xrightarrow[\mathcal{C}_\text{EM}^\text{max}\gg 1]{\Delta = -\omega_\text{m}} \eta_\text{e}^2 \eta_\text{o} 2N (1+(4\omega_\text{m}/\kappa_\text{o})^2),
\end{gather}
providing a relatively simple relationship between $\eta_\text{peak}$ and $N$ when optimizing the latter under the stipulated conditions in the regime $\mathcal{C}_\text{OM}\ll 1 + \mathcal{C}_\text{EM}^\text{max}$. 
$\eta_\text{peak}$ in Eq.~\eqref{eq:eta-peak-Nmatch} can be further optimized by choosing the optical outcoupling rate $\kappa_\text{ext}$ that strikes the right balance between $\eta_\text{o}$ and $\mathcal{L}_\pm^2$ (see Appendix~\ref{sec:kappa_ext_opt}).

\subsection{Discussion}

\begin{figure}
\includegraphics[width=1\linewidth]{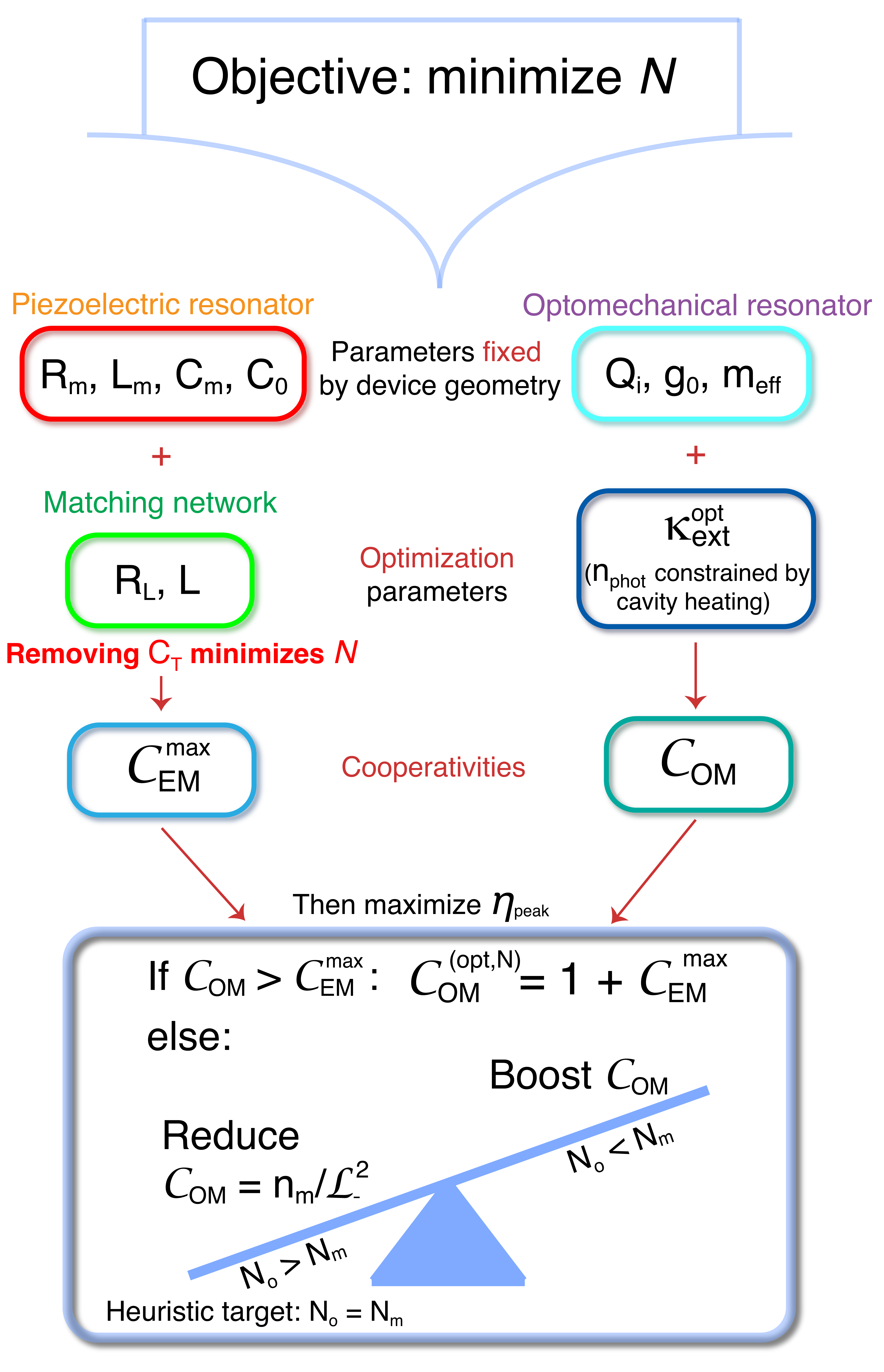}
\caption{Flow chart (from top to bottom) detailing some of the important dependencies for microwave-to-optical transduction in order to minimize added noise $N$.}
\label{fig:MinimizeN}
\end{figure}

\noindent The minimization of noise $N$ is based principally on realizing the maximum potential of the piezoelectric coupling with assistance from the matching inductor $L$ to reach $\mathcal{C}_\text{EM}^\text{max}$. Once reached, it only remains to optimize $\mathcal{C}_\text{OM}$ to achieve a reasonable level of efficiency depending on the noise regime as illustrated in Fig.~\ref{fig:MinimizeN}.

In the rare case that we can achieve $\mathcal{C}_\text{OM} > \mathcal{C}_\text{EM}^\text{max}$, the most judicious choice of $\mathcal{C}_\text{OM}$ is $\mathcal{C}_{\text{OM}}^{(\text{opt},N)}$ [Eq.~\eqref{eq:C-OM-optN}], indicated by the white ridge in Fig.~\ref{fig:etaNoise}, provided that the optical noise $N_{\text{o}}$ does not dominate.  
If optical noise is dominant, then both $N_\text{o}$ and $\mathcal{C}_\text{OM}$ should be scaled back to the heuristic target of $\mathcal{C}_\text{OM}=n_\text{m}/\mathcal{L}_{-}^2$ and $N_\text{o} = N_\text{m}$. 
Otherwise, if added noise is dominated by thermal noise, $\mathcal{C}_\text{OM}$ should be maximized to achieve as large $\eta$ as possible. 

We note that operating at $\mathcal{C}_{\text{EM}}=\mathcal{C}_{\text{EM}}^{\text{max}}$, as considered in the present section, typically implies being in the regime $2g_\text{EM} > \kappa_\text{e}$, where the efficiency spectrum $\eta(\omega)$ exhibits electromechanical normal-mode splitting. In this case, $\eta_{\text{peak}}\equiv\eta(\omega_{\text{m}})$ is no longer a peak value of $\eta(\omega)$, but, crucially, it remains the value of $\eta$ at the Fourier frequency $\omega=\omega_{\text{m}}$ where $N(\omega)$ is minimal (see Appendix~\ref{sec:deriveEtaN} for details). On a related note, we refrain in this regime from discussing the transducer bandwidth $\Delta \omega$ as it is not uniquely defined (see plots in Appendix~\ref{sec:deriveEtaN}).

\section{Application to specific material platforms}
\label{sec:ApplicationMaterials}

\noindent Gallium arsenide (GaAs), aluminum nitride (AlN), and lithium niobate (LiNbO$_3$) are materials currently used in integrated piezoelectric devices. AlN and LiNbO$_3$ exhibit strong piezoelectric effect and are also natural platforms on which to build photonic integrated devices. On the other hand, GaAs exhibits weak piezoelectric effect compared to the other two materials. Its piezoelectric constant $e_{14} = -0.16$\,C/m$^2$ is about an order of magnitude smaller than that of AlN ($e_{33} = 1.55$\,C/m$^2$) and LiNbO$_3$ ($e_{33} = 1.77$\,C/m$^2$). Therefore, as developed in previous sections, an electrical matching network would be beneficial to compensate for lower $k^2$,  and raise $\mathcal{C}_\text{EM}$ through resonant enhancement. On the other hand, GaAs optomechanical devices have been demonstrated with $g_0/(2\pi) = $ 1.1\,MHz~\cite{balram_coherent_2016,balram_moving_2014}, which is nearly an order of magnitude larger than that achieved in the other piezoelectric materials~\cite{vainsencher_bi_2016, jiang_lithium_2019}, due to its higher linear refractive index and larger photoelastic coefficients. This is important given the potential optical-absorption-induced heating expected in a millikelvin environment~\cite{meenehan_pulsed_2015, forsch_microwave_2018, ramp_elimination_2018}, which would restrict $n_{\text{phot}}$ so that appreciable $\mathcal{C}_\text{OM}$ requires large $g_0$. 

Contrary to previous works, our approach considers the optimization of the transduction chain as a whole. In this section, we start with device-level simulations and results from recent experiments to obtain a better perspective on what performance might be realizable in the near term and if certain parts of the system can be further optimized in various material platforms. We use state-of-the-art experimental demonstrations from the literature to extract parameters for a potential piezo-optomechanical transducer while keeping operating frequencies and the overall structure similar to our example in GaAs. The following parameters and design choices are used to mimic realistic constraints in fabrication and experimentation as much as possible
\begin{itemize}
\item A thin film plate made of piezoelectric material with electrodes on top only;
\item The mechanical series resonance frequency is set to $\omega_\text{s}/(2\pi) \approx $ 2.4\,GHz;
\item The piezoelectric resonator is coupled directly to a photonic crystal nanobeam optomechanical cavity with their mechanical frequencies matched;
\item The optical wavelength of the optical cavity is set near 1,550$\,$nm;
\item The effective cryogenic temperature is set to $T$ = 100$\,$mK leading to cold input and superconducting metal circuitry (lossless matching circuit) such that $R_\text{L} = 0\,\Omega$ \cite{harabula_measuring_2017} and hence $\eta_{e} = 1$;
\item Room-temperature values of $k^2$ are maintained here due to lack of data on piezoelectric coefficients in cryogenic environments. In general, the piezoelectric constant $e$ decreases at low temperatures but the level of reduction is material dependent~\cite{bukhari_shear_2014}.
\end{itemize}

\subsection{Piezo-optomechanical transducer in GaAs}

\begin{figure}
\includegraphics[width=1\linewidth]{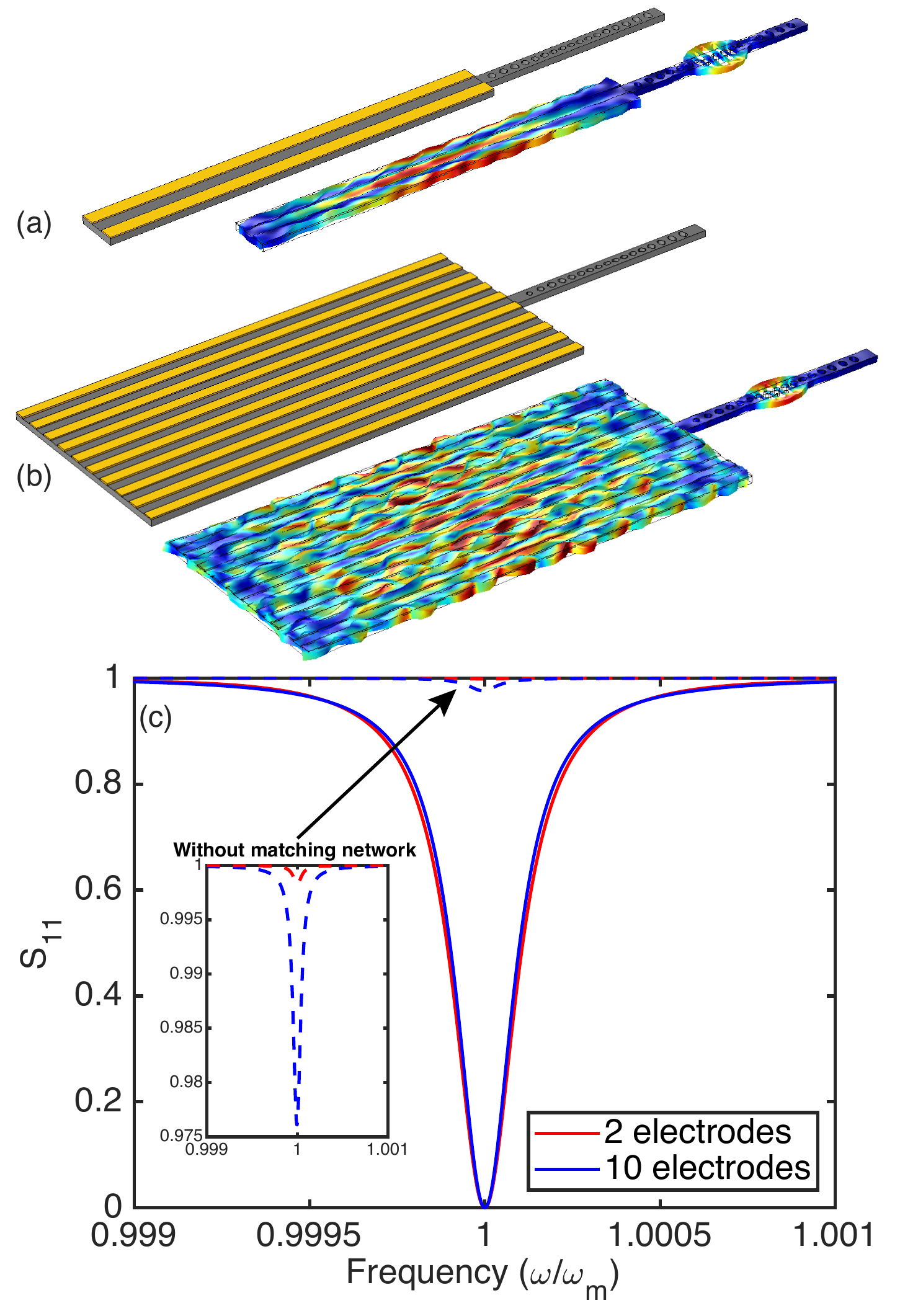}
\caption{Device schematic and mechanical displacement of the target supermode for the (a) two-electrode and (b) ten-electrode designs in GaAs. (c) Reflection spectra $S_\text{11}$ of the piezoelectric frequency response of the coupled resonator (with optomechanical loading) calculated via admittance from numerical simulations for the device alone (dotted lines) and with matching $RLC$ network (solid lines).}
\label{fig:S11}
\end{figure}

\noindent In our specific example for GaAs, we develop a shear-mode piezoelectric resonator in which the mechanical mode is driven piezoelectrically by a row of electrodes (see Fig.~\ref{fig:S11}(a) and (b)). This resonator is directly coupled to a photonic crystal nanobeam optomechanical cavity developed in an earlier work~\cite{balram_moving_2014}. The localized breathing mode of the nanobeam hybridizes with the shear mode in the piezoelectric resonator to form a mechanical supermode. Strong coupling between the vibrations in the plate and the breathing mode in the nanobeam can be achieved as long as the mode splitting is larger than the mechanical decay rates of the individual modes. Our piezo-optomechanical device is simulated using finite-element analysis with a numerical software~\footnote{COMSOL Multiphysics. Certain commercial items are identified to foster understanding and do not constitute an endorsement by the National Institute of Standards and Technology.} to extract piezoelectric circuit parameters and optomechanical properties (see Appendix~\ref{sec:piezo_modelling}).

\begin{table*}[]
\begin{tabular}{l|l|llllll}
\hline
Parameter & Symbol & GaAs (2 el.) & GaAs (10 el.) & GaAs (pot.) & AlN & LiNbO$_3$ & AlN-on-Si\\
\hline
\textbf{Common parameters} & &   &   &   &   &   &  \\
Series mechanical frequency & $\omega_\text{s}/(2\pi)$ & 2.328 GHz & 2.329 GHz & 2.4 GHz & 2.4 GHz & 2.4 GHz & 2.4 GHz \\
Effective temperature & $T$ & 100 mK & 100 mK & 100 mK & 100 mK & 100 mK & 100 mK  \\
Effective mass & $m_\text{eff}$ & 4.5 pg & 30 pg & 4.5 pg & 4.5 pg & 4.5 pg & 4.5 pg \\
Motional resistance & $R_\text{m}$ & 55,000 $\Omega$ & 4,100 $\Omega$ & 5,300 $\Omega$ & 1,300 $\Omega$ & 8.07 $\Omega$ & 9.3788 $\Omega$ \\
Motional inductance & $L_\text{m}$ & 36.47 mH & 2.827 mH & 17.58 mH & 870.73 $\upmu$H & 19.79 $\upmu$H & 621.95 $\upmu$H \\
Motional capacitance & $C_\text{m}$ & 0.128 aF & 1.652 aF & 0.250 aF & 5.051 aF & 222.2 aF & 7.071 aF \\
Static capacitance & $C_0$ & 0.6 fF & 5.7 fF & 0.5 fF & 0.5 fF & 2 fF & 0.7 fF \\
Piezoelectric coupling coefficient & $k^2$ & 0.022 \% & 0.029 \% & 0.05 \% & 1 \% & 10 \% & 1 \% \\
Load resistance & $R_\text{L}$ & 0 $\Omega$ & 0 $\Omega$ & 0 $\Omega$ & 0 $\Omega$ & 0 $\Omega$ & 0 $\Omega$ \\
Acoustic loss rate & $\gamma_\text{m}/(2\pi)$ & 240 kHz & 231 kHz & 48 kHz & 240 kHz & 65 kHz & 2.4 kHz \\
Mechanical quality factor & $Q_\text{m}$ & $\approx$ 10,000 & $\approx$ 10,000 & $\approx$ 50,000 & $\approx$ 10,000 & $\approx$ 38,000 & 10$^6$\\
Optical quality factor (intrinsic) & $Q_\text{i}$ & 77,000 & 77,000 & 700,000 & 130,000 & 10$^6$ & 10$^6$ \\ 
Optomechanical coupling rate & $g_0/(2\pi)$ & 300 kHz & 100 kHz & 300 kHz & 38.333 kHz & 40 kHz & 333 kHz \\
Intra-cavity photon number & $n_\text{phot}$ & 280 & 280 & 1000 & 1000 & 1000 & 1000 \\
Enhanced optomech. coupling rate & $g_\text{OM}/(2\pi)$ & 5 MHz & 1.7 MHz & 9.5 MHz & 1.2 MHz & 1.3 MHz & 10.5 MHz \\
Electrical coupling efficiency & $\eta_\text{e}$ & 1 & 1 & 1 & 1 & 1 & 1 \\
\hline
Electrical decay rate & $\kappa_\text{tx}/(2\pi)$ & 218 Hz & 2.8 kHz & 453 Hz & 9.14 kHz & 402 kHz & 12.8 kHz\\
Optical coupling efficiency & $\eta_\text{o}$ & 0.52 & 0.50 & 0.84 & 0.5 & 0.55 & 0.93 \\
Electromechanical cooperativity &  $\mathcal{C}_\text{EM}$ & 9$\times 10^{-4}$ & 0.0122 & 0.0094 & 0.0381 & 6.2 & 5.33 \\
Optomechanical cooperativity & $\mathcal{C}_\text{OM}$ & 0.08 & 0.0096 & 4.3 & 0.0082 & 0.2287 & 64.4 \\
Impedance (real part) & $\text{Re}(Z)$ & 47 k$\Omega$ & 3,700 $\Omega$ & 26 k$\Omega$ & 1.3 k$\Omega$ & 9.9 $\Omega$ & 563 $\Omega$ \\
Impedance (imaginary part) & $\text{Im}(Z)$ & 24 k$\Omega$ & 1,300 $\Omega$ & 5,400 $\Omega$ & 13 $\Omega$ & 0.003 $\Omega$ & 3.4 $\Omega$ \\
Reflection & $S_\text{11}(\omega_\text{m})$ & 0.9983 & 0.976 & 0.9964 & 0.9272 & 0.6693 & 0.837 \\
\hline
\textbf{Peak transfer efficiency} & $\eta_\text{peak}$ & 0.013 \% & 0.023 \% & 0.51 \% & 0.057 \% & 5.67 \% & 30 \% \\
Transduction bandwidth & $\Delta \omega /(2\pi)$ & 255 kHz & 236 kHz & 248 kHz & 250 kHz & 480 kHz & 157 kHz \\
\textbf{Added total noise} & $N$ & 557 & 40 & 64 & 12 & 0.075 & 1.08 \\
Added optical noise & $N_\text{o}$ & 21 & 0.18 & 15 & 0.019 & 7.4$\times 10^{-5}$ & 0.99 \\
Added mechanical noise & $N_\text{m}$ & 536 & 40 & 49 & 12 & 0.075 & 0.09 \\
\hline
\end{tabular}
\caption{\textbf{Common parameters and performance of bare piezo-optomechanical transducers.} This is the first in a series of tables outlining parameters for potential piezo-optomechanical transducers and comparing different device types and materials. The first three columns are GaAs devices. Columns 1 and 2 are two-electrode and ten-electrode devices, respectively, with parameters from our simulations plus Refs.~\cite{forsch_microwave_2018, ramp_elimination_2018}. Column 3 is a potentially optimized two-electrode device using the best $Q_\text{i}$ achieved in GaAs photonic crystal cavities~\cite{combrie_GaAs_2008} and best $Q_\text{m}$ for isolated nanobeam optomechanical crystals~\cite{forsch_microwave_2018, ramp_elimination_2018}. Columns 4 and 5 are devices in AlN and LiNbO$_3$ with parameters from Refs.~\cite{fan_AlN_OMC} and~\cite{jiang_lithium_2019}, respectively. The last column is a hybrid AlN-on-Si device assuming the best optomechanical performance in Ref.~\cite{macCabe_phononic_2019}. The values for  cooperativities, efficiency and noise (bottom two sections) are calculated based on a BVD-optomechanical circuit without matching network.}
\label{table:parameters}
\end{table*}

\begin{table*}[]
\begin{tabular}{l|l|llllll}
\hline
Parameter & Symbol & GaAs (2 el.) & GaAs (10 el.) & GaAs (pot.) & AlN & LiNbO$_3$ & AlN-on-Si\\
\hline 
\textbf{Maximize efficiency $\eta$ (RLC circuit)} &   &  &  &   &  & \textbf{RC circuit} &  \\
Effective mechanical frequency & $\omega_\text{m}/(2\pi)$ & 2.328 GHz & 2.329 GHz & 2.4 GHz & 2.4 GHz & 2.4 GHz & 2.4 GHz \\
Thermal phonon number & $n_\text{m}$ & 0.4864 & 0.4859 & 0.46 & 0.462 & 0.462 & 0.462\\
Tuning capacitance & $C_\text{T}$ & 39.41 fF & 144.5 fF & 56.164 fF & 257.35 fF & 2.666 pF & 394.4 fF\\
Matching inductance & $L$ & 117 nH & 31 nH & 77.6 nH & 17.05 nH & - & 11.13 nH\\
Piezoelectric coupling rate & $g_\text{EM}/(2\pi)$ & 2.1 MHz & 3.9 MHz & 2.5 MHz & 5.3 MHz & - & 5.1 MHz \\
Electrical coupling rate & $\kappa_\text{e}/(2\pi)$ & 68 MHz & 256 MHz & 102 MHz & 467 MHz & 80 kHz & 715 MHz \\
Optical coupling rate & $\kappa_\text{ext}/(2\pi)$ & 2.8 GHz & 2.54 GHz & 1.47 GHz & 1.5 GHz & 238 MHz & 2.68 GHz \\
Optical decay rate & $\kappa_\text{o}/(2\pi)$ & 5.3 GHz & 5.05 GHz & 1.74 GHz & 3 GHz & 431 MHz & 2.88 GHz \\
Optical quality factor (loaded) & $Q_\text{o}$ & 36,700 & 38,000 & 111,000 & 64,700 & 449,000 & 67,300 \\
Optical coupling efficiency & $\eta_\text{o}$ & 0.52 & 0.50 & 0.87 & 0.5 & 0.55 & 0.93\\
Electromechanical cooperativity &  $\mathcal{C}_\text{EM}$ & 1.06 & 1.007 & 5.166 & 1.0075 & 1.2283 & 60.1 \\
Optomechanical cooperativity & $\mathcal{C}_\text{OM}$ & 0.08 & 0.0096 & 4.3 & 0.0082 & 0.2287 & 64.4 \\
Impedance (real part) & $\text{Re}(Z)$ & 50 $\Omega$ & 50 $\Omega$ & 50 $\Omega$ & 50 $\Omega$ & 50 $\Omega$ & 50 $\Omega$ \\
Impedance (imaginary part) & $\text{Im}(Z)$ & 0 $\Omega$ & 0 $\Omega$ & 0 $\Omega$ & 0 $\Omega$ & 0 $\Omega$ & 0 $\Omega$\\
Reflection & $S_\text{11}(\omega_\text{m})$ & 0 & 0 & 0 & 0 & 0 & 0\\
\hline
\textbf{Peak transfer efficiency} & $\eta_\text{peak}$ & 3.9 \% & 0.48 \% & 70.1 \% & 0.41 \% & 10.3 \% & 99.95 \%\\
Transduction bandwidth & $\Delta \omega /(2\pi)$ & 510 kHz & 466 kHz & 496 kHz & 484 kHz & 159 kHz & 288 kHz \\
\textbf{Added total noise} & $N$ & 0.48 & 0.49 & 0.12 & 0.46 & 0.38 & 0.096 \\
Added optical noise & $N_\text{o}$ & 0.02 & 0.0022 & 0.027 & 0.0007 & 0.0004 & 0.088 \\
Added mechanical noise & $N_\text{m}$ & 0.46 & 0.48 & 0.09 & 0.46 & 0.377 & 0.0077 \\
\hline
\end{tabular}
\caption{\textbf{Maximizing efficiency via input electrical network to match cooperativities and impedances.} The values in this table are computed with the goal of maximizing efficiency $\eta$ using an $RLC$ matching circuit, except the LiNbO$_3$ column where an $RC$ circuit is used due to the low impedance of the bare device.}
\label{table:parameters2}
\end{table*}

\begin{table*}[]
\begin{tabular}{l|l|llllll}
\hline
Parameter & Symbol & GaAs (2 el.) & GaAs (10 el.) & GaAs (pot.) & AlN & LiNbO$_3$ & AlN-on-Si\\
\hline
\textbf{Minimize noise $N$ (RL circuit)} & &   &   &   &   &   &  \\
Effective mechanical frequency & $\omega_\text{m}/(2\pi)$ & 2.3279 GHz & 2.3293 GHz & 2.4006 GHz & 2.4121 GHz & 2.5298 GHz & 2.4121 GHz \\
Thermal phonon number & $n_\text{m}$ & 0.4863 & 0.4858 & 0.4619 & 0.458 & 0.422 & 0.4582 \\
Matching inductance & $L$ & 7.909 $\upmu$H & 814.3 nH & 8.791 $\upmu$H & 8.7073 $\upmu$H & 1.979 $\upmu$H & 6.220 $\upmu$H\\
Piezoelectric coupling rate & $g_\text{EM}/(2\pi)$ & 17 MHz & 19.77 MHz & 26.84 MHz & 121 MHz & 400 MHz & 121 MHz\\
Electrical coupling rate & $\kappa_\text{e}/(2\pi)$ & 1 MHz & 9.77 MHz & 905 kHz & 914 kHz & 4 MHz & 1.28 MHz\\
Optical coupling rate & $\kappa_\text{ext}/(2\pi)$ & 2.5 GHz & 2.51 GHz & 278 MHz & 1.49 GHz & 194 MHz & 96.8 MHz\\
Optical decay rate & $\kappa_\text{o}/(2\pi)$ & 5.03 GHz & 5.03 GHz & 553 MHz & 2.98 GHz & 387 MHz & 290 MHz \\    
Optical quality factor (loaded) & $Q_\text{o}$ & 38,500 & 38,500 & 350,000 & 65,000 & 500,000 & 667,000 \\
Optical coupling efficiency & $\eta_\text{o}$ & 0.5 & 0.5 & 0.5 & 0.5 & 0.5 & 0.333\\
Electromechanical cooperativity &  $\mathcal{C}_\text{EM}$ & 4,860 & 691 & 66,315 & 265,260 & 2.45 $\times 10^{6}$ & 1.89 $\times 10^{7}$ \\
Intra-cavity photon number & $n_\text{phot}$ & 280 & 280 & 1000 & 1000 & 1000 & 794 \\
Optomechanical cooperativity & $\mathcal{C}_\text{OM}$ & 0.08 & 0.0096 & 14 & 0.0082 & 0.2549 & 507 \\
Impedance (real part) & $\text{Re}(Z)$ & 228 k$\Omega$ & 34 k$\Omega$ & 228 k$\Omega$ & 13 M$\Omega$ & 98 M$\Omega$ & 1.87 M$\Omega$ \\
Impedance (imaginary part) & $\text{Im}(Z)$ & 0 $\Omega$ & 0 $\Omega$ & 0 $\Omega$ & 0 $\Omega$ & 0 $\Omega$ & 0 $\Omega$ \\
Reflection & $S_\text{11}(\omega_\text{m})$ & 0.9994 & 0.9971 & 0.9996 & $\approx 1$ & $\approx 1$ & $\approx 1$ \\
\hline
\textbf{Peak transfer efficiency} & $\eta_\text{peak}$ & 0.0034 \% & 0.0028 \% & 0.041 \% & 6$\times 10^{-6}$ \% & 2$\times 10^{-6}$ \% & 0.0036 \%\\
Transduction bandwidth & $\Delta \omega/(2\pi)$ & - & - & - & - & - & - \\
\textbf{Added total noise} & $N$ & 1$\times 10^{-4}$ & 7.1$\times 10^{-4}$ & 7.6$\times 10^{-6}$ & 1.73$\times 10^{-6}$ & 1.7$\times 10^{-7}$ & 4.8$\times 10^{-8}$ \\
Added optical noise & $N_\text{o}$ & 3.9$\times 10^{-6}$ & 3.1$\times 10^{-6}$ & 6.8$\times 10^{-7}$ & 2.7$\times 10^{-9}$ & 1.5$\times 10^{-10}$ & 2.4$\times 10^{-8}$ \\
Added mechanical noise & $N_\text{m}$ & 1$\times 10^{-4}$ & 7$\times 10^{-4}$ & 7$\times 10^{-6}$  & 1.73$\times 10^{-6}$ & 1.7$\times 10^{-7}$ & 2.4$\times 10^{-8}$ \\
\hline
\end{tabular}
\caption{\textbf{Minimizing added noise via maximizing $\mathcal{C}_\text{EM}$.} The values in this table are computed for minimizing noise $N$ by using an input $RL$ circuit. The  first five columns fit the case where $N_\text{o} < N_\text{m}$, therefore, $\mathcal{C}_\text{OM}$ is maximized. In the last column, $N_\text{o} > N_\text{m}$, therefore the target is set to equate both sources of noise. The transduction bandwidth is not defined here due to normal-mode splitting in $\eta(\omega)$, a consequence of $2g_\text{EM}>\kappa_e$.}
\label{table:parameters3}
\end{table*}

From simulations, we find that the piezoelectric coupling coefficient in GaAs devices is low ($k^2 \approx 0.02\,\%-0.03\,\%$). We assume $Q_\text{m} = \omega_\text{m}/\gamma_\text{m} \approx 10^4$, based on recent demonstrations of isolated GaAs nanobeam optomechanical crystal resonators with $Q_\text{m}=20,000-30,000$ at $T<100\,$mK~\cite{forsch_microwave_2018, ramp_elimination_2018} and the potential increase in losses due to hybridization with the shear-mode resonator. We also assume a modest intrinsic quality factor $Q_\text{i} =$ 77,000 for the optical cavity and $n_\text{phot} = 280$ intracavity photons, consistent with recent experiments demonstrating ground-state operation of GaAs nanobeam optomechanical crystals~\cite{forsch_microwave_2018,ramp_elimination_2018}. 

The first two columns of Table~\ref{table:parameters} show two possible configurations of a GaAs piezo-optomechanical transducer with two electrodes and ten electrodes, corresponding to Fig.~\ref{fig:S11}(a) and (b), respectively. In the two-electrode case, the small capacitance $C_\text{m}$ leads to high motional resistance $R_\text{m}$ and, without matching network (see Appendix~\ref{sec:RCcircuit}), high impedance $Z$. The optomechanical coupling $g_0/(2\pi) = 300\,$kHz is determined by numerical simulations and takes into account the whole supermode with effective mass $m_\text{eff}$ = 4.5\,pg, an order of magnitude larger than the standalone breathing mode in the nanobeam. With the aforementioned low intracavity photon number to prevent heating ($n_\text{phot} = 280$), the optomechanical cooperativity is $\mathcal{C}_\text{OM}$ = 0.08. Note that GaAs nanobeam optomechanical crystals have reached $\mathcal{C}_\text{OM}>1$ at room temperature~\cite{balram_moving_2014, forsch_microwave_2018} due to both the larger $g_0$ of the isolated nanobeam geometry and the increased $n_\text{phot}$ available when no restriction on heating of the mechanical resonator is imposed. Reducing the motional resistance $R_\text{m}$ can be of interest to simplify impedance-matching requirements, and can be achieved by increasing the capacitance of the device by expanding the area of the piezoelectric resonator and adding electrodes. Using a ten-electrode geometry as an example, $R_\text{m}$ is reduced by an order of magnitude with a corresponding reduction in $Z$. However, the corresponding increase in effective mass to 30\,pg causes $g_0/(2\pi)$ to drop even further to 100$\,$kHz which reduces $\mathcal{C}_\text{OM}$ below 0.01. Moreover, the larger size of the ten-electrode piezoelectric resonator may lead to possible spurious modes that act as effective loss channels if made close in frequency to our mode of interest. In practice, fabrication nonidealities and asymmetries may lead to their mechanical coupling. In the end, efficiencies $\eta_\text{peak} \approx 0.01~\%$ achieved by our currently simulated devices without employing a matching network are higher than some traveling-wave schemes, but remain low for efficient quantum transduction purposes.

In Table~\ref{table:parameters2}, a matching $RLC$ circuit is added for maximizing efficiency. With cooperativity matching and $\eta_\text{e} = 1$, reflection is reduced to zero at the effective frequency $\omega_\text{m}$ in Fig.~\ref{fig:S11}(c) compared to the case of mismatched impedance (see inset). This is due to a large boost in $\mathcal{C}_\text{EM}$ which is tuned to match $\mathcal{C}_\text{OM}$ according to Eq.~\ref{eq:Copt} and, simultaneously, a reduction of the impedance of the device via $L$ and $C_\text{T}$ with respect to the input $Z_\text{tx}$. Peak transfer efficiencies $\eta_\text{peak} \approx 1~\%$ are now achievable due to impedance matching.

We now turn to minimizing noise $N$ in table~\ref{table:parameters3}. The introduction of a large inductor $L$ (no tuning capacitor $C_\text{T} = 0\,$F) releases the full potential of the electromechanical coupling so that the added noise reaches a level around $N \approx 10^{-4}$, limited by thermal noise, in our two examples. The thermal-noise contribution $N_\text{m}$ is 2 orders of magnitude larger than $N_\text{o}$, as expected from Eq.~\eqref{eq:Nopt1} for systems with adequate sideband resolution and small $\mathcal{C}_\text{OM}$. However, the two-electrode device demonstrate higher electromechanical potential with $\mathcal{C}_\text{EM}$ an order of magnitude larger than the ten-electrode device (due to smaller $C_0$), and thus exhibits lower $N_\text{m}$ and, in turn, lower $N$. The low noise $N\ll 1$ in this case comes at the expense of low efficiency $\eta \ll 1$.

Increasing the transduction efficiency appreciably requires an increase in $\mathcal{C}_\text{OM}$ while being able to maintain cooperativity matching, i.e., $\mathcal{C}_\text{EM} = 1 + \mathcal{C}_\text{OM}$ in the limit of adequate sideband resolution. Assuming $g_0/(2\pi)$ = 300$\,$kHz as in the targeted two-electrode device, improvements in $\mathcal{C}_\text{OM}$ can be realized through improved $n_\text{phot}$, $Q_\text{o}$, and $Q_\text{m}$, with the latter, along with potentially higher $k^2$, also resulting in increased $\mathcal{C}_\text{EM}$. This would ensure that cooperativity matching can be achieved without requiring exceedingly small values of $C_\text{T}$ (i.e., $C_\text{T}$ can remain substantially larger than any expected parasitic capacitance). Improvements in $Q_\text{o}$, $Q_\text{m}$, and $k^2$ should be possible through improved design and fabrication, for example, incorporating optimized photonic and phononic shielding. In fact, GaAs-based nanophotonic devices have exhibited much higher $Q_\text{o}$ than that assumed so far, with intrinsic $Q_\text{i} \approx~7\times10^5$ and $\approx~6\times10^6$ demonstrated in two-dimensional photonic crystals~\cite{combrie_GaAs_2008} and microdisks~\cite{Guha_High_2017}, respectively, and our numerical simulations indicate that $Q_\text{i}>10^6$ is achievable in our system from a radiation-loss perspective. Optical absorption is expected to be reduced for such high-$Q$ geometries, suggesting that, together with improved thermalization~\cite{qiu_high-fidelity_2019}, larger $n_\text{phot}$ = 1,000 can potentially be achieved.  Finally, as mentioned earlier, $Q_{\text{m}}~\approx~20,000-30,000$ has already been observed for GaAs optomechanical crystals at $T<100\,$mK, and the achievement of ultra-high $Q_{\text{m}}$ values in silicon-based devices~\cite{meenehan_silicon_2014, macCabe_phononic_2019} will help inform approaches to further increase $Q_\text{m}$ in GaAs.

Taking these improved parameters ($Q_\text{o}$ = 94,000, $Q_\text{i}$ = 700,000, $Q_\text{m}$ = 50,000, and $n_\text{phot}$ = 1,000) into account, we arrive at the predicted performance for a more optimized GaAs device in the third column of Tables~\ref{table:parameters}-\ref{table:parameters3}. Here, we find that an efficiency $\eta_\text{peak} \approx 70\,\%$ is possible in the maximal $\eta$ case (corresponding $N\approx0.1$) and $N \approx 10^{-5}$ when minimizing $N$ (corresponding $\eta\approx0.04\,\%$). While challenging, these outstanding transducer performance metrics appear to be within reach of current technology. 

Given the importance that $\mathcal{C}_\text{OM}>1$ (for $n_\text{phot}$ small enough to avoid heating) plays in realizing efficient transduction, increasing $g_{0}$ could be of particular benefit. The designs presented above are not necessarily optimal in this regard.  In Appendix~\ref{sec:trade-off}, we discuss how $g_{0}$ can be increased if a shorter (and smaller motional mass) piezoelectric resonator is employed, but that this choice necessitates an impedance-matching network with a larger tuning inductance and smaller tuning capacitance to impedance match to the microwave input.

\subsection{Piezo-optomechanical transducer in AlN and LiNbO$_3$}

\noindent Stronger piezoelectric materials such as AlN and LiNbO$_3$ have been used as piezo-optomechanical platforms in the context of optical modulation~\cite{fan_AlN_OMC, tadesse_acousto-optic_2015, ghosh_laterally_2016} and microwave-to-optical conversion~\cite{bochmann_nanomechanical_2013, vainsencher_bi_2016, jiang_lithium_2019}. In purely piezoelectric resonators, the effective piezoelectric coupling coefficient $k^2$ can reach 3\,\% to 7\,\% in AlN~\cite{cassella_super_2017} and as high as 30\,\% in LiNbO$_3$~\cite{Yang_5_2017, Yang_1.7_2018, Wang_Design_2015}. Since $k^2$ is reduced when electrodes are placed solely on the top surface (as is the case for the geometries we consider), conservative numbers of $k^2$ = 1\,\% and 10\,\% are chosen for our examples, respectively. State-of-the-art photonic crystal nanobeam cavities now exhibit excellent optomechanical performance, for which the following parameters are extracted: $Q_\text{i}$ = 130,000, $Q_\text{m} = $ 10,000 (reached at $T$ = 2.5\,K), $g_0/(2\pi)$ = 115$\,$kHz for AlN~\cite{vainsencher_bi_2016, fan_AlN_OMC}; $Q_\text{i}$ = $10^6$, $Q_\text{m} = $ 37,000, and $g_0/(2\pi)$ = 120\,kHz for LiNbO$_3$~\cite{li_photon-level_2019, jiang_lithium_2019}. For comparison, we also assume that the optomechanical coupling $g_0$ is reduced by about a factor of 3 when adopting a similar device geometry ($m_\text{eff}\approx 4.5\,$pg) as in GaAs such that $g_0/(2\pi)$ = 38.3 kHz and 40 kHz for AlN and LiNbO$_3$, respectively. The static capacitance $C_0$ of devices built in AlN and LiNbO$_3$ are taken from numerical simulations as 0.5\,fF and 2\,fF, respectively, while $C_\text{m}$ is calculated from the aforementioned values of $k^2$.

These parameters are employed to showcase potential performance of monolithic AlN and LiNbO$_3$ piezo-optomechanical transducers as shown in the fourth and fifth columns, respectively, of all three tables. The photon number $n_\text{phot}$ = 1,000 is left the same as in the final GaAs example for comparison purposes. For the bare device (without a matching network) with parameters listed in Table~\ref{table:parameters}, the efficiency $\eta_\text{peak}$ reached for AlN is at the $0.1\,\%$ level, while LiNbO$_3$ is significantly higher with $\eta_\text{peak} \approx 6\,\%$.  When $\mathcal{C}_\text{EM}$ is instead matched to $\mathcal{C}_\text{OM}$ via Eq.~\eqref{eq:Copt} (as in Table~\ref{table:parameters2}), higher optimal efficiencies of $\eta_\text{peak}$ = 0.4\,\% and 10\,\% can be attained for AlN and LiNbO$_3$, respectively. This performance is lower than the potential device in GaAs (third column in Table~\ref{table:parameters2}) and highlights the bottleneck of low $g_0$. Due to the low initial impedance of the LiNbO$_3$ example ($Z_\text{BVD}<Z_\text{tx}$, see Appendix~\ref{sec:piezo_modelling}), the $RLC$ matching network should not enhance the signal (i.e., it should have $Q_\text{LC} < 1$) and hence a simpler $RC$ matching network is used since a resonance is not required (see Appendix~\ref{sec:RCcircuit} for details).

When minimizing $N$, the strong piezoelectric performance of these materials is prominent with its added noise $N$ lower than that of GaAs. From Eq.~\eqref{eq:Cem_max}, we can deduce that $\mathcal{C}_\text{EM}^\text{max}$ is 1 to 2 orders of magnitude larger thanks to higher $k^2$.

\subsection{Piezo-optomechanical transducer in AlN on Si}

\noindent One solution to escape the mismatched cooperativity conundrum in monolithic piezoelectric materials is to consider hybrid systems. Potential combinations include AlN-on-Si or LiNbO$_3$-on-Si platforms. The obvious appeal of such systems is the potential to combine the outstanding performance of Si optomechanical crystal devices, in which $g_0/(2\pi) \approx$ 1~MHz, $Q_{\text{o}} \approx 10^6$, and $Q_\text{m} > 10^9$ have been achieved~\cite{chan_optimized_2012, meenehan_silicon_2014, macCabe_phononic_2019}, with the aforementioned electromechanical performance of AlN and LiNbO$_3$.  Of course, the development of hybrid platforms comes with its own challenges, including those related to physical integration of the different materials and the extent to which hybridization (e.g., of mechanical modes across the materials) reduces the performance observed in the individual platforms.  

For comparison, we take AlN on Si as our example of a hybrid platform. We assume a piezo-optomechanical device with a base layer made of silicon on which the piezoelectric section is patterned on top with a layer of AlN before finishing with electrodes ($C_0 = $ 0.7\,fF according to simulations). Incorporating two different materials can result in a combination of their best assets but also of their drawbacks; their effects on the joint performance are not quantifiable \textit{a priori}. The parameters used in this exercise are therefore assumed to be slightly lower than a pure silicon optomechanical device with $g_0/(2\pi) =$ 333\,kHz (reduced due to larger device), $Q_{\text{i}} = 10^6$, and $Q_\text{m} = 10^6$.  With high overall performance, it is evident that devices based on AlN on Si can reach high efficiencies $\eta_\text{peak} \approx 30\,\%$ even without a matching network to equalize the cooperativities (see Table~\ref{table:parameters}). The addition of a matching network takes the efficiency to near unity. As for minimal added noise, AlN-on-Si can reach a groundbreaking level of $N \approx 10^{-8}$ with reduced $n_\text{phot}$ to lower $N_\text{o}$ to the level of $N_\text{m}$.

\subsection{Optical-to-microwave conversion}

In previous sections, we alluded to the bidirectional nature of these transducers in their ability to operate in the forward and reverse directions. The overall transfer efficiency $\eta$ is identical in both directions. However, the noise terms are different depending on the choice of input and output ports. In particular, in the optical-to-microwave direction, the following substitutions for $N$ must be made: $\eta_\text{e} \rightarrow \eta_\text{o}$ and $\mathcal{C}_\text{EM} \rightarrow \mathcal{C}_\text{OM}$ in Eqs.~\eqref{eq:Nopt1} and~\eqref{eq:Nm} along with other replacements detailed in Appendix~\ref{sec:optical2electrical}. From this, one can conclude that low noise in the reverse direction relies heavily on high optical performance, including sideband resolution for optical noise and optomechanical cooperativity for thermal noise. In current devices where $\mathcal{C}_\text{OM}$ seems to be the bottleneck, noise in the optical-to-microwave transduction direction has been observed to be higher than in the forward direction~\cite{vainsencher_bi_2016}. This motivates the current focus on microwave-to-optical conversion as long as the devices exhibit relatively small $\mathcal{C}_\text{OM} < 1$. Overall, good bidirectional operation requires $\mathcal{C}_\text{OM} \approx \mathcal{C}_\text{EM} \gg \text{max}\{1,n_\text{m}/\eta_\text{o}, n_\text{m}/\eta_\text{e} \}$ and $\mathcal{L}_-^2/\eta_\text{o} \ll 1$ to ensure low noise in both directions and large $\eta$. This bidirectional regime is within reach of the potential device in GaAs (third column in Table~\ref{table:parameters2}) and fully achieved in our AlN-on-Si example (last column of the same table).

\subsection{Discussion}

\begin{table}[]
\begin{tabular}{llll}
\hline
Material & Current & Supermode & Supermode+network \\
\hline
GaAs     & $10^{-10}$ \% & 0.5 \%  & 70 \% \\
AlN      & 0.01 \% & 0.06 \% & 0.4 \% \\
LiNbO$_3$ & $10^{-6}$ \% & 6 \% & 10 \% \\
AlN on Si & - & 30 \% & $\approx$ 100 \% \\
\hline
\end{tabular}
\caption{Comparison table of efficiency $\eta_\text{peak}$ between various device types in different piezoelectric materials. The first column represents current experimental values of nanobeam optomechanical devices with mechanical excitation driven by IDTs in GaAs~\cite{forsch_microwave_2018}, focused IDTs in AlN~\cite{vainsencher_bi_2016}, and electrodes at each end in LiNbO$_3$~\cite{jiang_lithium_2019}. The second and third columns show potential devices implementing our proposed mechanical supermode concept and the same with a matched input electrical network, respectively.}
\label{table:comparison}
\end{table}

\noindent The target for an optimized microwave-to-optical transducer is to achieve high transduction efficiency $\eta$ and low added noise $N$, which can be realized in the limit of large, matched electromechanical and optomechanical cooperativities.  Reaching this regime is quite challenging, however. On the electromechanical side, developments within the electromechanics community on platforms such as thin-film LiNbO$_3$ suggest that large $\mathcal{C}_\text{EM}$ can be achieved, as we see in Table~\ref{table:parameters}. However, realizing a large and matched $\mathcal{C}_\text{OM}$ is difficult~\cite{han_high_2018}, both because of the relatively low $n_{\text{phot}}$ required to eliminate adverse heating effects, and the comparatively small $g_{\text{0}}$ that has been achieved in AlN and LiNbO$_3$ in comparison to materials like GaAs. As a result, the high $\mathcal{C}_\text{EM}$ that is achievable in LiNbO$_3$ is in some sense 'wasted' by the difficulty in reaching a correspondingly high $\mathcal{C}_\text{OM}$ if the goal is to reach high efficiency $\eta$. However, high-$\mathcal{C}_\text{EM}$ systems might be ideal for lowering added noise $N$. In this case, large piezoelectric coupling, low mechanical loss, and low $C_0$ are desirable. Reducing $N$ as much as possible essentially insulates the electromechanical subsystem from any external coupling, and therefore results in almost perfect reflection of the input signal at the transducer (Table~\ref{table:parameters3}), yielding a low $\eta$.

\begin{table}[]
\begin{tabular}{lll}
\hline
Material & EM coupling & OM coupling \\
Metric & $k_\text{EM}^2 = \frac{e^2}{\epsilon c}$ & $M = \frac{n^6 p^3}{\rho v^3}\times 10^{16}$ \\
Units & \% & s$^3$/kg \\
\hline
Silicon   & 0   & 300 \\
Quartz    & 1   & 17 \\
GaAs      & 0.4 & 2000 \\
GaP       & 0.2 & 630 \\
GaN       & 1.3 & 1.3 \\
GaPO$_4$  & 1.7 & 500 \\
AlN       & 7   & 0.2 \\
LiNbO$_3$ & 17  & 26 \\
BaTiO$_3$ & 60  & 1200 \\
\hline
\end{tabular}
\caption{Table comparing the bulk electromechanical and optomechanical strengths of some commonly used materials (adapted from Ref.~\cite{delsing_the_2019}). The electromechanical coupling coefficient ($k_\text{EM}^2$, material only) is defined in terms of the piezoelectric coefficient ($e$), the dielectric constant ($\epsilon$), and the elastic coefficient ($c$). The optomechanical figure of merit ($M$) is defined ($\lambda$=1.55~$\upmu$m) in terms of the refractive index ($n$), the photoelastic coefficient ($p$), density ($\rho$), and the speed of sound ($v$). Displayed values are based on the maximum piezoelectric and photoelastic coefficients for the materials.}
\label{table:materials}
\end{table}

In GaAs, heating due to the optical field is also an issue, restricting $n_{\text{phot}}$, but the significantly larger $g_{0}$ means that appreciable $\mathcal{C}_\text{OM}$ can more readily be achieved, particularly considering its squared dependence on $g_0$. Moreover, geometries that allow for better thermal dissipation such as two-dimensional photonic crystals~\cite{ren_quasi-2D_2019} or higher bandgap materials such as gallium phosphide~\cite{schneider_optomechanics_2019} might mitigate the heating problem. On the other hand, the electrical and mechanical resonance enhancement enables $\mathcal{C}_\text{EM}>1$ even with low $k^2$.  As a result, high-performance piezo-optomechanical transducers in GaAs seem to be within reach --- achieving $\mathcal{C}_\text{OM}\approx 4$ and matched $\mathcal{C}_\text{EM}$ (= 5) with adequate sideband resolution and at $T<100$\,mK results in $\eta\approx70\,\%$. Moving to higher $\eta$ and lower $N$ requires improvements in optical and mechanical loss (i.e., higher $Q_\text{o}$ and $Q_\text{m}$), with the latter providing benefit to both the electromechanical and optomechanical subsystems, and the former ideally occurring together with reduced thermo-optic heating, enabling larger $n_{\text{phot}}$ to be used. Alternatively, $g_0$ can be increased even further by reducing the motional mass $m_\text{eff}$ of the supermode by, for example, reducing the length of the piezoelectric resonator (see Appendix~\ref{sec:trade-off}), at the cost of a more impractical implementation of the matching network.

Overall, mechanical hybridization into a supermode is a key step in realizing efficient transduction between the microwave and optical domains, with its impact on the efficiency relative to current piezo-optomechanical transducer devices summarized in Table~\ref{table:comparison}. When combined with a suitably tailored matching network, this approach offers the possibility to reach high transduction efficiency and low added noise in low piezoelectric materials such as GaAs, representing a vast improvement relative to the current state of the art. Alternately, AlN on Si seems to offer the best of both worlds (piezoelectric and optomechanical performance), assuming no degradation in performance when creating the hybrid platform. Other materials such as gallium orthophosphate (GaPO$_4$) and barium titanate (BaTiO$_3$), which can simultaneously support strong optomechanical and electromechanical effects, are worth consideration (see Table~\ref{table:materials} and Refs.~\cite{hamidon_finite_2009, gao_recent_2017}).

\section{Conclusions}
\label{sec:Conclusions}

\noindent In summary, we propose an approach for microwave-to-optical transduction by hybridizing the mechanical modes of piezoelectric and optomechanical resonators into a mechanical supermode. An $RLC$ matching network is incorporated to engineer the electromechanical interaction and impedance match to the input microwave transmission line. Each part of the transducer is analyzed and optimized via an equivalent circuit model in which device-level parameters are linked to figures of merit for conversion efficiency and added noise. Using data from recent experiments in platforms such as GaAs, AlN, LiNbO$_3$, and AlN-on-Si as a guide, our analysis shows that high efficiencies $>50\,\%$ and low added noise at the level of $10^{-6}$ photons are achievable by optimizing for high optomechanical and electromechanical coupling, respectively. These transducers can enable new quantum applications such as remote entanglement of superconducting quantum nodes and state-transfer protocols~\cite{Zeuthen_Figures_2016}.

\section{Acknowledgments}
\noindent M.W.~and E.Z.~contributed equally to this work. We thank Simon Gr\"{o}blacher, Vikrant Gokhale, and Christian Haffner for useful discussions.

M.W. acknowledges support under the Cooperative Research Agreement between the University of Maryland and NIST-CNST/PML, Award No. 70NANB10H193. M.W. and K.S. acknowledge support from the ARO/LPS CQTS program. E.Z.~acknowledges funding from the Carlsberg Foundation. K.C.B. acknowledges support from the European Research Council (StG SBS 3-5, 758843).

\newpage

\appendix
\numberwithin{equation}{section} 
\begin{center}
\textbf{APPENDIX}
\end{center}

\section{Design and modeling of coupled piezoelectric and optomechanical resonators}
\label{sec:piezo_modelling}
\hspace{0.1in}

\noindent In this Appendix, we give a brief discussion of the design and modeling of the piezoelectric resonator.  Our current device design in GaAs follows the simple geometry of interdigitated electrodes on a rectangular suspended plate~\cite{Wang_Design_2015,cassella_super_2017,Yang_5_2017,Yang_1.7_2018}. The examples in the main text assumes a 220-nm GaAs film in the \{100\} crystal orientation loaded with 50-nm-thick aluminum electrodes with width and spacing of 475\,nm which piezoelectrically drive the mechanical mode. The piezoelectric section is $16\,\upmu \text{m}$ long and is directly attached in-line to a $7$-$\upmu \text{m}$-long optomechanical nanobeam cavity~\cite{balram_moving_2014}. 

\begin{figure}
\includegraphics[width=0.95\linewidth]{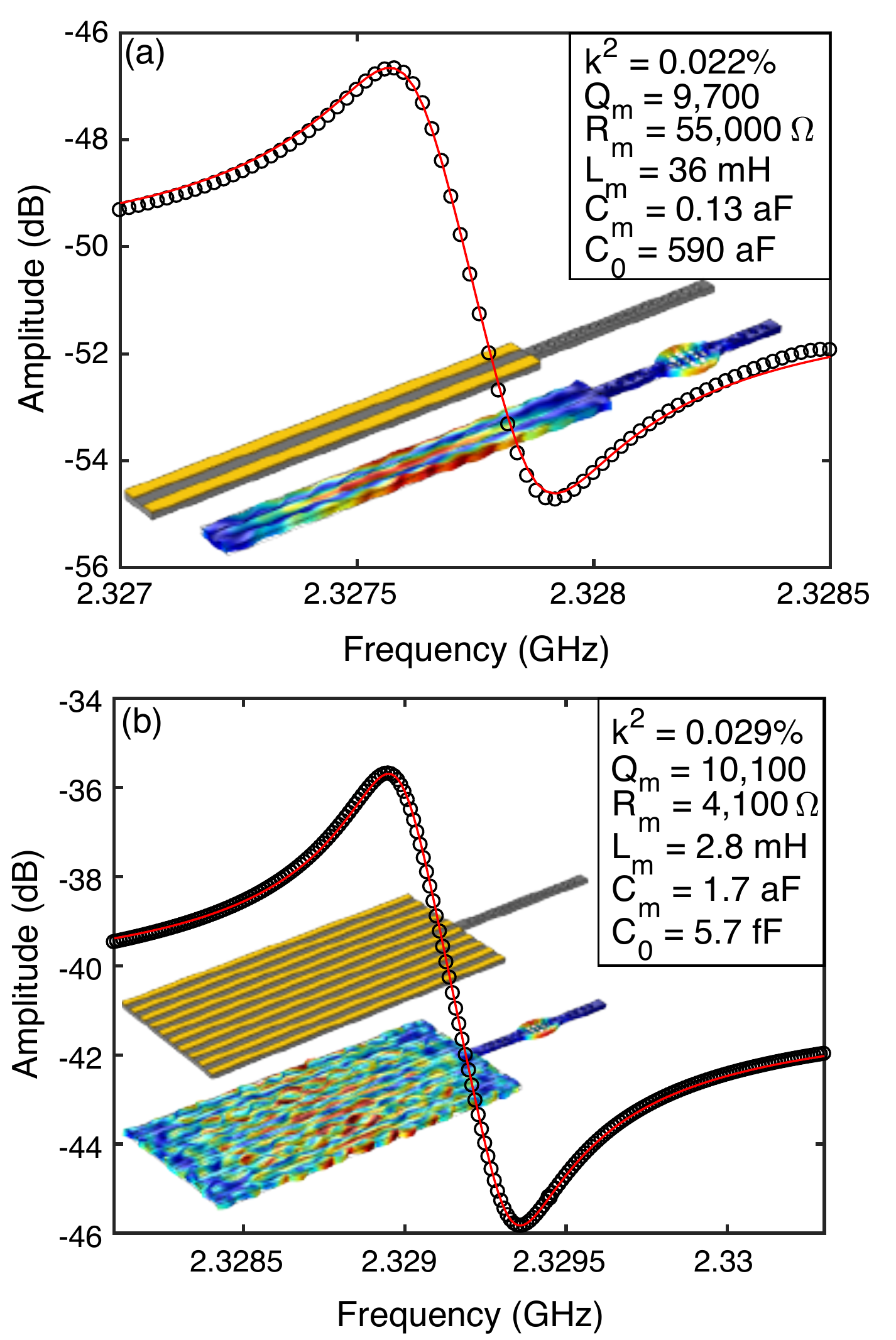}
\caption{Coupled piezoelectric and optomechanical resonators in GaAs. Piezoelectric frequency responses of the coupled resonator calculated via admittance from numerical simulations (black circles) fitted to the Butterworth-van Dyke model (red line). In the background are the schematic of the device and the mechanical displacement of the supermode for the (a) two-electrode and (b) 10-electrode designs.}
\label{fig:CoupledResonators}
\end{figure}

Due to the anisotropy of GaAs, only shear modes are piezoelectrically active, examples of which are shown in the background of Fig.~\ref{fig:CoupledResonators}. Unlike the case in IDTs on bulk material, the acoustic energy of the thin-film shear mode is mostly confined within the center of the coupled resonator where the acoustic leakage can be controlled by support tethers and phononic shielding as shown in Fig.~\ref{fig:coupled_res_example}(a). Coupling between the shear mode and the breathing mode in the optomechanical cavity is executed by engineering the holes in the nanobeam while tuning the frequency of the piezoelectric resonator via variations in electrode pitch. When the two mechanical modes are tuned to the same resonance frequency, their modes hybridize and a mode anticrossing can be observed. In our numerical simulations, the formation of a supermode is further verified by observing a fixed phase relationship between the two parts of the mode as they oscillate collectively. Other modalities of operation are possible using coupled mechanical modes (i.e., detuning the piezoelectric resonator away from the optomechanical resonator) but is not further explored in this work. 

The piezoelectric response of the coupled resonator is computed via finite-element method with its admittance ($Y_\text{BVD} \equiv Z^{-1}_\text{BVD}$) fitted to the BVD model as~\cite{garcia-rodriguez_low_2010,OConnell_piezo-quantum-limit_2014}
\begin{equation}
    Z_\text{BVD} = \frac{1}{-i\omega C_0} \frac{(\omega_\text{s}^2-\omega^2)-i\omega R_\text{m}/L_\text{m}}{(\omega_\text{p}^2-\omega^2)-i\omega R_\text{m}/L_\text{m}},
\end{equation}
where $\omega_\text{s}$ ($\omega_\text{p}$) is the series (parallel) resonance frequency as described in Section~\ref{sec:matchingNetwork} of the main text. The piezoelectric parameters are then extracted (in particular, $R_{\text m}$, $L_{\text m}$, $C_{\text m}$, $C_0$, and $k^2$) as shown in Fig.~\ref{fig:CoupledResonators} and used in Table~\ref{table:parameters}.

The geometries presented here illustrate the main features exploited using the supermode approach, but further optimization may be possible to, for example, increase $g_{0}$. In general, there are a number of different supermode designs~\cite{Sridaran_Electrostatic_2011,ghosh_piezoelectric_2015} that one can consider as a starting point.

\section{Piezo-optomechanical equivalent circuit}
\label{sec:equiv-circ}
\hspace{0.1in}

\noindent In this Appendix, we introduce in more detail the piezo-optomechanical equivalent circuit in Fig.~\ref{fig:BVD-LCcircuit-Tx-Opto}. While a rigorous derivation can be found in Ref.~\cite{Zeuthen_Electrooptomechanical_2018}, here we confine ourselves to a mainly qualitative account that emphasizes how the circuit captures the physical effects expected from the transducer. Moreover, we provide the equations needed to derive the results presented in the main text. A derivation of the transducer figures of merit $\eta$ and $N$ is given in Appendix~\ref{sec:deriveEtaN}.

First, here are some general remarks. The equivalent circuit description of a piezoelectric system, the BVD circuit, is well established. The less-familiar elements of our treatment are as follows: (1) the equivalent circuit in the presence of an optomechanical coupling to the piezoelectric element, and (2) the accounting of quantum noise. Regarding (1), it is not particularly surprising that the linearized optomechanical dynamics is amenable to an impedance formulation. The main nontrivial aspect is the active nature of linearized optomechanical systems, i.e., the fact that the laser pump field provides and absorbs energy to bridge the mechanical and optical frequency scales. In terms of mathematical description, this entails that the optical fields are most naturally represented in the rotating frame with respect to the pump frequency. Below we describe how the coupling to such a rotating-frame variable can be incorporated in the BVD circuit. Regarding (2), the quantum mechanics of our linear transducer is accounted for simply by suitably quantizing the itinerant input and output fields. There is no need to explicitly quantize the internal degrees of freedom of the transducer insofar as only the input and output fields are of interest; for linear systems the scattering matrix linking those fields is the same quantum mechanically as it is classically.

\begin{figure*}
\includegraphics[width=0.6\linewidth]{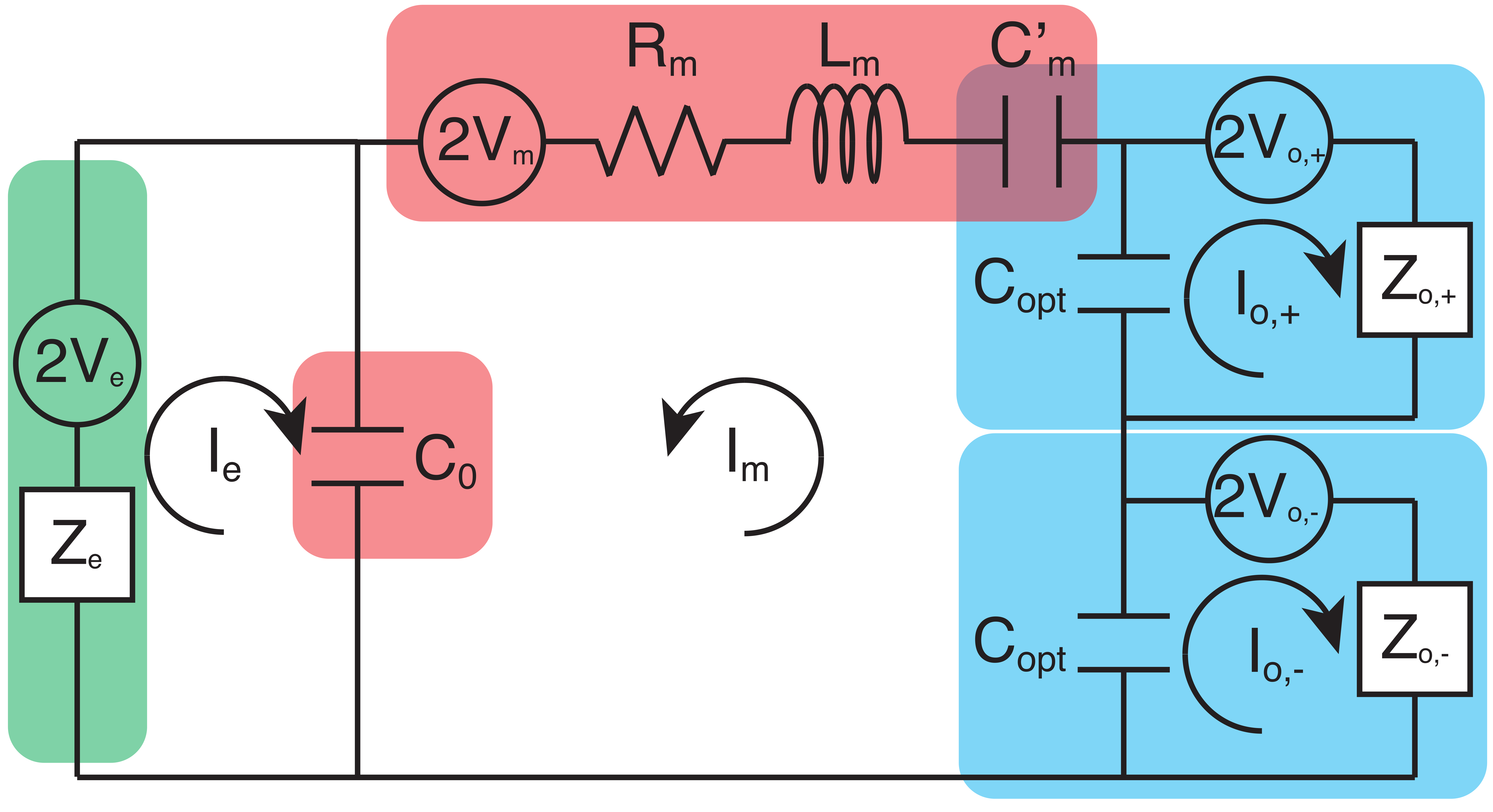}
\caption{Exact piezo-optomechanical equivalent circuit including an arbitrary linear matching network with loop current $I_\text{e}$ parametrized by Th\'{e}venin impedance $Z_\text{e}$ and voltage $2V_\text{e}$. The central mechanical loop $I_\text{m}$ is the well-established BVD circuit for a piezoelectric element except that here the electrical capacitor $C_0$ coupling $I_\text{m}$ to $I_\text{e}$ is supplemented by optical equivalent capacitors $C_\text{opt}$ entailing coupling to the optical loop currents $I_{\text{o,}\pm}$. The latter represent the two optical sidebands generated by the optomechanical interaction. The loops are governed by the impedances $Z_{\text{o,}\pm}$ and voltages $V_{\text{o,}\pm}$, which are essentially the Lorentzian susceptibilities of the optical cavity mode at the respective sidebands and the associated optical input fields. Note the modified mechanical capacitance $1/C_\text{m}' \equiv 1/C_\text{m} - 2/C_\text{opt}$ in the mechanical loop.}
\label{fig:Exact-equiv-circ}
\end{figure*}

\subsection{Piezoelectric subcircuit}

As our starting point, we consider the equivalent circuit in Fig.~\ref{fig:Exact-equiv-circ}, which is more general than that in Fig.~\ref{fig:BVD-LCcircuit-Tx-Opto}, and is the exact equivalent circuit for the linearized dynamics of a piezoelectric system in which the mechanical element is dispersively coupled to a single optical cavity mode. The simpler circuit in Fig.~\ref{fig:BVD-LCcircuit-Tx-Opto} emerges from this in the limit of adiabatic coupling to the optical cavity.

Consider first the leftmost part of the circuit, consisting of the current loops $I_\text{e}$ and $I_\text{m}$, setting $C_\text{opt}\rightarrow \infty$; this is exactly the standard BVD circuit connected to a matching network parametrized by its Th\'{e}venin impedance $Z_\text{e}$ and voltage $2V_\text{e}$. Let us henceforth specialize to the $RLC$ matching network considered in Fig.~\ref{fig:BVD-LCcircuit-Tx-Opto}, for which $Z_\text{e}(\omega)=[-i\omega C_\text{T}+1/(-i\omega L + Z_\text{tx}+R_\text{L})]^{-1}$ and $V_\text{e}=(V_\text{tx}+V_\text{L})(-i\omega C_\text{T})^{-1}/[Z_\text{e}+(-i\omega C_\text{T})^{-1}]$.
The incoming transmission line signal can be quantized $V_\text{tx}\rightarrow\hat{V}_\text{tx}$ by expanding it on a set of bosonic quantum operators $[\hat{a}_\text{in}(\omega),\hat{a}_\text{in}^{\dagger}(\omega')]=\delta (\omega-\omega')$ as,
\begin{equation}
    \hat{V}_\text{tx}(t) = \int_{0}^{\infty}\frac{d\omega}{\sqrt{2\pi}}\sqrt{\frac{\hbar \omega Z_\text{tx}}{2}}\left[\hat{a}_\text{in}(\omega)e^{-i\omega t}+\text{H.c.}\right];\label{eq:V-tx-expansion}
\end{equation}
analogous expansions for the Ohmic Johnson noise $V_\text{L}\rightarrow \hat{V}_\text{L}$ and the mechanical thermal noise $V_\text{m}\rightarrow \hat{V}_\text{m}$ hold with $Z_\text{tx}$ replaced by $R_\text{L}$ and $R_\text{m}$, respectively. This allows us to calculate the normal-ordered mechanical noise variance in the Fourier domain ($
\omega>0$),
\begin{equation}
    \langle\hat{V}_\text{m}^{\dagger}(\omega)\hat{V}_\text{m}(\omega')\rangle = \frac{\hbar \omega R_\text{m}}{2}n_\text{m}(\omega)\delta(\omega-\omega'),\label{eq:V-m_var}
\end{equation}
having assumed a thermal state for the mechanical bath $\langle \hat{a}_\text{m,in}^{\dagger}(\omega)\hat{a}_\text{m,in}(\omega')\rangle=n_\text{m}(\omega)\delta(\omega-\omega')$, where $n_\text{m}$ is the bath occupancy. By the same token, if the electrical circuit is in the ground state in thermal equilibrium, we have $ \langle\hat{V}_\text{L}^{\dagger}(\omega)\hat{V}_\text{L}(\omega')\rangle = 0$.

\subsection{Optomechanical subcircuit}

\noindent We now consider the optomechanical coupling of the piezoelectric element~\cite{aspelmeyer_cavity_2014}; this is accounted for by the two current loops $I_{\text{o,}\pm}$ seen in the rightmost part of Fig.~\ref{fig:Exact-equiv-circ} (now taking $C_\text{opt}$ to be finite). These loops represent the anti-Stokes ($I_{\text{o,}+}$) and Stokes ($I_{\text{o,}-}$) sidebands arising from the beam splitter ($\propto \hat{b}^\dagger \hat{c} + \text{H.c.}$) and two-mode squeezing ($\propto \hat{b} \hat{c} + \text{H.c.}$) interactions, respectively, that arise from the standard radiation-pressure Hamiltonian $\propto (\hat{b}+\hat{b}^\dagger)(\hat{c}+\hat{c}^\dagger)$. The relative strength of these two types of interaction can be controlled by the pump detuning and the sideband resolution $(4\omega_\text{m}/\kappa_\text{o})^2$, which together determine the parts of the optical-cavity Lorentzian $\mathcal{L}$ [Eq.~\eqref{eq:Lorentzian}] being sampled by the sidebands generated by the mechanical system. The cavity Lorentzian is encoded in the optical sideband impedances $Z_{\text{o,}\pm}$ whereas the optomechanical interaction strength is encoded in the absolute scale of $C_\text{opt}$ and $Z_{\text{o,}\pm}$. Dissipation of energy in these loops due to $\text{Re}[Z_{\text{o,}\pm}]\neq 0$ simply corresponds to the emission of photons into the optical output channel. Note that the lower sideband impedance has $\text{Re}[Z_{\text{o,}-}] < 0$, reflecting the amplification induced by the Stokes process.

Having motivated qualitatively the features of the general equivalent circuit in Fig.~\ref{fig:Exact-equiv-circ}, we now address how the simplified circuit employed in the main text (Fig.~\ref{fig:BVD-LCcircuit-Tx-Opto}) arises as a limiting case. To the end of determining the mechanical current $I_\text{m}$ in Fig.~\ref{fig:Exact-equiv-circ}, it is clear that we may \emph{algebraically} eliminate the optical currents $I_{\text{o,}\pm}$ by applying Kirchhoff's voltage law (KVL). The resulting KVL for $I_\text{m}$ includes the effective load impedances and voltage sources from the optical loops. Now if the (Fourier) frequency dependence of these quantities is weak over the signal bandwidth of interest, we may neglect it by evaluating them at $\omega=\omega_\text{m}$; this constitutes \emph{adiabatic} elimination of the optical cavity (in the Fourier domain). The real parts of $Z_{\text{o,}\pm}(\omega_\text{m})$ result in the resistances $R_{\text{OM},\pm}$ [Eq.~\eqref{eq:R-OM-def}] in the simplified circuit (Fig.~\ref{fig:BVD-LCcircuit-Tx-Opto}), whereas the imaginary parts amount to an effective frequency shift of the mechanical resonance (however, we assume this to be negligible as is indeed the case for the parameter sets considered in this work).

Whether adiabatic elimination is performed or not, the equivalent circuits in Figs.~\ref{fig:BVD-LCcircuit-Tx-Opto} and~\ref{fig:Exact-equiv-circ} must be supplemented with input-output relations relating the incoming and outgoing itinerant fields to the currents in the circuit. Generically, for a signal port with resistance $R$ and current $I$, the input-output relation reads
\begin{equation}
    \hat{V}_\text{out} = -R\hat{I} + \hat{V}_\text{in},\label{eq:electrical-IO}
\end{equation}
where $\hat{V}_\text{in(out)}$ can be decomposed into bosonic frequency components as in Eq.~\eqref{eq:V-tx-expansion} in order to achieve a scattering relation of the type seen in Eq.~\eqref{eq:scattering_relation}. 

We henceforth specialize to the regime of adiabatic optics described by the simplified circuit in Fig.~\ref{fig:BVD-LCcircuit-Tx-Opto}. The effective optomechanical input-output relation, specifying how the mechanical motion is mapped onto the outgoing itinerant light field associated with the upper sideband (assumed to be the target channel of our transducer), is
\begin{multline}
    \hat{b}_\text{out}(\omega_\text{pump}+\omega) = \sqrt{\eta_\text{o}}\sqrt{\frac{2R_\text{OM,+}}{\hbar\omega_\text{m}}}\frac{\omega_\text{m}}{\omega}\hat{I}_\text{m}(\omega) \\+ \text{vacuum terms};\label{eq:b-o-out}
\end{multline}
the omitted vacuum terms vanish when calculating the normal-ordered expectation values associated with photon counting as considered here.
Finally, the optical voltage responsible for the amplification noise is
\begin{equation}
    \hat{V}_\text{o}(\omega) = \sqrt{\frac{\hbar\omega_\text{m} R_{\text{OM,}-}}{2}}\hat{b}_\text{in}^{\dagger}(\omega_\text{pump}-\omega) + \text{vacuum term},
\end{equation}
where the vacuum term of the upper sideband does not contribute to the normal-ordered noise ($\omega>0$)
\begin{equation}
\langle\hat{V}_\text{o}^{\dagger}(\omega)\hat{V}_\text{o}(\omega') \rangle = \frac{\hbar\omega_\text{m} R_{\text{OM,}-}}{2}\delta(\omega-\omega').\label{eq:V-o_var}
\end{equation}
The above equations suffice to derive the expressions for $\eta$ and $N$ given in the main text as detailed in Appendix~\ref{sec:deriveEtaN}.

\section{Derivation of $\eta$ and $N$}
\label{sec:deriveEtaN}
\hspace{0.1in}

\noindent In this Appendix, we derive the scattering relation in Eq.~\eqref{eq:scattering_relation} for the itinerant fields linked by our piezo-optomechanical transducer, as parametrized by the signal transfer efficiency $\eta$ and the added noise $N$ (referenced to the input) for electrical-to-optical transduction. The elements required to do so are laid out in Appendix~\ref{sec:equiv-circ}: the equivalent circuit (Fig.~\ref{fig:BVD-LCcircuit-Tx-Opto}), the input-output relation for the target port of the transducer, and the thermal statistics of the noise sources. 

We start by determining the mechanical response $I_\text{m}$ to the various inputs by means of the equivalent circuit (Fig.~\ref{fig:BVD-LCcircuit-Tx-Opto}), as can be achieved either by using standard impedance rules or by algebraically solving the KVLs of the circuit. We find
\begin{equation}
    2V_\text{m} + 2V_\text{o} + \frac{2V_\text{tx}+2V_\text{L}}{-i\omega(C_0+C_\text{T})Z_\text{e}(\omega)}
    = I_\text{m} Z_{\text{m,eff}}(\omega),
    \label{eq:KVL-Im}
\end{equation}
where the effective impedance governing the mechanical loop current $I_\text{m}$ is
\begin{equation}
    Z_\text{m,eff}(\omega) \equiv Z_{\text{OM}}(\omega)+\frac{1}{-i\omega(C_0+C_\text{T})}+\frac{\left(\frac{1}{\omega (C_0+C_\text{T})}\right)^{2}}{Z_\text{e}(\omega)},
    \label{eq:Z-m-eff}
\end{equation}
in terms of the electrical $LC$ impedance
\begin{equation}
   Z_\text{e}(\omega)\equiv -i\omega L + Z_{\text tx} + R_\text{L} + \frac{1}{-i\omega (C_0+C_\text{T})} ,
\end{equation}
and the impedance of the optically loaded mechanical arm
\begin{equation}
Z_\text{OM}(\omega)\equiv -i\omega L_\text{m} + R_\text{m} + R_\text{OM,+} - R_{\text{OM,}-} + \frac{1}{-i\omega C_\text{m}}.
\label{eq:Z-OM}
\end{equation}
Even before arriving at $\eta$ and $N$, several important conclusions can be extracted from Eqs.~\eqref{eq:KVL-Im}-\eqref{eq:Z-OM}. From Eq.~\eqref{eq:KVL-Im} we find, unsurprisingly, that maximal electrical signal enhancement occurs at resonance $\omega\approx\omega_{\text{LC}}$ (assuming $Q_{\text{LC}}\gg 1$); by evaluating the last term in Eq.~\eqref{eq:Z-m-eff} at this frequency, the resonant impedance-transformed electromechanical load $R_{\text{EM}}$ is found to be real and given by Eq.~\eqref{eq:R-EM_LC} in the main text. Next, by considering the first two terms in Eq.~\eqref{eq:Z-m-eff}, we see that a joint electromechanical resonance, where the maxima of the signal enhancement and effective mechanical susceptibility coincide, is achieved by tuning the electrical resonance to $\omega_{\text{LC}}=\omega_{\text m}$, where $\omega_{\text m}$ is the effective mechanical resonance stated in Eq.~\eqref{eqn:effectiveResonance}.

To continue our derivation of $\eta$ and $N$, we combine $I_{\text m}$ as given by Eq.~\eqref{eq:KVL-Im} with Eq.~\eqref{eq:V-tx-expansion} and the optical input-output relation in Eq.~\eqref{eq:b-o-out} to arrive at the scattering relation for the optical output port (upper sideband)
\begin{multline}
    \hat{b}_\text{out}(\omega_\text{pump}+\omega) = \sqrt{\eta_{\text e}\eta_{\text o}}\frac{\sqrt{4(Z_{\text tx}+R_{\text L})R_{\text{OM,}+}}\sqrt{\omega_{\text m}/\omega}}{-i\omega(C_0+C_{\text T})Z_{\text e}(\omega)Z_{\text{m,eff}}(\omega)}\\
    \times\left(\hat{a}_\text{in}(\omega)+\frac{-i\omega(C_0+C_{\text T})Z_{\text e}(\omega)}{\sqrt{\eta_{\text e}\hbar\omega(Z_{\text{tx}}+R_{\text L})/2}}(\hat{V}_\text{m} + \hat{V}_\text{o})\right)\\
    +\text{vacuum terms},
    \label{eq:b-o-out_omega}
\end{multline}
written in a manner suggestive of the transducer relation, Eq.~\eqref{eq:scattering_relation}; we introduce the electrical coupling efficiency $\eta_\text{e}$ using Eq.~\eqref{eq:eta_electrical}. The vacuum terms omitted in Eq.~\eqref{eq:b-o-out_omega} now include both $V_\text{L}$ and optical contributions. 
This relies on the assumption of a ground-state electrical circuit (in thermal equilibrium) for which the Ohmic Johnson noise $V_\text{L}$ does not contribute to normal-ordered expectation values, as pointed out above, and hence can be ignored in the photon counting scenario considered here.
We identify the prefactor in the first line of Eq.~\eqref{eq:b-o-out_omega} with the square root of the (complex) signal transfer efficiency, $\sqrt{\eta(\omega)}$, for arbitrary Fourier frequency $\omega$; similarly $N(\omega)$ can be evaluated from the second term in the second line.

We now focus on the performance at the transducer resonance $\omega=\omega_\text{m}$. We observe that choosing $\omega_\text{LC}=\omega_\text{m}$ with the latter given by Eq.~\eqref{eqn:effectiveResonance}, we have $Z_\text{m,eff}(\omega_\text{m})=R_\text{m} + R_\text{EM} + R_{\text{OM},+} - R_{\text{OM},-}$ on account of Eq.~\eqref{eq:R-EM_LC}. Evaluating Eq.~\eqref{eq:b-o-out_omega} at $\omega=\omega_\text{m}$ yields
\begin{multline}
    \hat{b}_\text{out}(\omega_\text{pump}+\omega_\text{m}) =  \sqrt{\eta_\text{e}\eta_\text{o}}\frac{\sqrt{4R_\text{EM}R_\text{OM,+}}}{Z_\text{m,eff}(\omega_\text{m})}\\
    \times\left(\hat{a}_\text{in}(\omega_\text{m})+\sqrt{\frac{2}{\eta_\text{e}\hbar\omega_\text{m}R_\text{EM}}}(\hat{V}_\text{m} + \hat{V}_\text{o})\right)\\+\text{vacuum terms}.\label{eq:b-o-out_final}
\end{multline}
In view of Eq.~\eqref{eq:scattering_relation}, the formula for the peak value of the signal transfer efficiency $\eta_{\text{peak}}=\eta(\omega_{\text m})$, Eq.~\eqref{eq:peakEta}, can be directly read off from Eq.~\eqref{eq:b-o-out_final}. The added noise referenced to the input at the transducer resonance $N(\omega_\text{m})$, Eqs.~\eqref{eq:Nopt1} and~\eqref{eq:Nm}, follow from Eq.~\eqref{eq:b-o-out_final} in conjunction with the thermal expectation values, Eqs.~\eqref{eq:V-m_var} and~\eqref{eq:V-o_var}.

Having determined the transducer performance at resonance, $\eta(\omega_{\text m})$ and $N(\omega_{\text m})$, we now address the question of bandwidth. From Eq.~\eqref{eq:b-o-out_omega} it is clear that $\eta(\omega)$ and $N(\omega)$ are characterized by different bandwidths in general.
Assuming we can neglect the frequency dependence of the noise contribution from $\hat{V}_{\text m}$, Eq.~\eqref{eq:V-m_var}, the FWHM bandwidth of $1/N(\omega)$ is found to equal the electrical decay rate $\kappa_{\text e}=(Z_\text{tx}+R_{\text L})/L$ assuming $Q_{\text LC}\gg 1$ (so that a Lorentzian approximation of $1/[-i\omega(C_0+C_\text{T})Z_{\text LC}(\omega)]$ is warranted).

\begin{figure}
\begin{center}
\includegraphics[width=1\linewidth]{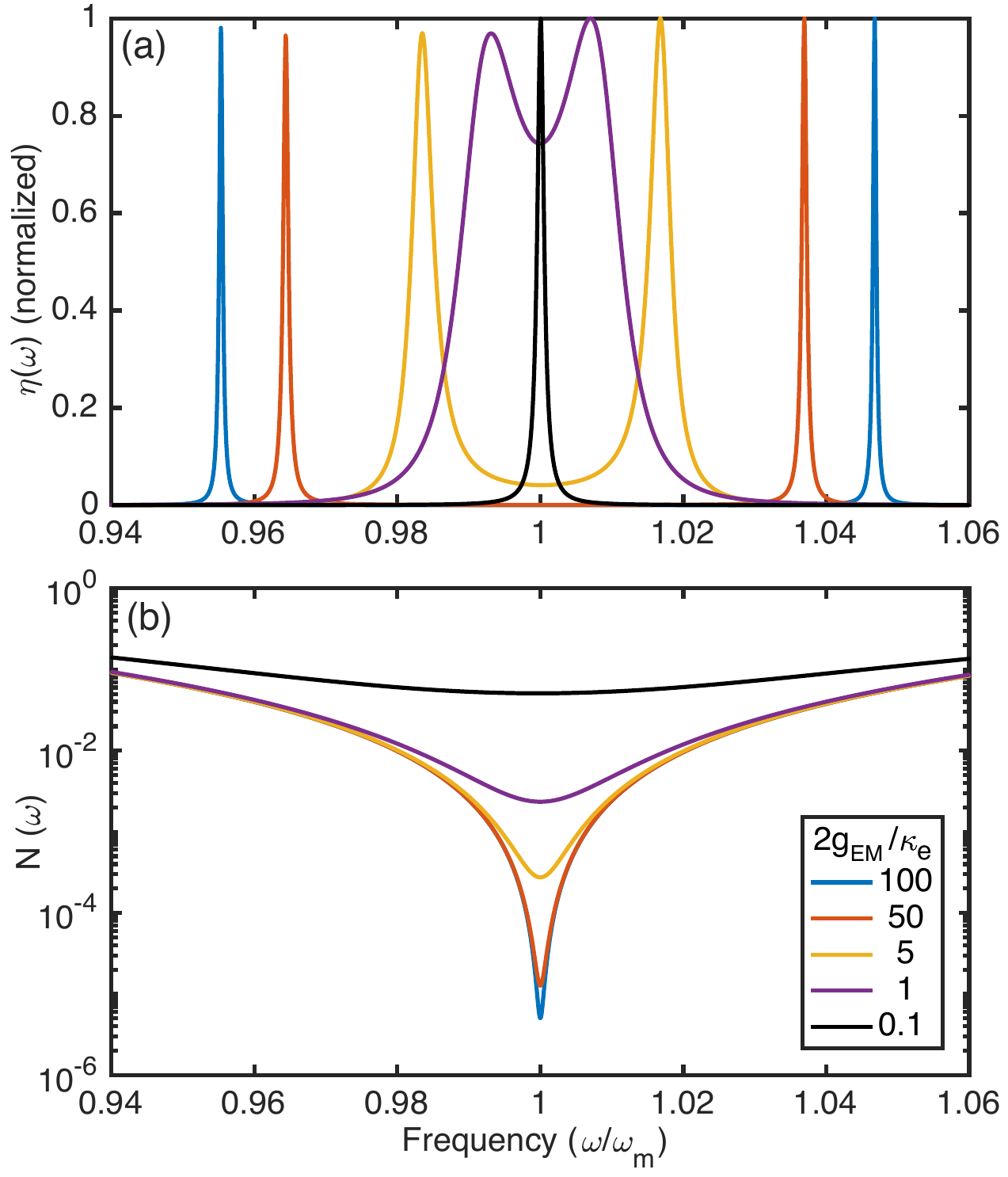}
\end{center}
\caption{Frequency dependence of (a) $\eta(\omega)$ and (b) $N(\omega)$ for various electromechanical coupling rates $g_\text{EM}$. Parameters used are: $\eta_\text{e}$ = $\eta_\text{o} = 1$, $T = 100\,$mK, $R_\text{L} = 0$, $C_0 = 1\,$fF, $k^2 = 1\,\%$, $Q_\text{m} = 10,000$, $\mathcal{C}_\text{OM} = 1$, $\mathcal{L}_+^2 = 1$, and $\mathcal{L}_-^2 = 0$. }
\label{fig:SpectralEtaN}
\end{figure}

However, in the regime of strong electromechanical coupling (2$g_\text{EM} > \kappa_\text{e}$), the frequency dependence of $\eta(\omega)$ and $N(\omega)$ becomes more pronounced leading to normal-mode splitting (colored peaks in Fig.~\ref{fig:SpectralEtaN}, plotting from Eq.~\ref{eq:b-o-out_omega}). This is the case when minimization of noise is the goal while sacrificing efficiency. The initial definition of bandwidth must thus be replaced by some choice suitable for the application at hand.

\section{Optical-to-electrical conversion}
\label{sec:optical2electrical}
\hspace{0.1in}

\noindent Our transducer has the ability to perform frequency conversion in both directions between the microwave and optical parts of the electromagnetic spectrum. However, only the electrical-to-optical noise analysis is considered in the main text for specificity. Here we complete the noise analysis in the reverse direction and also give a proof that the transduction efficiency is the same in the two directions.

In the reverse direction of transduction, that is for optical-to-electrical conversion, the input is at the optical port $\hat{b}$ and assumed to be localized near the upper sideband $\omega\approx \omega_\text{pump}+\omega_{\text m}$, while the output is on the electrical port $\hat{a}$ such that the analog of Eq.~\eqref{eq:scattering_relation} reads ($\omega>0$)
\begin{equation}
\hat{a}_\text{out}(\omega) = \sqrt{\eta(\omega)} [\hat{b}_\text{in}(\omega_\text{pump}+\omega) + \sqrt{N'(\omega)}].
\label{eq:scattering-relation-o2e}
\end{equation}
The signal transfer efficiency $\eta$ is the same in both directions. This hinges on the reciprocity theorem~\cite{pozar_microwave_1990} according to which the admittance of the mechanical arm to a voltage in the transmission line arm $Y_\text{m}^\text{{(tx)}}$ equals the admittance of the transmission line arm to a voltage in the mechanical arm $Y_\text{tx}^\text{{(m)}}$, i.e., $Y_\text{m}^\text{{(tx)}}=Y_\text{tx}^\text{{(m)}}\equiv Y$. The corresponding current responses are $I_\text{m}^\text{{(tx)}}=2V_\text{tx}Y$ and $I_\text{tx}^\text{{(m)}}=2V_\text{o}Y$. The electrical-to-optical peak efficiency $\eta_{e\rightarrow o}$ can then be expressed, using Eqs.~\eqref{eq:V-tx-expansion} and~\eqref{eq:b-o-out},
\begin{equation}
    \hat{b}_\text{out}(\omega_\text{pump}+\omega_\text{m}) = \overbrace{2\sqrt{\eta_\text{o}}\sqrt{R_\text{OM,+}Z_\text{tx}}Y}^{\sqrt{\eta_{e\rightarrow o}}} \hat{a}_\text{in}(\omega_\text{m}) + \text{noise}.
\end{equation}
The analogous expression for optical-to-electrical conversion follows knowing that the signal part of the optical voltage is $V_\text{o}(\omega_\text{m}) = \sqrt{\eta_\text{o}}\sqrt{\hbar\omega_\text{m} R_\text{OM,+}/2}\hat{b}_\text{o,in}(\omega_\text{pump}+\omega_\text{m})$ along with Eq.~\eqref{eq:V-tx-expansion} and the electrical input-output relation, Eq.~\eqref{eq:electrical-IO},
\begin{equation}
    \hat{a}_\text{out}(\omega_\text{m}) = \overbrace{2\sqrt{\eta_\text{o}}\sqrt{R_\text{ OM,+}Z_\text{tx}}Y}^{\sqrt{\eta_{o\rightarrow e}}} \hat{b}_\text{in}(\omega_\text{pump}+\omega_\text{m}) + \text{noise},
\end{equation}
showing that $\eta_{e\rightarrow o}=\eta_{o\rightarrow e}\equiv\eta$.

In contrast, the added noise of the transducer for optical-to-electrical conversion $N'$ differs in general from that of electrical-to-optical conversion $N$ analyzed in the main text~\cite{Zeuthen_Electrooptomechanical_2018}. Applying the approach laid out in Appendices~\ref{sec:equiv-circ} and~\ref{sec:deriveEtaN} to determine the current in the transmission line arm and, in turn, its itinerant output, we calculate the added noise for optical-to-electrical transduction $N'=N_{\text o}' + N_{\text m}'$, where $N'(\omega)\delta(\omega-\omega')=(1/\eta_\text{peak}) \langle \hat{a}^\dagger_\text{out} (\omega) \hat{a}_\text{out} (\omega') \rangle$, in the following subsections, thereby complementing Section~\ref{sec:N} in the main text.

\subsection{Optical amplification noise $N_{\text{o}}'$ (Raman noise)}

\noindent For transduction from the upper optical sideband into the electrical transmission line, the added noise flux per unit bandwidth referenced to the input signal is
\begin{equation}
N_{\text{o}}' =   \frac{1}{\eta_\text{o}} \frac{\mathcal{L}^2_-}{\mathcal{L}^2_+}
\end{equation}
which does not depend on the cooperativities. Assuming again a red-detuned laser drive, $\Delta = -\omega_\text{m}$, we have $\mathcal{L}_+ = 1$ and
\begin{eqnarray}
N_{\text{o}}' = \frac{1}{\eta_\text{o}} \frac{ \left(\frac{\kappa_\text{o}}{4\omega_\text{m}} \right)^2}{\left(\frac{\kappa_\text{o}}{4\omega_\text{m}}\right)^2+1} \xrightarrow{(4\omega_\text{m}/\kappa_\text{o})^2 \gg 1} \frac{1}{\eta_\text{o}} \left(\frac{\kappa_\text{o}}{4\omega_\text{m}} \right)^2
\end{eqnarray}
where the last expression is valid for good optomechanical sideband resolution.

\subsection{Mechanical thermal noise $N'_\text{m}$}

\noindent The mechanical thermal noise in the electrical output from the upper optical sideband input is
\begin{eqnarray}
N_{\text{m}}' = \frac{1}{\eta_\text{o}} \frac{n_\text{m}}{\mathcal{L}^2_+ \mathcal{C}_\text{OM}}.
\end{eqnarray}
The quantity $\mathcal{C}_\text{OM}/n_\text{m}$ is known as the optomechanical quantum cooperativity; it is (approximately) the ratio of coherent optomechanical coupling to the thermal decoherence induced by the mechanical bath. Obviously, quantum-level transduction requires $\mathcal{L}^2_+ \mathcal{C}_\text{OM}/n_\text{m} \gtrsim 1$ (and $\eta_\text{o} \approx 1$).

\section{Choice of $C_\text{T}$ and $L$ in the $RLC$ matching network}
\label{sec:network}
\hspace{0.1in}

\noindent Suitable values of $C_\text{T}$ and $L$ must be chosen in order to impedance match the piezo-optomechanical circuit to the input transmission line for which the impedance is assumed to be $Z_\text{tx} = 50\, \Omega$.

At the effective resonance $\omega_\text{m}$, the total resistance of the optomechanical branch (right arm of the circuit in Fig.~\ref{fig:BVD-LCcircuit-Tx-Opto}) reduces to a resistor with total resistance $R_\text{EM}^\text{opt} = R_\text{m} \mathcal{C}_\text{EM}^\text{opt} = R_\text{m}+R_{\text{OM,}+}-R_{\text{OM,}-}$ in parallel with the static capacitor $C_0$. The tuning capacitor $C_\text{T}$ will uptransform $Z_\text{tx}$ to $R_\text{EM}=R_\text{EM}^\text{opt}$ in order to match this typically larger resistance, provided that we choose the value
\begin{equation}
C_\text{T} = \frac{1}{\omega_\text{m}} \sqrt{\frac{1}{R_\text{EM}^\text{opt} (Z_\text{tx}+R_\text{L})}} - C_0,
\label{eq:CT-eta}
\end{equation}
assuming frequency matching $\omega_\text{m} = \omega_\text{LC} \equiv 1/\sqrt{L(C_0+C_\text{T})}$; the desired impedance transformation is possible if a solution $C_\text{T}\geq 0$ exists. Next, the matching inductor $L$ is chosen to counter the capacitance $C_0+C_\text{T}$ (i.e., to have $\text{Im}[Z]=0$), 
\begin{equation}
L = \frac{1}{\omega_\text{m}} \sqrt{R_\text{EM}^\text{opt} (Z_\text{tx}+R_\text{L})}.
\label{eq:L-eta}
\end{equation}
Choosing $C_\text{T}$ and $L$ according to the above equations, the electromechanical and optomechanical cooperativities are optimized for efficiency at the matching condition Eq.~\eqref{eq:Copt}. 
Moreover, as discussed earlier, the conditions $R_\text{EM}=R_\text{EM}^{\text{opt}}$ and $\omega_\text{MW}=\omega_\text{LC}=\omega_\text{m}$ cause the piezo-optomechanical transducer to be perfectly impedance matched to the transmission line at $\omega_\text{m}$. Impedance matching to the mechanical serial resonance $\omega_\text{s}$ within a matching network has been discussed previously~\cite{garcia-rodriguez_low_2010}. Note that Eqs.~\eqref{eqn:effectiveResonance}, \eqref{eq:CT-eta}, and \eqref{eq:L-eta} are not on closed form, but must be solved self-consistently. An analytical solution can be obtained if the dependence on $\omega_\text{m}$ of the optomechanical contributions in $R_\text{EM}^\text{opt}$ can be neglected (e.g., by evaluating them using $\omega_\text{m}\approx \omega_\text{s}$) as is warranted in the typical scenario $k^2 \ll \kappa_\text{o}/\omega_\text{s}$. In that case a solution with $C_\text{T}\geq 0$ exists provided that $R_\text{EM}^\text{opt} \omega_{\text s}^2 C_0 (C_\text{m} + C_0)(Z_\text{tx}+R_\text{L})\ez{\leq} 1$ and is given by
\begin{equation}
    C_\text{T} =  \frac{C_\text{m}}{2}\left( \sqrt{1+\frac{4}{R_\text{EM}^\text{opt} \omega_\text{s}^2 C_\text{m}^2(Z_\text{tx}+R_\text{L})}} -1 \right)- C_0,
\end{equation}
from which $\omega_\text{m}$ [Eq.~\eqref{eqn:effectiveResonance}] and $L$ [Eq.~\eqref{eq:L-eta}] can be evaluated.

\begin{figure}
\begin{center}
\includegraphics[width=1\linewidth]{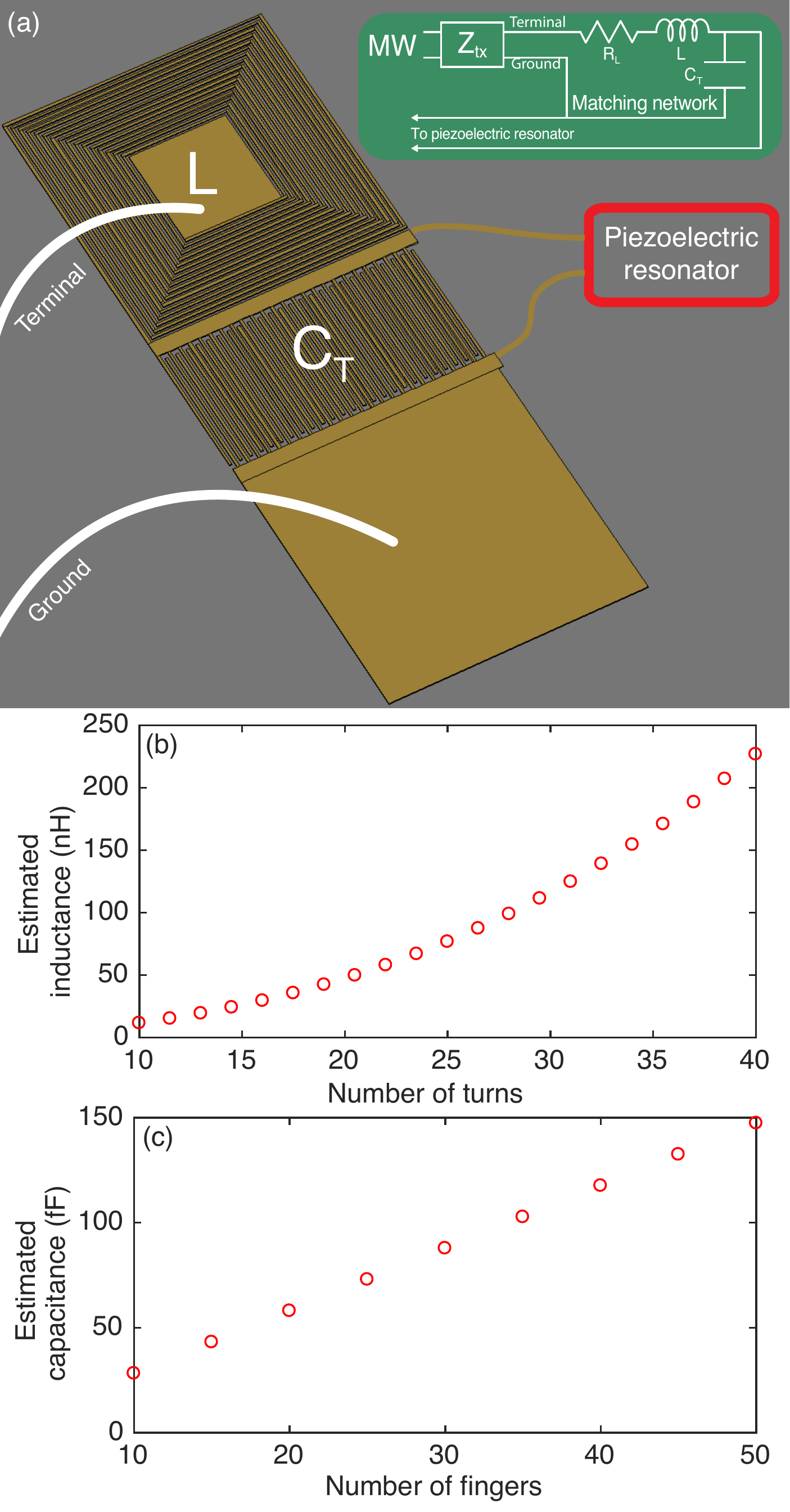}
\end{center}
\caption{Representative matching network using a chip-integrated approach. (a) Illustration of a possible physical implementation of a matching network consisting of a planar spiral square inductor $L$ connected to a planar IDT-like capacitor $C_\text{T}$ in parallel with the piezoelectric resonator depicted in Fig.~\ref{fig:CoupledResonators}. The gray region corresponds to a GaAs layer while the gold traces depict metal deposited on top. The $LC$ circuit (overview schematic in green inset) includes electrical pads that can be connected to the input microwave drive via rf probes for testing or an air bridge (white lines) to the rest of the on-chip circuit for signal routing. (b) Estimated inductance of the spiral inductor calculated with square pad size of 50 $\times$ 50 $\upmu$m$^2$. (c) Estimated capacitance of the planar capacitor with finger length of 50 $\upmu$m. Common parameters to the inductor and capacitor are metal width and pitch of 1 $\upmu$m. }
\label{fig:MatchingNetwork}
\end{figure}

The implementation of a matching network can take various forms as discussed in the main text. Here, we propose one practical design based on a co-planar approach, using the same metal layer as the electrodes for the piezoelectric resonator. Our suggested circuit is illustrated in Fig.~\ref{fig:MatchingNetwork} and is composed of a planar spiral square inductor with inductance $L$ and a planar IDT-like capacitor with capacitance $C_\text{T}$. We perform calculations to determine that $L$ can range from 20 to 230~nH using Ref.~\cite{mohan_simple_1999} while $C_\text{T}$ varies from 30 to 150~fF using finite-element analysis. These values are consistent with the targeted values required to effectively match the transducer geometries we propose (see Table~\ref{table:parameters2}).

\section{Device design trade-offs: optomechanical coupling and microwave impedance matching}
\label{sec:trade-off}
\hspace{0.1in}

In the main text we present a few specific designs of piezo-optomechanical devices suitable for microwave-to-optical transduction. However, these designs are not necessarily optimized, as there are trade-offs that must be considered based on the performance of the rest of the transducer. One of these trade-offs is in the optomechanical coupling rate $g_0$, which may suffer relative to a bare nanobeam optomechanical crystal as a larger overall geometry (higher effective mass $m_{\text{eff}}$) results from use of the hybridized mechanical supermode. If a transducer system specification requires $g_0$ to be as large as possible, then one can consider reducing the size of the supermode by, for example, decreasing the length of the piezoelectric portion of our proposed design. Doing so, and the accompanying reduction in $m_\text{eff}$, enhances $g_{0}$. This scheme actually has the effect of improving the peak efficiency $\eta_\text{peak}$ of the overall transducer as shown in Table~\ref{table:trade-off}. Note that all the scenarios in the table assume the lower $Q_\text{m}$ = 10,000 of our current device. If $Q_\text{m}$ = 50,000 of the potential device is used instead, the maximum efficiency $\eta_\text{peak}$ can reach 99.9\% in the high-performance scenario, on par with the performance of the AlN-on-Si platform presented in the main text.

\begin{table}[]
\begin{tabular}{l|llll}
\hline
Length & 5 $\mu$m & 8 $\mu$m & 11 $\mu$m & 16 $\mu$m \\
\hline
 & \multicolumn{2}{c}{$T$ = 100 mK}  & \multicolumn{2}{c}{$Q_\text{m}$ = 10,000} \\
$m_\text{eff}$ & 430 fg & 720 fg & 2 pg & 4.5 pg \\
$g_0$ & 820 kHz & 640 kHz & 375 kHz & 300 kHz \\
\hline 
\textbf{Low} & \multicolumn{2}{c}{\textbf{$Q_\text{i}$ = 77,000}}  & \multicolumn{2}{c}{\textbf{$n_\text{phot}$ = 280}} \\
$C_\text{T}$ & 16.2 fF & 18.1 fF & 33.9 fF & 39.4 fF\\
$L$  & 243.8 nH & 215.8 nH & 117.0 nH & 116.7 nH\\
$C_\text{OM}$ & 0.46 & 0.30 & 0.12 & 0.08 \\
$C_\text{EM}$ & 1.32 & 1.22 & 1.09 & 1.06 \\
$\eta_\text{peak}$ & 21.4 \% & 14.1 \% & 5.7 \% & 3.9 \%\\
\hline
\textbf{Medium} & \multicolumn{2}{c}{\textbf{$Q_\text{i}$ = 200,000}}  & \multicolumn{2}{c}{\textbf{$n_\text{phot}$ = 500}} \\
$C_\text{T}$ & 12.6 fF & 14.4 fF & 29.5 fF & 35.4 fF\\
$L$  & 311.7 nH & 269.8 nH & 134.3 nH & 129.7 nH\\
$C_\text{OM}$ & 1.34 & 0.99 & 0.46 & 0.33 \\
$C_\text{EM}$ & 2.16 & 1.91 & 1.44 & 1.31 \\
$\eta_\text{peak}$ & 47.2 \% & 36.3 \% & 19.4 \% & 14.4 \% \\
\hline
\textbf{High} & \multicolumn{2}{c}{\textbf{$Q_\text{i}$ = 700,000}}  & \multicolumn{2}{c}{\textbf{$n_\text{phot}$ = 1,000}} \\
$C_\text{T}$ & 8.7 fF & 9.4 fF & 20.4 fF & 25.5 fF\\
$L$  & 449.8 nH & 409.0 nH & 192.5 nH & 179.0 nH\\
$C_\text{OM}$ & 3.77 & 3.49 & 1.98 & 1.51 \\
$C_\text{EM}$ & 4.49 & 4.38 & 2.95 & 2.50 \\
$\eta_\text{peak}$ & 75.8 \% & 67.5 \% & 50.7 \% & 43.7 \% \\
\hline
\end{tabular}
\caption{\textbf{Transducer performance with varying length of the piezoelectric resonator} The values in this table are computed for maximizing efficiency $\eta$ in a two-electrode device coupled to a $RLC$ matching circuit. Fixed parameters common to all designs are temperature $T$ and mechanical quality factor $Q_\text{m}$. Three scenarios shown in the second, third, and fourth rows of the table are considered based on the assumption of low, medium, and high device performance, respectively, depending on the optical quality factor $Q_\text{i}$ and intracavity photon number $n_\text{phot}$. Additional improvement in $Q_\text{m}$ (e.g., $Q_\text{m}$=50,000) can yield $\eta_\text{peak}$=99.9~$\%$.}
\label{table:trade-off}
\end{table}

Despite this advantage, the smaller scale of the piezoelectric resonator can potentially lead to engineering challenges with the matching network. For our proposed on-chip implementation, large inductances ($>$ 200 nH) will necessitate planar inductors with a larger footprint while the required smaller capacitances ($<$ 10 fF) will start to be at the same level as the typical parasitic capacitance of these same inductors~\cite{fink_quantum_2016}. Thus, optimizing the resonator for higher $g_0$ to improve efficiency must be balanced against potential difficulties in the matching network.

\section{Electromechanical coupling rate $g_{EM}$}
\label{sec:gEM}
\hspace{0.1in}

\noindent In this Appendix, we derive the electromechanical coupling between a mechanical mode and an electrical $LC$ resonance to arrive at the coupling rate $g_{EM}$ used in the main text. 

To start, the electromechanical energy is given by~\cite{taylor_laser_2011}
\begin{align}
H_\text{EM} = {} & \frac{\hat{p}^2}{2m_\text{eff}} + \frac{m_\text{eff}\omega_\text{m}^2\hat{x}^2}{2} \nonumber\\ & + \frac{\hat{\phi}^2}{2L} + \frac{\hat{q}^2}{2(C_0+C_{\text T})} + G \hat{x} \hat{q},\label{eq:H-EM_xq}
\end{align}
where $\hat{p}$, $\hat{x}$, $\hat{\phi}$, and $\hat{q}$ are the momentum, position, electrical flux, and charge operators, respectively. $G$ is the electromechanical coupling strength introduced in Ref.~\cite{Zeuthen_Electrooptomechanical_2018} and rewritten for the piezoelectric case as
\begin{eqnarray}
G = \omega_\text{m} \sqrt{k_\text{T}^{2}} \sqrt{\frac{m_\text{eff}}{C_0+C_{\text T}}}.
\end{eqnarray}
The reduced piezoelectric coupling strength is expressed as 
\begin{equation}
k_\text{T}^{2} = \frac{C_\text{m}}{C_{\text m}+ C_0 +C_{\text T}}=\frac{C_{\text m}+ C_0}{C_{\text m}+ C_0 +C_{\text T}}k^2,
\end{equation}
where the final expression contains the nominal ($C_{\text T}=0$) value of the coupling strength $k^2$, Eq.~\eqref{eq:k-square-def}.

Expressing the Hamiltonian [Eq.~\eqref{eq:H-EM_xq}] in terms of bosonic annihilation operators $\hat{a}$ ($\hat{c}$) for the $LC$ circuit (mechanical mode), it can be written
\begin{align}
H_\text{EM} = {} & \hbar\omega_\text{LC}\hat{a}^\dagger\hat{a} + \hbar\omega_\text{m}\hat{c}^\dagger\hat{c} \nonumber\\ 
& + \hbar g_\text{EM} (\hat{a}+\hat{a}^\dagger)(\hat{c} + \hat{c}^\dagger).
\end{align}
The interaction Hamiltonian represents the coupling of the electrical resonator to the mechanical part of the equivalent circuit represented by the BVD model. $k_\text{T}^{2}$ can be related to the electromechanical coupling rate $g_\text{EM}$ in the presence of an electrical $LC$ resonance such that
\begin{equation}
g_\text{EM} = \frac{1}{\sqrt{2L\omega_\text{LC}}} \frac{1}{\sqrt{2m_\text{eff}\omega_\text{m}}} G = \frac{\sqrt{\omega_\text{m}\omega_\text{LC}}}{2} \sqrt{k_\text{T}^{2}},
\label{eq:gem}
\end{equation}
where $\omega_\text{LC} = 1/\sqrt{L(C_\text{T}+C_0)}$. When $\omega_\text{LC} = \omega_\text{m}$, Eq.~\eqref{eq:gem} reduces to the equation for $g_\text{EM}$ in Section~\ref{sec:matchingNetwork}.

\section{Amplification and optical broadening}
\label{sec:amplification}
\hspace{0.1in}

\noindent In Section~\ref{sec:maxEta}, we restrict the efficiency in the regime $\eta < 1$. However, when the optical broadening dominates the mechanical linewidth, $\mathcal{C}_\text{OM} (\mathcal{L}^2_+ - \mathcal{L}^2_-) \gg 1 $, such as in the unresolved sideband regime, then the peak maximum efficiency from Eq.~\eqref{eq:etaCom} saturates at the limiting value:
\begin{align}
\eta_\text{peak}^\text{opt} & \xrightarrow{\mathcal{C}_\text{OM} (\mathcal{L}^2_+ - \mathcal{L}^2_-) \gg 1 } \eta_\text{e} \eta_\text{o} \frac{\mathcal{L}^2_+}{ \mathcal{L}^2_+ - \mathcal{L}^2_- } \nonumber\\
& \xrightarrow{\Delta \rightarrow-\omega_\text{m}} \eta_\text{e} \eta_\text{o}  \left[ \left( \frac{ \kappa_\text{o}}{4\omega_\text{m}} \right)^2 +1 \right]
\end{align}
where in the last expression we consider the laser drive to be red detuned from the cavity resonance by $\omega_{\text{m}}$. In the general case that $\mathcal{L}_- > 0$, $\eta_\text{peak}^\text{opt}$ can thus exceed $\eta_\text{e} \eta_\text{o}$ (and hence potentially unity) by as much as the optomechanical gain factor $\mathcal{L}^2_+ / (\mathcal{L}^2_+ - \mathcal{L}^2_-)$, leading to amplification. This amplification happens at the price of increased transducer (amplification) noise $N$ via the optical noise $N_\text{o}$. Assuming evaluation at $\mathcal{C}^\text{opt}_\text{EM}$ and $\mathcal{C}_\text{OM}(\mathcal{L}_+^2 - \mathcal{L}_-^2) \gg 1$, we find
\begin{align}
N_\text{o} & \xrightarrow{\mathcal{C}_\text{OM}(\mathcal{L}_+^2 - \mathcal{L}_-^2) \gg 1} \frac{1}{\eta_\text{e}} \frac{\mathcal{L}^2_-}{\mathcal{L}_+^2 - \mathcal{L}_-^2} \nonumber\\
& \xrightarrow{\Delta = -\omega_\text{m}} \frac{1}{\eta_\text{e}} \left(\frac{\kappa_\text{o}}{4\omega_\text{m}} \right)^2.
\end{align}
These expressions are valid regardless of the degree of sideband resolution.

\section{Optimal external optical coupling $\kappa_\text{ext}^\text{opt}$}
\label{sec:kappa_ext_opt}
\hspace{0.1in}

\noindent One of the major tuning knobs on the optical side is the coupling between the optical cavity and an external waveguide, given by $\kappa_\text{ext}$. This parameter can be found in $\eta_\text{o}$ [Eq.~\eqref{eq:eta_optical}] and $\mathcal{C}_\text{OM}$ [Eq.~\eqref{eq:Com}], which together contributes to the peak efficiency $\eta_\text{peak}$ while only the latter contributes to the optical noise $N_\text{o}$ in electrical-to-optical conversion. Here, we derive approximate analytical relations to optimize the figures of merit.

\subsection{$\kappa_\text{ext}^\text{opt}$ for maximal $\eta$}

In the case where maximal efficiency is key, higher $\kappa_\text{ext}$ increases optical coupling efficiency $\eta_\text{o}$ but lowers the optomechanical cooperativity $\mathcal{C}_\text{OM}$. This trade-off points to an optimal $\kappa_\text{ext}^\text{opt}$. In the regime of negligible signal amplification, $\mathcal{C}_{\text{OM}}\mathcal{L}_{-}^2\ll 1$, the $\kappa_\text{ext}^\text{opt}$ that maximizes $\eta_\text{peak}$ in Eq.~\eqref{eq:etaCom_resolved_sideband} is approximately (assuming the intracavity photon number $n_\text{phot}$ to be fixed and choosing $\Delta=-\omega_\text{m}$ for specificity):
\begin{equation}
    \kappa_\text{ext}^\text{opt} = \kappa_\text{i}\sqrt{1+\mathcal{C}_\text{OM,i}},
\label{eq:kappa-ext-opt}
\end{equation}
where $\mathcal{C}_\text{OM,i}\equiv 4g_\text{OM}^2/(\gamma_\text{m}\kappa_\text{i})$ is the maximal $\mathcal{C}_\text{OM}$ that can be achieved by letting $\kappa_\text{ext}\rightarrow 0$ while keeping $n_\text{phot}$ constant. 
The resulting optimal value of $\mathcal{C}_\text{OM}$ is hence
\begin{equation}
    \mathcal{C}_\text{OM}^{\text{opt}}=[1-\eta_\text{o}^{\text{opt}}]\mathcal{C}_{\text{OM,i}} =\sqrt{1+\mathcal{C}_{\text{OM,i}}}-1.
\end{equation}
Evaluating $\eta_\text{peak}$ [Eq.~\eqref{eq:etaCom_resolved_sideband}] at $\kappa_\text{ext} = \kappa_\text{ext}^\text{opt}$ [Eq.~\eqref{eq:kappa-ext-opt}] we arrive at its maximally achievable value within the regime $\mathcal{C}_{\text{OM}}^{\text{opt}}\mathcal{L}_{-}^2\ll 1$ for a transducer in which the optical coupling is the bottleneck
\begin{equation}
    \left. \eta_\text{peak}^\text{opt} \right|_{\kappa_\text{ext} = \kappa_\text{ext}^\text{opt}} =\eta_\text{e} \frac{(\sqrt{1+\mathcal{C}_\text{OM,i}}-1)^2}{\mathcal{C}_\text{OM,i}}.
\end{equation}
These equations are valid within first approximation. For the exact solution, Eq.~\ref{eq:etaCom} must be solved analytically or numerically.

\subsection{$\kappa_\text{ext}^\text{opt}$ for minimal $N$}

For minimization of $N$, $\kappa_\text{ext}^\text{opt}$ depends on the competition between optical noise and thermal noise. If weak optomechanical interaction is the bottleneck, i.e., $\mathcal{C}_\text{OM} \mathcal{L}_{-}^2 < n_\text{m}$, the optical noise is much smaller than the thermal noise. In this case, the strategy is to maximize $\eta_\text{peak}$ via the product $\eta_{\text{o}}\mathcal{C}_\text{OM}\mathcal{L}_{-}^2$, seen in Eq.~\eqref{eq:eta-peak-smallCOM}. We find that the optimal outcoupling amounts to critical coupling $\kappa_\text{ext}^\text{opt}=\kappa_\text{i}\Leftrightarrow \eta_\text{o} = 1/2$ so that $\mathcal{C}_\text{OM}=2g_\text{OM}^2/(\gamma_\text{m}\kappa_\text{i})=\mathcal{C}_\text{OM,i}/2$, resulting in ($\Delta = -\omega_\text{m}$)
\begin{equation}
    \eta_\text{peak} \sim \eta_\text{e} \frac{\mathcal{C}_\text{EM}^\text{max} \mathcal{C}_\text{OM,i}}{(1+\mathcal{C}_\text{EM}^\text{max})^2}.
\end{equation}

On the other hand, for $\mathcal{C}_\text{OM} \mathcal{L}_{-}^2 > n_\text{m}$, the optical noise is heuristically matched to the thermal noise by decreasing $n_\text{phot}$. $\eta_\text{peak}$ from Eq.~\eqref{eq:eta-peak-Nmatch} can be further optimized by choosing the optical outcoupling rate $\kappa_\text{ext}$ that strikes the right balance between large $\eta_\text{o}$ and small $\mathcal{L}_{-}^2$ (thereby permitting larger $C_\text{OM}$ according to Eq.~\eqref{eq:N-o_bound}) under the assumption of fixed $\kappa_\text{i}$. In the limit $(4\omega_\text{m}/\kappa_\text{o})^2 \gg 1$ we find the optimum point to be $\kappa_\text{ext}^\text{opt} = \kappa_\text{i}/2\Leftrightarrow \eta_\text{o} =1/3$, resulting in ($\Delta = -\omega_\text{m}$)
\begin{equation}
    \eta_\text{peak} =\eta_\text{e} \frac{\mathcal{C}_\text{EM}^\text{max} n_\text{m} }{(1+\mathcal{C}_\text{EM}^\text{max})^2} \frac{2^8}{3^3}\frac{\omega_\text{m}^2}{\kappa_\text{i}^2}.
\end{equation}

\section{RC circuit for low-impedance piezoelectric resonators}
\label{sec:RCcircuit}
\hspace{0.1in}

\begin{figure}
\begin{center}
\includegraphics[width=1\linewidth]{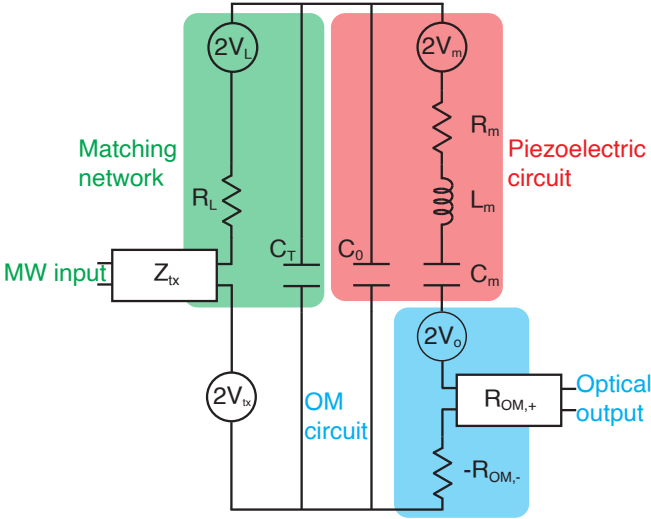}
\end{center}
\caption{Piezo-optomechanical circuit: a transmission line is piezoelectric coupled to a mechanical mode by a BVD circuit with a matching $RC$ network comprised of a tuning capacitor $C_\text{T}$ and resistor $R_\text{L}$.}
\label{fig:BVD-Tx-Opto}
\end{figure}

\noindent We consider an electrical $RLC$ circuit for impedance matching in the main text in the context of maximizing the transfer efficiency $\eta$. In some scenarios where the piezoelectric resonator exhibit low enough impedance such that $Z_\text{BVD} \lesssim Z_\text{tx}$, due to high $k^2$ and $C_0$ for example, a resonant matching circuit is not needed and a simpler circuit can be used instead, namely the $RC$ circuit resulting from letting the inductance $L \rightarrow 0$ in Fig.~\ref{fig:BVD-Tx-Opto}. In short, the inductor from the matching network is removed. One can retain the loading resistor $R_\text{L}$ in the analysis to account for incoupling losses, such as electrical signal routing. 

The relevance of turning to the $RC$ circuit in the regime $Z_\text{BVD} \lesssim Z_\text{tx}$, which requires $R_\text{EM}=R_\text{EM}^\text{opt} \lesssim Z_\text{tx}$ in order to fulfill Eq.~\eqref{eq:Copt}, is seen by considering our results for the $RLC$ network in Section~\ref{sec:matchingNetwork}. In particular, Eq.~\eqref{eq:R-EM_LC} implies $Z_\text{LC} \lesssim Z_\text{tx} + R_\text{L}\Leftrightarrow \omega_\text{LC} \lesssim \kappa_\text{e}$, where $\kappa_\text{e}$ is the loaded electrical decay rate as in the main text; hence, this amounts to a loaded quality factor of the electrical resonance less than unity. While in principle this can be engineered with a suitable small inductance $L \lesssim 1\,$nH , this is often impracticable and, more importantly, unnecessary for impedance matching, as shown in the following.

To proceed, we take the limit $L\rightarrow 0$ in Eq.~\eqref{eq:Z-m-eff} to find
\begin{equation}
    Z_\text{m,eff}(\omega) \equiv Z_{\text{OM}}(\omega)+\frac{Z_\text{tx}+R_\text{L}}{1-i\omega(Z_\text{tx}+R_\text{L})(C_0+C_\text{T})}.
    \label{eq:Z-m-eff-RC}
\end{equation}
Hence we see that the impedance-matching capability of the $RC$ circuit is to decrease the effective impedance $R_\text{EM}$ of the transmission line (plus incoupling losses) from the nominal value $Z_\text{tx} + R_\text{in}$ as seen from the point of view of the mechanical BVD circuit. As seen from Eq.~\eqref{eq:Z-m-eff-RC} the parameter responsible for controlling this impedance transformation is the ratio of the RC time of the circuit $\tau_\text{RC} = (Z_\text{tx} + R_\text{in}) (C_0+C_\text{T})$ to the oscillation period of the signal frequency $1/\omega_\text{MW}$. In the limit of short RC time $\tau_\text{RC} \ll 1/\omega_\text{MW}$ the transmission-line impedance retains its nominal value (from the point of view of the mechanical mode), $R_\text{EM}^\text{RC} \approx Z_\text{tx} + R_\text{L}$. For general $\tau_\text{RC}$ we find the effective electromechanical loading of the mechanical circuit
\begin{equation}
R^\text{RC}_\text{EM} = \frac{Z_\text{tx}+R_\text{L}}{1+(Z_\text{tx}+R_\text{L})^2/Z^2_\text{RC}} = 
\frac{Z_\text{tx}+R_\text{L}}{1+\tau^2_\text{RC}\omega_\text{MW}^2},\label{eq:R-EM_RC}
\end{equation}
where the characteristic RC impedance is $Z_\text{RC} \equiv 1/(\omega_\text{MW}(C_0+C_\text{T}))$. This leads to the electromechanical cooperativity
\begin{eqnarray}
\mathcal{C}^\text{RC}_\text{EM} \equiv \frac{R^\text{RC}_\text{EM}}{R_\text{m}} = \frac{Z_\text{tx}+R_\text{L}}{R_\text{m}(1+\tau^2_\text{RC}\omega_\text{m}^2)} = \frac{\kappa_\text{tx}}{\gamma_\text{m}}.
\end{eqnarray}
where the electrical coupling rate $\kappa_\text{tx} = R^\text{RC}_\text{EM}/L_\text{m}$. The electrical coupling efficiency is equivalent to that of the $RLC$ circuit discussed in the main text with $\eta_\text{e}^\text{RC} = Z_\text{tx}/(Z_\text{tx}+R_\text{L})$. To achieve impedance matching, Eq.~\eqref{eq:Copt}, i.e., $R_\text{EM}^\text{RC}=R_\text{EM}^\text{opt}$, the $RC$ time $\tau_\text{RC}$ can be adjusted by adding a suitable tuning capacitance $C_\text{T}$; from Eq.~\eqref{eq:R-EM_RC} we find
\begin{equation}
    C_\text{T} = \frac{\sqrt{\frac{Z_\text{tx}+R_\text{L}}{R_\text{EM}^\text{opt}}-1}}{\omega_\text{MW}(Z_\text{tx}+R_\text{L})} - C_0,
\end{equation}
which provides a valid result $C_\text{T}\geq 0$ provided that $\left. R_\text{EM}^\text{RC}\right|_{C_\text{T}=0}\geq R_\text{EM}^\text{opt}$.

In our discussion of the $RLC$ in the main text, the imaginary part of the joint circuit impedance seen by the transmission line is engineered to be zero by choosing the input signal frequency $\omega_\text{MW}=\omega_\text{LC}=\omega_\text{m}$. In the absence of the electrical inductor $L$ to cancel the imaginary impedance associated with the electrical capacitors of total capacitance $C_0 + C_\text{T}$, this cancellation can be achieved with the mechanical inductance $L_\text{m}$ instead. This is done by choosing the input signal frequency $\omega_\text{MW}=\omega_\text{m}^\text{RC}$, where the effective mechanical resonance $\omega_\text{m}^\text{RC}$ in the $RC$ scenario is given by the positive root of the second-order polynomial
\begin{equation}
    (\omega_\text{m}^\text{RC})^2 = \omega_\text{s}^2\left(1+\omega_\text{m}^\text{RC}R_\text{EM}^\text{opt}C_\text{m}\sqrt{\frac{Z_\text{tx}+R_\text{L}}{R_\text{EM}^\text{opt}}-1}\right).
\end{equation}
This equation is valid as long as $R_\text{EM}^\text{opt}$ varies slowly with $\omega_\text{m}^\text{RC}$.

Conveniently, the discussion of efficiency and added noise in Sections~\ref{sec:eta} and \ref{sec:N} carries over to the present case of the $RC$ circuit with the replacements $R_\text{EM} \rightarrow R^\text{RC}_\text{EM}$, $\eta_\text{e} \rightarrow \eta_\text{e}^\text{RC}$, and $\mathcal{C}_\text{EM} \rightarrow \mathcal{C}^\text{RC}_\text{EM}$. This allows a rather straightforward comparison between the two alternatives, $RLC$ versus $RC$. The absence of resonant enhancement in the $RC$ circuit means that in general only small cooperativities $\mathcal{C}^\text{RC}_\text{EM} < (Z_\text{tx}+R_\text{L})/R_\text{m}$ can be obtained using this circuit; however, if in this way one can achieve the value $\mathcal{C}_{\text{EM}}^\text{opt}$ [Eq. ~\eqref{eq:Copt}] required to impedance match with the optical system then the $RC$ is preferable. Note however that quantum-level suppression of mechanical noise, in the case of electrical-to-optical conversion for specificity, requires $\eta_\text{e}\mathcal{C}_\text{EM} > n_\text{m}$, which in the matched $RC$ case $\mathcal{C}_\text{EM}^\text{RC}=\mathcal{C}_\text{EM}^\text{opt}$ amounts to $\eta_\text{e}^\text{RC} R_{\text{EM}}^\text{opt} > n_\text{m}R_\text{m}$, which in general demands large piezoelectric coupling and/or near-ground-state bath temperatures. As a final remark, a potential benefit of the $RC$ circuit is that it generally has a higher electrical coupling efficiency than the $RLC$, $\eta_\text{e}^\text{RC} \leq \eta_\text{e}^\text{RLC}$, since presumably the first does not suffer from extra loss from inductor $L$.

The process described in this appendix can also be followed in the bare-circuit case where no matching circuit is present by setting $C_\text{T} = 0$ and $\omega_\text{m} = \omega_\text{s}$.


%

\end{document}